\documentclass[12pt]{article}
\usepackage[utf8]{inputenc}
\usepackage[T1]{fontenc}
\usepackage{amsmath,amssymb,amsthm}
\usepackage[letterpaper,margin=1in]{geometry}
\usepackage{graphicx}
\usepackage{booktabs}
\usepackage{fourier}
\usepackage{xcolor}
\usepackage[abspath]{currfile}
\usepackage{bigints}
\usepackage{dsfont}
\usepackage[longnamesfirst]{natbib}
\usepackage{threeparttable}
\usepackage{multirow}
\usepackage{bbm}
\usepackage{comment} 
\usepackage{setspace}
\usepackage{nicefrac}
\usepackage{framed}
\usepackage{booktabs}
\usepackage[colorlinks=true,linkcolor=blue,citecolor=blue]{hyperref}
\usepackage{appendix}
\usepackage{varwidth}
\usepackage{caption}

\newcommand{\cvgp}{\stackrel{p}{\rightarrow}}
\newcommand{\cvgd}{\stackrel{d}{\rightarrow}}

\newcommand{\pr}{\mathbb{P}}

\allowdisplaybreaks

\usepackage{algorithm}
\usepackage{algorithmicx}
\usepackage{algpseudocode}

\title{Selecting Experimental Sites for External Validity\footnote{This paper has previously circulated under the title ``Site Selection for External Validity: Theory and an Application to Mobile Money in South Asia.'' This work was supported by the Bill and Melinda Gates Foundation. Hirano received support from the National Science Foundation through grant SES-2117260. Ravindran is funded by a startup grant at the Lee Kuan Yew School of Public Policy, National University of Singapore. We are grateful for research assistance from Caio Figueiredo, Nihal Mehta, Preksha Jain and Shashank Sreedharan. We thank Seth Garz and participants of workshops at the Advances in Field Experiments Conference, Duke, Florida State, the Penn State - Cornell IO/Econometrics Conference, Tennessee, the World Bank, UN-FAO, Yale, and Y-RISE for helpful comments.}}
\author{
  Michael Gechter, Keisuke Hirano, Jean Lee, Mahreen Mahmud, Orville Mondal, \\
  Jonathan Morduch, Saravana Ravindran and Abu S.~Shonchoy\thanks{Gechter: Pennsylvania State University (email: mdg5396@psu.edu); Hirano: Pennsylvania State University (email: kuh237@psu.edu); Lee: The World Bank (email: jlee20@worldbank.org); Mahmud: University of Exeter (email: m.mahmud@exeter.ac.uk); Mondal: Bates White (email: orville.dm@gmail.com); Morduch: New Yorl University (email: jonathan.morduch@nyu.edu); Ravindran: National University of Singapore (email: saravana@nus.edu.sg); Shonchoy: Florida International University (email: shonchoy@fiu.edu).}
}
\date{\today}

\begin{document}
\maketitle

\begin{abstract}
Policy decisions often depend on evidence generated elsewhere. We take a Bayesian decision-theoretic approach to choosing where to experiment to optimize external validity. We frame external validity through a policy lens, developing a prior specification for the joint distribution of site-level treatment effects using a microeconometric structural model and allowing for other sources of heterogeneity. With data from South Asia, we show that, relative to basing policies on experiments in optimal sites, large efficiency losses result from instead using evidence from randomly-selected sites or, conversely, from sites with the largest expected treatment effects.

\medskip

\thispagestyle{empty}
\end{abstract}

\clearpage
\setcounter{page}{1}\allowdisplaybreaks
\newgeometry{margin=1in}

\clearpage
\section{Introduction}

With few exceptions, field experiments in economics are limited to one or a small number of sites, but researchers and policymakers seek to draw broader conclusions (\citealt{BatesGlennerster}, \citealt{viva:2020}, \citealt{Allcott2012}).\footnote{For example, \citet{Banerjee2015a} jointly design and analyze six RCTs, each in a different country, 
replicating a program that was first developed and analyzed in rural Bangladesh (\citealt{CGAP_ReachingThePoorest}; \citealt{Bandiera_etal_Graduation}). After the study, the program was scaled to more than 2.1 million households in Bangladesh and over 1.1 million households in 14 other countries \citep{JPALgraduation}.
}
Questions about external validity typically arise at the end of the research process when interpreting evidence: can estimates found in the experimental sites be generalized to other places?

We instead develop a framework that incorporates external validity throughout the research design. 
We start with a set of policy choices that will ultimately be made in different settings, and we then work backward to determine where experiments should take place to be most informative about those choices. In our application, where the question is whether and where to promote the adoption of a new financial technology, we show that the ability to generalize depends on the locations of the experiments, and we find large potential gains to selecting sites optimally at the start. The framework also elucidates tradeoffs from adding more experimental sites under a fixed budget, where gains in external validity are balanced against losses in statistical power.

We begin with a social planner who must use the results from a single experiment, or a limited set of coordinated experiments, to determine where and if a program should be implemented across a broader range of sites.
Following \citet{DeatonCartwright}, our notion of external validity departs from the  narrow question of whether a treatment effect from one site is expected to be similar elsewhere. 
Instead, we ask how informative a treatment effect estimated in a given site is about the expected treatment effects in other sites, taking into account knowledge about the structure of economic relationships, the broad range of economic contexts, and previous estimates.
Based on this notion of external validity, we consider the experimental design problem of choosing  the sites in which to run experiments.  
We develop a novel method to choose future experimental sites that
incorporates structural econometric modeling and estimation applied to a pilot data set in combination with other prior judgments about the likely similarity of treatment effects across sites. 
Formally incorporating this pilot data in our design problem leads to a multi-stage or adaptive framework, where past experimental data can influence, and improve, future site selection. 

In our application, we choose a small set of migration corridors from among hundreds of candidates across South Asia in which to experimentally evaluate a program that facilitates the adoption and use of a mobile banking technology.
\citet{Lee2018} report on the original intervention,  which was implemented in a single migration corridor in Bangladesh. The intervention made it easier for migrant workers to send and receive remittances, and an experiment showed an increased volume of urban-to-rural remittances and reduced rural poverty. It is unclear, however, when and where those results might generalize. The goal of the new experiments is to derive policy recommendations for adopting the intervention in all the candidate sites. Specifically, the experiments are chosen to inform a rule for each of the hundreds of migration corridors which maps each corridor's characteristics to an up-or-down recommendation for whether, ultimately, the technology adoption program should be implemented there.
The problem captures one of the essential challenges of external validity: the number of potential corridors is large and their characteristics vary substantially in ways that could alter the effect of the intervention.\footnote{\citet{mura:nieh:2017} discuss the scaleup from experimentation in a pilot experiment in one site to several sites, raising additional concerns about spillover effects and differences in small vs.~large scale interventions, which we do not address directly, although our approach does not preclude including either concern in experimental design. } 

Each new experiment takes place in a ``site,'' which in our context is a migration corridor consisting of an origin and destination district (a geographic area roughly equivalent to a US county). 
While the experiments will provide direct estimates of the average treatment effects (ATEs) for the chosen sites, we wish to be able to extrapolate those estimates to gain knowledge about the overall distribution of site-level ATEs across Bangladesh, India, and Pakistan. 
The ATEs of the intervention may vary across sites due to differences in the characteristics of migrants associated with the site, other site-level characteristics, and further unobserved factors.
In our analysis, we take advantage of the prior experimental study that was conducted in Bangladesh \citep{Lee2018}.
Since many field experiments have a pilot phase, or are replications or extensions of previous experiments in different contexts, such prior information may often be available to the designer of the new experiment. Given the large number of possible sites and the limited scope for experimentation, being able to incorporate this prior experimental evidence adds critical predictive power. The framework allows us to link the prior evidence and the effects at the chosen experimental sites to the effects at the remaining sites.

We propose a quasi-Bayesian framework for this experimental design problem that allows for various prior assumptions about the heterogeneity in site-level effects. 
A randomized experiment at a particular site will provide a noisy measure of the treatment effect at that site, and will also be informative about the treatment effects at other sites that are related to the experimental site through the prior distribution over \emph{all} site effects. The optimal design chooses sites to maximize the overall information gained, as measured by the improvement in policy decisions that they enable.  

One version of this approach uses a ``smoothing'' prior that 
encodes the intuition that sites with similar characteristics (such as origin--destination distance and average household income) should have similar treatment effects. 
If there are many site characteristics, however, the analysis can be sensitive to the inclusion and weighting of different characteristics. 

A second approach is structural. We introduce a model for migrant remittance decisions that provides a microeconomic foundation for site-level effect heterogeneity and allows us to link the original experimental data to the experimental design problem. 
In particular, we specify and estimate a structural model using the data from \cite{Lee2018}, which we regard as a pilot experiment, and we then use this to construct a joint prior for the ATEs at all the prospective experimental sites. 
This allows us to draw on the detailed individual-level data in the pilot experiment to assess the {\em a priori} similarity of different sites. 
While a number of researchers have used structural modeling methods to analyze data from randomized controlled trials in development economics (see, for example, the survey articles by \citet{todd:wolp:2010, todd:wolp:2023}), the potential for applying structural approaches to {\em experimental design} has been relatively unexplored.  

Incorporating structural modeling can greatly increase the prior information available for the experimental design step, but the resulting design could be fragile to misspecification of the structural model.  
For this reason, we also develop a hybrid joint prior for the potential experimental sites that combines the two approaches above, mixing between the prior generated by the structural model and the smoothing prior that allows sites with similar characteristics to have similar predicted treatment effects. 

We use the empirical setting to demonstrate how our procedure works and to compare it to alternatives.
We measure the effects of the intervention, and the overall implications of the site selection, through a constant elasticity of substitution social welfare function (c.f.~\citealt{deaton1997}, \citealt{alderman2019}) for home household members, where we proxy individual utility by per capita expenditure.

Experimentation is sometimes used to demonstrate the viability of a concept, often by implementing interventions in places where they are likely to succeed, but our approach highlights that welfare gains are instead concentrated among experimental sites where the suitability of the treatment is
{\em a priori} 
fundamentally uncertain.

We find that the sites selected through the welfare-optimizing approach are relatively centrally located in the space of characteristics of all sites, where our model specifies the appropriate notion of distance in characteristic space. 
This is because the optimal sites for experimentation are those that are most informative about policy decisions in the other sites---so the optimal sites will tend to be similar to other sites along relevant dimensions. 
The optimal choices can thus be very different from sites where the intervention is likely to have the highest expected treatment effect or the sites where the efficacy of the intervention can be best demonstrated.
Indeed, experimenting in the sites predicted to have the highest treatment effect under our prior and naively extrapolating the results can yield a substantial \emph{negative} welfare effect from experimentation. 

Relative to using only the information from the \citet{Lee2018} experiment, as embodied in the prior, there are large welfare gains to additional experimentation in other sites, but we show that the marginal gains diminish in the number of experimental sites. 
The data also show that rule-of-thumb procedures for selecting sites can be very inefficient relative to our welfare-optimizing approach. 
To demonstrate this, we implement a heuristic procedure (loosely modeled after a recommendation in \citealt{Allcott2012}) that selects sites randomly, computes the average of ATEs across sites, and makes a uniform recommendation to implement the program in all sites or none based on whether the average of ATEs exceeds the cost of the program. 
We show that experimenting in a single randomly selected site can result in almost no welfare benefit from experimentation.
Even with more sites, welfare is always well below the optimum.

\subsection{Background and Related Literature}

Our work contributes to a growing literature on external validity and generalizability in economics, statistics, and other fields. 
Some authors, such as \citet{prit:sand:2015}, \citet{viva:2020}, and \citet{meag:2019}, have primarily viewed external validity as holding when treatment effects are relatively similar across different contexts, and focus on measuring the extent of departure from homogeneity.
Other work, such as \cite{stua:etal:2011}, \citet{tipt:2013,tipt:2014},  and \citet{omui:hedges:2014},
focuses on generalizing across contexts by adjusting for differences in the distribution of covariates, under the assumption that conditional average treatment effects are stable. 
Unlike these papers, we allow for unobserved factors that could lead to differences across sites.

The problem of transporting causal estimates and inferences across contexts has also been studied by \citet{hotz:imbe:mort:2005}, 
\citet{gech:2023}, \citet{andr:etal:2023}, 
\citet{kuan:etal:2018}, \citet{dehe:pope:sami:2018}, \citet{Gechter2019}, \citet{adja:chri:2022}, \citet{kall:zhou:2021}, 
and \cite{menz:2023}, among others. 
Our setup is particularly closely related to the problem considered in 
\citet{ishihara2021evidence} and \citet{olea2023decision}. 
While most of these papers impose {\em a priori} bounds  on the differences between site-level populations to facilitate extrapolation, 
our strategy is to craft a Bayesian prior that allows us to model varying degrees of similarity across sites. 

\citet{glen:2017} and \citet{chas:kapo:2021}, among others, discuss the importance of considering external validity when designing experiments. 
\citet{tipt:2014} and \cite{omui:hedges:2014} discuss how experiments should be designed so that the distribution of covariates in the experiment is sufficiently rich for covariate-based extrapolation to other contexts. 
There has been relatively little work on experimental design that accounts for more  general forms of heterogeneity across sites. 
Our approach is Bayesian and sequential, in that we use data from a pilot experiment to inform our choice of future experimental sites. 
\citet{fina:pouz:2023} consider sequential experimentation in one site or context, using various sources of prior information including past data from other contexts to inform the experimental stopping rule and policy recommendation. 
Other recent work in sequential experimental design in economics includes \citet{hahn:hira:karl:2011}, \citet{tabo:2018}, \citet{kasy_adaptive_2021}, 
\citet{caria_adaptive_2021}, 
and \citet{athey_shared_2021}.
Unlike many of these other works, our key experimental design choice is in the selection of experimental sites.\footnote{Site selection is an important concern for clinical trials in medicine; see for example \citet{pott:etal:2011} and \citet{hurt:etal:2017}. However, this literature has primarily focused on choosing sites to ensure sufficient enrollment of experimental subjects, and timely and high-quality implementation of experimental treatments. Related issues are discussed in the context of social science field experiments by \citet{glen:2017}.}

We focus on settings with few experimental sites because even in very well-resourced trials the number of manageable sites is small relative to the number of potential sites.
The total number of trials evaluating the aforementioned program targeting extreme poverty was only 10.
Other examples of multi-site trials have similarly few sites.
\citet{Angrist5countrystudy2023} and \citet{AngristBergmanMatsheng} report on six RCTs of remote instruction for primary school children during the school shut-downs caused by COVID-19 and \citet{CusolitoDarovaMcKenzie} evaluate an intervention in six Balkan countries that provides skills to help small businesses compete in the export sector. 

In the next section, we explain our proposed methodology at a general level that could be ported to other applications. We set up experimental site selection as a formal decision-theoretic problem that embodies our notion of external validity, and propose a Bayesian approach that allows us to tractably incorporate prior experimental evidence and structural econometric analyses. Then, in Section \ref{sec:model}, we begin to focus on our specific application by developing a model for migrant labor supply and remittance choices. This model is taken to the prior experimental data in Section \ref{sec:estimation}. Section \ref{sec:application} then implements the full experimental design analysis for Bangladesh, Pakistan, and India.  

\section{Experimental Site Selection for External Validity}
\label{sec:experimental_design}
\subsection{Setup and Design Problem}

We first introduce a framework for analyzing experimental site choice with external generalizability as its key goal, by formalizing a policy choice problem across all potential experimental sites. 
We have a set of potential experimental sites
$s=1,\ldots,S$. 
In our application, each site $s$ represents a migration corridor consisting of a home--destination district pair.
For each site $s$, we observe site characteristics $V_s \in \mathcal{V}$.
The variable $V_s$ could be high-dimensional, for example to capture the distributions 
of demographic and other variables at the site. 

The true (but unknown) effect at site $s$ is $\tau_s$. Let $\tau = (\tau_1,\dots,\tau_S)^{\prime}$. 
If we select a site $s$ and run a randomized controlled trial at that site, we will obtain an estimate  $\hat{\tau}_s$ of the site-level effect, where 
\[\hat{\tau}_s = \tau_s + \epsilon_s, \quad \epsilon_s \stackrel{\text{ind}}{\sim} N(0,\sigma_{\epsilon,s}^2).\] 
We assume that the $\sigma^2_{\epsilon,s}$ are known (or user-specified and fixed in advance) for the remainder of the analysis. In our application, this assumption is plausible, as we fix the sample size at each site in advance, and we have prior data that is informative about individual-level treatment effects.\footnote{Note that any clustering in the sampling design of the trial will be reflected in $\sigma_{\epsilon,s}$ but does not affect the estimand $\tau_s$.}

The experimental design problem is to choose a {\em subset} of sites, 
$\mathcal{S} \subset \{1,\dots,S\}$, on which to run experiments. 
We are given a feasible set of subsets $\mathcal{A}$, and must choose 
$\mathcal{S} \in \mathcal{A}$.
In the case where we are choosing a single site, the choice set $\mathcal{A}$ would consist of all the singleton sets $\{1\}, \{2\},\dots,\{S\}$. 
More generally, the feasible set $\mathcal{A}$ can incorporate various constraints; for example it may limit the total number of experimental sites, or rule out certain combinations of sites. 
After choosing $\mathcal{S}$ we will observe $\hat{\tau}_s$ only for $s \in \mathcal{S}$. 
Let $\hat{\tau}_{\mathcal{S}} := \left\{ \hat{\tau}_s : s \in \mathcal{S} \right\}$
denote the observed experimental estimates for the sites in $\mathcal{S}$.

Although we are constrained to experiment on a limited subset of the sites, we want to choose the experimental sites to maximize the information gained about the effects across all sites. 
We operationalize this goal 
by focusing on the entire vector of site-level effects $\tau = (\tau_1,\dots,\tau_S)^{\prime}$ and posing a hypothetical policy choice problem as a sequential decision problem. 
The social planner, after observing $\hat{\tau}_s$ for the subset of sites $s \in \mathcal{S}$, chooses treatments for every site $s \in \{1,\dots,S\}$, with the goal of maximizing aggregate welfare. 
This sequential decision problem is 
outlined in Figure \ref{fig:decisionproblem1}. 

\begin{figure}[ht]
    \centering
        \caption{Basic Site Selection Problem}

\begin{framed}
\begin{enumerate}
	\item Observe all the site characteristics $V = (V_1,\dots,V_S)$.
	\item Choose $\mathcal{S}  \in \mathcal{A}$. 
	\item Observe $\hat{\tau}_{\mathcal{S}}$. 
	\item Choose a vector of binary treatments $T = (T_1,\dots,T_S)$. 
	\item Evaluate $T$ by a social welfare function $W(\tau,T)$. 
\end{enumerate}
\end{framed}
    \label{fig:decisionproblem1}
\end{figure}

The key design decision is in Step 2. From the standpoint of the experimental designer, Steps 3--5 are prospective, and serve to quantify the designer's objectives. 
Note that, in Step 4, we are choosing whether or not to implement the intervention separately across {\em all} sites, not just the experimental sites in $\mathcal{S}$. 
Thus a site-selection subset $\mathcal{S}$ is desirable if it will allow the social planner to make good policy decisions across observed and unobserved sites.
This prospective policy choice problem embodies our concern for the external validity or generalizability of the choice of a small number of experimental sites.  
As a consequence, we are interested in the full vector $\tau = (\tau_1,\dots,\tau_S)^{\prime}$ of site-level treatment effects, not some aggregate such as the average effect across all sites. 
We will discuss the specification of the social welfare function $W(\tau,T)$ in more detail below, but before doing so, we will extend this basic framework to allow us to incorporate prior information in the form of 
a pilot experimental study, and develop a solution concept for the extended problem.\footnote{We assume the vector of site-specific treatment effects $\tau$ is invariant with respect to the vector of implementation decisions $T$. It is in this sense that we abstract from one dimension of ``scaling up,'' where, say, the number of treated sites may modify treatment effects through general equilibrium effects. This consideration could in principle be added to our framework.}

The basic decision problem outlined in Figure \ref{fig:decisionproblem1} is difficult in part because the decision-relevant parameter $\tau$ may be high-dimensional, yet the experimental process will provide direct information on only a limited number of its elements. 
This reflects the general challenge of external validity in empirical studies, but in cases where the total number of experimental sites is limited, it will be especially helpful to incorporate additional sources of information and modeling assumptions. 
We will do so by using prior experimental evidence in conjunction with structural modeling techniques that facilitate extrapolation of likely effects across all sites. 

In our application to evaluating the impacts of mobile money training, we can take advantage of the 
previous experimental study conducted by \citet{Lee2018} on a different experimental site.  
We can view this as an additional site $s=0$. For this site, we have observed not only its simple 
treatment effect estimate $\hat{\tau}_0$, but also a rich micro-level data set with individual characteristics, assigned treatments, and outcomes which permit more detailed analysis and extrapolation to other sites.
We therefore augment the original decision problem with an initial step in which the previous randomized control trial, which we will refer to as the pilot RCT, is made available to the designer. This extended site selection problem is outlined in Figure \ref{fig:decisionproblem2}. 
\begin{figure}[ht]
    \centering
        \caption{Extended Site Selection Problem}

\begin{framed}
\begin{enumerate}
    \item[0.] Pilot RCT: for site $s=0$, observe individual characteristics, randomized treatments, and experimental outcomes. 
	\item Observe all the site characteristics $V = (V_1,\dots,V_S)$.
	\item Choose $\mathcal{S} \in \mathcal{A}$. 
	\item Observe $\hat{\tau}_{\mathcal{S}}$. 
	\item Choose a vector of binary treatments $T = (T_1,\dots,T_S)$. 
	\item Evaluate $T$ by a social welfare function $W(\tau,T)$. 
\end{enumerate}
\end{framed}
    \label{fig:decisionproblem2}
\end{figure}

\subsection{Bayesian Solution} 

We propose to solve the extended experimental design problem by adopting a (quasi-) Bayesian ``preposterior'' analysis, in which a prior distribution for the unknown vector $\tau$ is used to evaluate the expected welfare gains from alternate possible choices of the site selection subset $\mathcal{S}$. 
Specifically, after Steps 0 and 1 in Figure \ref{fig:decisionproblem2}, we will construct a prior distribution $\Pi$ for the $S$-dimensional vector $\tau$. 
The prior $\Pi$ will be based on the preliminary experimental analysis and other modeling assumptions, as well as the vector of site characteristics $V = (V_1,\dots,V_S)$. 
(For notational ease, we suppress the dependence of $\Pi$ and other quantities on the characteristics $V$ in the sequel.) 
We will discuss the specification of the prior in more detail in the next subsection, but for now we allow for general forms of $\Pi$. 

We can solve the problem using backward induction. 
Specifically, 
given a choice for $\mathcal{S} \in \mathcal{A}$ and observation of $\hat{\tau}_{\mathcal{S}}$, we update the prior $\Pi(\tau)$ for the vector of treatment effects to its posterior
$\Pi\left(\tau \mid \hat{\tau}_{\mathcal{S}} \right)$.
The vector of treatments that maximize posterior expected welfare is given by 
\begin{equation}\label{eq:postoptreatment}
T^* = T^*(\Pi, \hat{\tau}_{\mathcal{S}}) = \arg\max_{T} \int W(\tau,T) d\Pi\left( \tau \mid \hat{\tau}_{\mathcal{S}} \right),
\end{equation}
where the maximization is over all $S$-vectors $T$ of binary treatments. 

The optimal choice  $\mathcal{S}^*$ of the site-selection subset can then be obtained by solving: 
\begin{equation}
\label{eq:preposterior} 
\mathcal{S}^* = \arg\max_{\mathcal{S} \in \mathcal{A} } \int \left[ \int  W(\tau, T^*(\Pi, \hat{\tau}_{\mathcal{S}}) )  dF\left( \hat{\tau}_{\mathcal{S}} \mid \mathcal{S}, \tau \right) \right]  d\Pi(\tau).
\end{equation}
To understand this equation, it may help to consider how to evaluate the right hand side of Equation (\ref{eq:preposterior}) through simulation. 
We could generate draws for the site effects $\tau$ from the prior $\Pi(\tau)$, reflecting the outer integration. For each draw of $\tau$, we can then simulate the observed site-level estimates $\hat{\tau}_{\mathcal{S}}$ from the sampling distribution $F(\hat{\tau}_{\mathcal{S}}\mid \mathcal{S},\tau)$ that appears in the inner integral, find the optimal treatment choice $T^*$ based on the observed estimates, and then evaluate the welfare of that choice. Repeating this process and averaging over the draws will approximate the {\em ex ante} or ``preposterior''\footnote{See \citet{berg:1993}, Chapter 7, for an introduction to  preposterior Bayesian analysis.} 
expected welfare that the policymaker can anticipate for a particular choice of experimental sites $\mathcal{S}$, incorporating both the prior uncertainty about the true $\tau$ and the sampling variability of the experimental estimates $\hat{\tau}_\mathcal{S}$.\footnote{See Appendix \ref{sec:algorithm} for a step-by-step walkthrough of the 8-pointed algorithm used in our application.} \footnote{In principle we can also allow for randomization across site-selection subsets, but given our solution concept and specifications for the priors and payoffs, the optimal choice will be generically nonrandomized.} 
To make this computationally feasible, the posterior distribution $\Pi(\tau\mid\hat{\tau}_{\mathcal{S}})$, which appears in the definition of $T^*$ in Equation (\ref{eq:postoptreatment}), should be easy to calculate. 

While the actual solution to Equation (\ref{eq:preposterior})  will depend on the specific choice of the prior $\Pi$ and the welfare function $W(\cdot)$, some intuition is possible even in this general form. 
Consider including a potential site $s^\prime$ in the set $\mathcal{S}$ of experimental sites. 
Observing $\hat{\tau}_{s^\prime}$ will typically improve the decision $T_{s^\prime}$ for whether or not to treat that site based on the precision of the estimate $1/\sigma_{\epsilon,s^\prime}^2$. It can also improve the decision for other sites $s \ne s^\prime$ if there is sufficient predicted similarity  between $\tau_s$ and $\tau_{s^\prime}$ , represented as cross-site dependence in the joint prior $\Pi$, so that there is learning (posterior updating) about $\tau_s$ from $\hat{\tau}_{s^\prime}$. 
However, such cross-site learning is only relevant for sites $s$ that are ``marginal'' in the sense that it is not already clear from the prior whether or not the treatment is effective at that site. 
Taking these considerations together, we expect the solution in (\ref{eq:preposterior}) to choose a subset of sites that, together, maximize learning about the welfare-relevant aspects of $\tau_s$ across all the {\em a priori} marginal sites.

\subsection{Prior Specifications and Updating Rules} 
\label{subsec:priorspec}
Next, we discuss the specification of the prior $\Pi$. 
To solve Equations (\ref{eq:postoptreatment}) and (\ref{eq:preposterior}) above, a key 
object is the posterior $\Pi(\tau \mid \hat{\tau}_{\mathcal{S}})$, which represents the updated beliefs about the entire vector $\tau = (\tau_1,\ldots,\tau_S)$ based on the observation of a limited set of sites $\mathcal{S}$. 
If the prior $\Pi$ is independent across the elements $(\tau_1,\ldots,\tau_S)$, then it will not be updated for any components that are not selected in $\mathcal{S}$. 
And if we are restricted to choose a small number of sites so that the number of elements in $ \mathcal{S}$ is much smaller than $S$, then the specification of the prior $\Pi$ will be important for updating beliefs about unobserved sites. 
This motivates using the observed site characteristics $V_s$, combined with modeling assumptions about the relationship between $V_s$ and $\tau_s$,  to generate a prior with dependence across elements of $\tau$. 
We will consider a number of priors, but all of them can be expressed as an $S$-dimensional multivariate normal distribution 
\[ \Pi(\tau) = N_S\left( \mu_{\tau} , \Sigma_{\tau}\right),\]
where the choice of $\mu_{\tau}$ and $\Sigma_{\tau}$ will determine how the posterior extrapolates from observed to unobserved sites.  

\subsubsection{Smoothing Prior}
\label{subsubsec:smoothingprior}
One class of priors we consider places smoothness restrictions on the relationship between $V_s$ and $\tau_s$, which in turn implies that sites with similar characteristics will be expected to have similar treatment effects. 
Specifically, let 
\[ 
	\tau \mid V \sim  N_S\left( \mu, \Sigma(V) \right) = \Pi_1, 
\] 
where $\mu$ is some $S\times 1$ vector of prior means, and $\Sigma(V)$ is a covariance matrix with elements 
\[ 
	\mathbb{Cov}(\tau_s,\tau_{s'}) = c_0 \cdot \exp\left( - c_1^{-1} \left\| h(V_s) - h(V_{s'})\right\|^2  \right),
\]  
where $h(V_s)$ is some low-dimensional function of the characteristics $V_s$.
This will result in the posterior for $\tau_s$ being a weighted average of estimates $\hat{\tau}_{s'}$ based on the distance of $h(V_s)$ to $h(V_{s'})$, similarly to conventional kernel regression estimators.  
Here $c_0>0$ and $c_1>0$ are tuning parameters for the prior;
$c_1$ controls the amount of local smoothing, and $c_0$ controls the overall scale of the variance matrix. 

\subsubsection{Structural Prior}
The smoothing prior above is a standard one in the literature on Gaussian smoothing methods (e.g., \citet{rasm:will:2006}). 
However, this conventional approach may be sensitive to the choice of the function $h(V_s)$ and the tuning parameters, especially in settings like ours where the dimension of $V_s$ is high. 
Moreover, we would like to take full advantage of the rich individual-level data in the pilot RCT, and insights from microeconomic modeling, to incorporate as much prior information as possible to aid in the choice of sites. 
To do so, we will construct an alternate prior based on a structural model of a migrant's choice of labor, consumption, and remittances. 
This model is fully detailed in Section \ref{sec:model}. 

For now, to explain how the structural analysis fits into our design problem, suppose we have an economic model for the underlying outcomes, where the model parameters are given by a vector $\theta \in \Theta$. 
For every site $s$, the structural model predicts average treatment effect 
\[ \tau_s = g(\theta,V_s),\]  
and we use $g(\theta,V)$ to denote the $S$-vector of predicted average treatment effects. 

Based on the pilot experiment (in Step 0 of Figure \ref{fig:decisionproblem2}), we have a posterior distribution for $\theta$, which we will denote $p(\theta)$. 
The distribution $p(\theta)$ combined with the mapping $g(\theta,V)$ generates a distribution for the vector $\tau$. This could be implemented by drawing $\theta \sim p(\theta)$ and then forming $\tau = g(\theta,V)$. 

To maintain tractability, we will assume that the induced prior for $\tau$ can be approximated well by a multivariate normal distribution with mean $\tilde{\mu}$ and variance $\tilde{\Sigma}$: 
$$\tau\mid V \sim N_S\left( \tilde{\mu},\tilde{\Sigma}\right) = \Pi_2.$$

\subsubsection{Mixed Prior}
\label{subsubsec:mixed_prior_intro}
The structural prior imposes the structural model exactly across all sites, which is likely to be unrealistic. We would like to allow for some discrepancy between the structural model $g(\theta,V)$ and the true effects $\tau_s$.
We do so by mixing the smoothing and structural priors. 
Specifically, for some value $b \in (0,1)$, we take the prior to be normal with mean $\tilde{\mu}$ (given by the structural model) and variance matrix a convex combination\footnote{An alternative would be to start with the structural prior and add some fraction of the smoothing variance to represent deviations from the structural model, leading to a prior of the form $N_S(\tilde{\mu},\tilde{\Sigma} + \tilde{b}\Sigma(V))$. More generally, we could consider various linear functions of the two variance components.}  of the smoothing and structural variances: 
\[ \Pi_{3} = N_S\left( \tilde{\mu}, b \tilde{\Sigma} + (1-b) \Sigma(V) \right). \] 

\subsubsection{Posterior Updating}\label{sec:postupdate}

All of the choices for the prior we outlined above are multivariate normal distributions for the vector $\tau$: for some mean vector $\mu_{\tau}$ and variance matrix $\Sigma_{\tau}$, the priors specify that $\tau \sim N_S(\mu_{\tau},\Sigma_{\tau})$.
Recall that every $\hat{\tau}_s$ is also normally distributed and centered at $\tau_s$. 
This joint multivariate normality implies that the posterior distribution of the entire vector $\tau$ given the observed estimates at the {\em selected} sites, $\hat{\tau}_{\mathcal{S}}$, will also be multivariate normal. 
In Appendix \ref{sec:posteriorexample}, we provide explicit expressions for these posterior means and variances, and illustrate the posterior updating formulas in a simple example. 

\subsection{Specification of Welfare Function}

To complete the specification of the experimental design problem, we need to specify  the welfare function $W(\tau,T)$.
The general solution method outlined above can be applied to any reasonable function $W$. 
Our analysis below will be based on the following welfare function: 
\[
    W(\tau,T) = \sum_{s=1}^S T_s \left( \tau_s - \text{cost}_s \right),
\]
where $\text{cost}_s$ is the cost associated with implementing the intervention at site $s$, in the same units 
as the treatment effect $\tau_s$. 
We give more detail on how we define treatment effects and specify costs for our application below. 
Note that, given any predictive distribution for $\tau_s$ with mean $\bar{\tau}_s$, the optimal treatment is 
given by 
\begin{align}
    \label{eq:optimal_site_treatment}
    T_s^* = {\bf 1}( \bar{\tau}_s \ge {\tt cost}_s).
\end{align}
An ``oracle'' treatment rule that uses knowledge of the true value of $\tau_s$ would simply assign treatment if the treatment effect $\tau_s$ is greater than or equal to $\text{cost}_s$.

\section{A Model of Remittances}\label{sec:model}

Section \ref{sec:experimental_design} has proposed a decision-theoretic formulation of the adaptive site-selection problem, and outlined its solution using a tractable class of prior distributions. 
A key element of our approach is to incorporate structural modeling to link the pilot experiment to the full set of potential future experimental sites. 
In this section, we develop an economic model of optimal labor supply, consumption, and remittance choices by a migrant whose cost of remitting income to their home household may be affected by the experimental intervention. 
We will then estimate this model in Section \ref{sec:estimation} and use it in our experimental design process. 

\subsection{The Migrant's Problem}
\label{subsec:model-given-pr}

Consider a worker who has migrated from his or her home (in a rural area) to a city to work. 
Some of the earnings from their work will be consumed by the migrant, and the rest sent back to their family for consumption at home.
The migrant must choose how much they work, how much of their earnings to consume, and how much to remit back home to their family. 
Remittances may be sent using traditional means, such as via an agent or a friend who may be returning to the same village as where the migrant's family lives. 
Alternatively, the migrant can remit using a new method: ``mobile money.''
The cost of sending remittances by mobile money does not vary with the distance between sender and recipient, and may be cheaper than traditional means of remittance. 
But the migrant may not be aware of the mobile money option, or know how to use it. 

Suppose there is now a ``treatment'' that makes migrants aware of the potential benefits of mobile money and helps them learn how to use it. 
We wish to study the effects of such interventions on outcomes such as the amount remitted or consumption per capita of the migrant's family members back home. 

Let $C_m$ and $C_h$ represent consumption by the migrant and  per-capita consumption at home, respectively, 
and let $L_m$ be the leisure of the migrant (in hours per day). 
We use $w_m$ to denote the migrant's hourly wage, $y_h$ for total income at home, and $R$ for the remittance sent by the migrant. 
There are multiple prices: $p_h$ is the price level at home relative to the city where the migrant works; 
and $p_r$ gives the units of consumption that must be remitted by the migrant in order for their home family to receive one unit of consumption.
Finally, let $a_h$ denote the size (number of adults and children) of the migrant's family at home.
In this section, we omit an  ``$i$'' subscript indexing migrants. The $i$ subscript will be introduced later when discussing the data and estimator. We will generally capitalize choice variables, and use lower case for non-choice variables.

The migrant's utility needs to possess a few key properties. 
We require the migrant to explicitly account for per capita expenditure by their family back home and we need the model to produce a remittance decision which leads the migrant to optimally remit nothing under appropriate conditions on the price to remit, while never remitting their entire income, both of which are features of the data.
The utility specification below has these features and also leads to closed form expressions for all optimal quantities, which facilitates computation.
The migrant chooses their consumption, leisure, and remittances to maximize utility, given by 
\begin{align*}
    \ln\left(C_m^\alpha L_m^{1-\alpha}\right)+\lambda a_h \ln\left(C_h^\eta\right),
\end{align*}
subject to the following constraints: 
\begin{align*}
	C_m, L_m, R &\ge 0, \\
	L_m & \le 24, \\
	C_m &= w_m (24-L_m) - p_r R,  \\
	a_h\cdot C_h&=y_h+\frac{R}{p_h}.
	\end{align*}
Here $\lambda$, $\alpha$, and $\eta$ are preference parameters.  The migrant's utility depends on both their own consumption $C_m$ and leisure $L_m$, as well as home consumption $C_h$. The parameter $\lambda$, which controls the relative weight placed on home utility vs.~migrant's utility, will be specified in more detail below.

There are two key budget constraints in this model. 
The migrant receives income $w_m(24-L_m)$ based on wages and hours worked 
(recall that $L_m$ is leisure of the migrant).  
This income can be allocated to the migrant's consumption, or sent home as remittance $R$ with price $p_r$. The price to remit a single unit of consumption, $p_r$, is  assumed to be at least one, i.e.~$p_r\ge 1$. This accounts for any transaction fees a migrant may have to incur when remitting. The home consumption $C_h$ depends on home income, plus any received remittance which has been normalized by the price index $p_h$, where typically $p_h \le 1$.
In Appendix \ref{subsec:model-solution-given-pr} we provide the optimal remittance amount $R^*$, as well as optimal migrant consumption and leisure $(C_m^*, L_m^*)$ given $p_r$.
Below, we develop a stochastic specification for $p_r$ which will lead to a distribution for remittances. 
 
\subsection{Stochastic Specifications and Remittance Mode Choice}
\label{subsec:stochastic-spec}

Here we take the basic model and build an empirical specification, which specifies the distributions of preferences and some other components of the model, and extends the model to include the choice of remittance mode. 

The preference parameter $\lambda$, which determines the relative weighting of utility derived from migrant and home consumption, is specified as: 
\[
\lambda =\exp(\phi_0+\phi_1 \text{male}),
\]
where ``$\text{male}$'' is an indicator for the migrant's gender, discussed as a potential source of heterogeneity in treatment effects on remittance sending in \cite{Lee2022}.

Next we specify the remittance price $p_r$. 
We suppose that the migrant can send remittances either through traditional means, or using the mobile money service, and that there is a treatment (the training intervention) $T \in \{ 0,1\}$, which can affect the non-monetary cost of using mobile money. 
Let $p_{r,\text{trad}}$ be the price of traditional remittance, and $p_{r,\text{mm}}(t)$ for $t \in \{0,1\}$ be the price of mobile money remittance after intervention $t$.
For the price of traditional remittance, we specify that 
\[ p_{r,\text{trad}} = 1+d\epsilon,\]
where $d$ is the distance between the migrant and home locations, and $\epsilon \ge 0$ is a individual-specific stochastic shock.\footnote{When estimating the model as described in Section \ref{sec:estimation}, we set $d=1$ but account for other distances in the design stage.}
Thus, the cost of remitting one unit is equal to one (the remittance itself) plus a positive amount $d \epsilon$ that increases with distance, capturing the increased risk of loss or direct cost of sending money a long way using traditional means.
The shock $\epsilon$ reflects individual heterogeneity in the cost of remitting by traditional means with a low shock arising, for example, if the migrant can send money with a trusted friend who happens to be visiting home. 
We assume that $\epsilon$ follows an exponential distribution with mean $\bar{\epsilon}$.  
This ensures that $\epsilon$ is nonnegative in accordance with its interpretation as the increase in remittance cost with an additional unit of distance. 

For the price of mobile money remittance, we specify that 
\[	
    p_{r,\text{mm}}(t) = 1 + \gamma + \exp(\delta o + \psi t) \xi. 
\]
Here, $\gamma \ge 0$ is the known monetary per-currency-unit-sent cost of remitting using mobile money.
We describe the source of $\gamma$ in each country in Appendix \ref{subsec:admindata}.
In addition to the formal price of mobile money, we assume there is a positive hassle cost represented by the term $\exp(\delta o + \psi t)\xi$. 
The price of remittance by mobile money depends on operator density $o$, which captures the ease with which mobile money transactions may be implemented. 
Use of mobile money to send remittances requires a physical agent at origin to withdraw funds.
The greater the availability of such agents or operators, the easier it is to remit using a mobile money service.
The parameter $\delta$ captures the effect of greater operator density on the price to remit using mobile money.
The term $\psi$ measures the change in the effective cost of mobile money remittance if the migrant receives the training intervention. 
Finally, we specify an individual-specific stochastic term $\xi$ that generates variation in the hassle cost of mobile money remittance. 
By analogy with the specification for $\epsilon$, we assume that $\xi$ is exponentially distributed with mean $\bar{\xi}$, and that $\xi$ is independent of $\epsilon$.\footnote{When we estimate the model parameters, $\bar{\xi}$ will not be separately identifiable from $\delta$ and $\psi$. In the experimental data, there is no variation in the operator density $o$, so we drop the term $(\delta o)$ for estimation purposes, and then calibrate the value of $\delta$ as described in more detail below.}
The stochastic terms $\epsilon$ and $\xi$ generate variation across individuals in their choice of remittance mode, as well as amount optimally remitted. 

The migrant will choose whichever remittance mode is cheapest, so the effective 
price of remittance will be 
\[ \min \{ p_{r,\text{trad}}, p_{r,\text{mm}}(t)\}. \]
Letting $M$ denote the remittance mode choice, with $M=1$ for mobile money and $M=0$ for traditional remittance, we have $M^* = {\bf 1}\left\{ p_{r,\text{trad}}  \ge  p_{r,\text{mm}}(t)\right\}$.

After augmenting the basic model with the remittance mode choice, and specifying the stochastic determinants of preferences and remittance prices, the complete structural model implies a joint distribution for remittance amount $R$ and remittance mode choice $M$ given household/migrant characteristics (and given model parameters).\footnote{We provide the specific expressions in Appendices \ref{subsec:model-solution-stochastic} and \ref{subsec:model-solution-m}.}
This joint distribution will be the basis for the estimation routine described in Section \ref{sec:estimation} when we use data from a randomized trial to fit the model described here. 

\section{Fitting the Structural Model to the Pilot Experiment}
\label{sec:estimation}

\cite{Lee2018} conducted an experiment in Bangladesh in 2014--2016 to evaluate an intervention that trained migrant workers in  Dhaka and their family members at home to use a digital money transfer service. 
The study tracked 813 migrant-home family pairs whose home families were located in the district of Gaibandha and measured multiple outcomes of interest for both the migrants and their family members. The families chosen for the study were all ``ultra poor'' as defined by local administrative standards, and regularly received remittances from migrant family members in Dhaka. The experiment randomly assigned half the sample to a treatment arm that was given detailed training on how to use the mobile money application bKash to remit money. Members of the control arm were not given any training. 

We use data from this experiment to fit the model from Section \ref{sec:model}, as the first step in defining a Bayesian prior for the experimental design process described in Section \ref{sec:experimental_design}.
Since we do not model migrant unemployment, we drop the 34 households with migrants who report no earnings.
We also drop the 116 households reporting no income at home because we believe these are in fact misreports.
A large majority of these households have high consumption levels that greatly exceed the amount of remittances they receive, but do not report any debt.
The loss in sample size is reflected in the uncertainty about the structural parameters as well as the implied site-level treatment effects.
Appendix Table \ref{table:bkash_summary_stats} displays summary statistics for the sample used for model estimation. 

To take the model developed in Section \ref{sec:model} to the prior experimental data, we assume that the 
households $i=1,\dots,n$ in the experimental data are a random sample from the relevant population at that site. 
For each household we observe characteristics 
\[ X_i = (p_{hi}, a_{hi}, w_{mi}, y_{hi}, \text{male}_{mi}, o_{i}, d_{i}, T_i), \] 
and we observe the household's amount remitted, and mode choice conditional on positive remittance: 
\[ Y_i = (R_i, M_i \cdot {\bf 1}(R_i > 0)).\] 
The full set of model parameters is $\left\{ \psi, \phi_0,\phi_1, \bar{\epsilon},\bar{\xi}, \alpha, \gamma, \eta, \delta \right\}$.  
While most of these parameters will be estimated by making explicit use of the model, as described below, $\gamma, \eta$, and $\delta$ will be normalized or estimated separately as described in Appendix \ref{sec:normalized_params}.
We collect the remaining parameters into a vector $\theta$: 
\[ \theta = \left( \psi, \phi_0,\phi_1, \bar{\epsilon},\bar{\xi}, \alpha \right)^{\prime}, \]
and we will suppress the dependence of model-implied quantities on the other parameters  for notational simplicity in what follows. 

For the experimental site, there is a distribution of the covariates $F_X$, i.e.~$X_i \stackrel{\rm iid}{\sim} F_X$. 
The distribution $F_X$ will in turn induce a distribution of outcomes $Y_i$ through the structural model $F_{Y|X}(Y_i | X_i,\theta)$.\footnote{We will also sometimes suppress the dependence 
of quantities on $F_X$ in the expressions below for notational convenience.}
At the other sites, the distribution of covariates will differ from $F_X$, and this will lead to variation in the predicted site-level average treatment effects generated by the structural model. 

\subsection{Minimum Distance Estimator}\label{sec:mdestimator}\label{sec:mdestimatorspec}

We estimate the parameter vector $\theta$ by matching a vector of model-implied quantities $q(\theta)$ to corresponding sample conditional moments $\hat{\pi}$. 
First, we consider the conditional distributions of remittances $R$ and remittance mode $M$. 
We fix a set of remittance values 
\[ \mathcal{R} = \left\{ r_1, r_2,\dots, r_{d_r} \right\} = \left\{0, 50, 100, 125, 150, 175, 200 \right\}.\] 
For each combination of $(\text{male}, T) = (m, t)\in \{0,1\}^2$ and $r \in \mathcal{R}$ we consider the following probabilities:
\[\pr(R \leq r \mid \text{male} = m, T = t), \hspace{3mm} \pr(M =1 , R>0\mid \text{male} = m, T = t).
\]
These probabilities can be estimated directly by their sample analogs and compared to the values derived from the structural model evaluated at a parameter vector $\theta$. 
The \cite{Lee2018} experiment asked migrants for the mode and amount of remittances sent over multiple months between the baseline and endline surveys. 
We first aggregate these up to the household level. 
The variable $R_i$ is therefore the average monthly remittances sent by a migrant over the course of the survey, and the mobile money usage variable is aggregated to a probability of remitting via mobile money (bKash) conditional on remitting a positive amount. 

We also include a component of $\hat{\pi}$ and $q(\theta)$ to estimate the parameter $\alpha$.
Solving the migrant's decision problem yields the following optimality condition:
\[
\frac{\alpha}{1-\alpha} = \frac{C_m}{L_m\cdot w_m}.
\]
We observe expenditure by a migrant $E_m$ and consider it to be a noisy observation for consumption $C_m$ as in:
\[
E_m=C_m+\tilde{\epsilon},\hspace{2mm}\text{where}\hspace{2mm}\mathbb{E}[\tilde{\epsilon}]=0.
\]
Substituting $E_m$ in the optimality condition and rearranging terms results in the following expression for the parameter $\alpha$: 
\begin{align}
	\label{eq:alpha_def}
\alpha= \frac{\mathbb{E}[E_m]}{\mathbb{E}[E_m]+\mathbb{E}[L_m\cdot w_m]}
\end{align}
We set one component of $q(\theta)$ as $q_j(\theta)=\alpha$, and the corresponding $\pi_j$ equal to the sample analog of the expression above. 

We  estimate the parameter vector $\theta$ by classical minimum distance, following \citet{newe:mcfa:1994}.\footnote{In our application we have fully specified the conditional likelihood function for outcomes given covariates and parameters, so it would be possible to instead estimate $\theta$ by conditional maximum likelihood or calculate its full posterior distribution through Bayesian estimation.
Under model misspecification, the probability limits of such likelihood-based estimates could differ from those of the minimum distance estimator.
We take a minimum distance approach to provide a template for future applications adopting our approach where likelihood-based estimation is not possible.}
The estimator solves 

\[ \min_{\theta} \left( \hat{\pi} - q(\theta) \right)^{\prime} {W}_n 
\left( \hat{\pi} - q(\theta) \right),\]
where ${W}_n$ is a weighting matrix with ${W}_n \cvgp W$ for some $W$. 
In our application we set $W_n$ equal to the identity matrix. 
Under standard regularity conditions, the estimator $\hat{\theta}$ will converge in probability to a
pseudo-true value $\theta_0$.\footnote{If the structural model does not hold exactly, as we generally take to be the case in our analysis, the pseudo-true parameter value $\theta_0$ depends on the choice of the limiting weighting matrix $W$. We suppress this dependence in the notation.}
Its asymptotic distribution is given by
\[ 
	\sqrt{n}(\hat{\theta}-\theta_0) \cvgd N(0,\Sigma_\theta),
\]  
where the asymptotic variance matrix $\Sigma_\theta$ can be consistently estimated. 
Appendix \ref{sec:asymptotics_appendix} provides further details on the estimation method.

Table \ref{table:estimates} displays estimated parameters, with standard errors in parentheses.
Appendix \ref{subsec:modelfit} details the model fit by comparing treatment effects on the amount of daily remittances sent and the probability of remitting via mobile money, as well comparing the observed empirical distribution of remittances in the experimental data with the model generated distributions.  

\begin{table}[h]
\renewcommand{\arraystretch}{1.5}
\caption{Model Parameter Estimates}
	\label{table:estimates}
	\centering
	\begin{tabular}{cccccc}
		\toprule  
		$\psi$&$\phi_0$&$\phi_1$&$\overline{\epsilon}$&$\overline{\xi}$&$\alpha$\\ \hline 
		-0.5116&		-2.2641&		0.1012&		0.2522&		0.1341&		0.3176\\
		(0.2369)&		(0.0772)&		(0.042)&		(0.2478)&		(0.1618)&		(0.0048)\\
		\bottomrule
	\end{tabular}
\begin{minipage}{0.6\textwidth}
\vspace{1pt}
{\footnotesize Notes: standard errors in parentheses.
\par}
\end{minipage}
\end{table}

The negative estimated value for $\psi$ shows that the treatment effectively reduces the hassle cost of remitting via mobile money.
The negative estimate for $\phi_0$ implies that migrants place a low weight on per-capita consumption across home household members relative to their own index of consumption and leisure.
Specifically, the value $\hat{\phi}_0 =-2.26$ implies that female migrants put a weight of 0.905 on their own consumption-leisure index and the rest on home consumption.
Our estimate of $\phi_1$ has male migrants putting a weight of 0.897 on their consumption-leisure index.
Based on \citet{Lee2022}, we conjecture that this could be a result of migrant women having less bargaining power relative to their male co-migrants so that they are less able to remit to family members at home.
Our estimate of $\hat{\alpha} \approx  1/3$ is in line with typical values for the consumption weight in consumption-leisure indices, e.g.~\citet{kydl:pres:1982}.

\subsection{Predicted Site-Level Effects Based on the Structural Model}
\label{subsec:prediction}

The structural model can be used to predict the average effect of the intervention for every prospective site. 
In conjunction with a posterior distribution for the structural model parameters $\theta$ obtained from the pilot RCT data, we can thereby generate a joint prior for the vector of site-level effects $\tau = (\tau_1,\dots,\tau_S)$. 
In our application, we face some additional computational and data limitations. 
To explain our approach, we first outline an idealized version of our simulation algorithm that abstracts from some of these complications, and then discuss the practical implementation that addresses the remaining issues. 

Let $p(\theta)$ denote the posterior distribution of the structural parameters, based on fitting the structural model to the pilot bKash experimental data as discussed earlier in this section. 
To generate draws for the average effect $\tau_s$ at a site $s$, we need the values of the site-level characteristics 
$(o,d,p_h)$.
We also require the joint distribution of the characteristics $(w_{mi},a_{hi},y_{hi},\text{male}_i)$ of households in corridor $s$.
To simplify the notation, let $F_{X,s}$ denote the joint distribution of the site- and individual-level covariates at site $s$.\footnote{This joint distribution will be degenerate for the site-level characteristics, putting probability one on their known values.} 

We can generate draws for $\tau_s$ as follows. First, take a draw for the structural parameter: $\theta \sim p(\theta)$.
Then, generate many draws $b=1,\dots,B$ for the site- and individual-level covariates: $ X_{s,b} \sim F_{X,s}$,
and also generate $B$ draws for the model shocks $\epsilon$ and $\xi$ 
from their exponential distributions, with means depending on the draw for $\theta$.
Then, for each simulated individual $b$, we can generate their counterfactual optimal remittances and mode choice under both treatment and control, by setting $t=0$ and $t=1$ and using the formulas in Appendix \ref{sec:model-solution}.

This results in a large number of draws $R_{s,b}(t),M_{s,b}(t), b=1,\dots,B$, for the outcome variables under both treatment and control at site $s$.
We can then calculate 
the (approximate) value for $\tau_s$. 
For example, if we are interested in the average effect of the treatment on remittances, we can set
\[ \tau_s = \frac{1}{B}\sum_{b=1}^B \left[ R_{s,b}(1) -  R_{s,b}(0) \right] .\]
(The actual definition of the treatment effect we work with in Section \ref{sec:outcome} is slightly more complicated, but the same idea can be applied to any notion of a treatment effect for which the structural model generates counterfactuals.) 
This generates one draw for $\tau_s$,  
and repeating the entire process many times will generate a large number of draws for $\tau_s$ from the structural prior. 

The full algorithm for simulating the structural prior extends this intuition in a number of ways. 
First, our minimum-distance estimation strategy described in Sections \ref{sec:mdestimator}-\ref{sec:mdestimatorspec} does not provide the full Bayesian posterior distribution for $\theta$.
We instead adopt a limited-information perspective and interpret the minimum distance estimation strategy as providing an approximate or quasi-posterior for $\theta$ given by 
\[ \theta \sim N\left( \hat{\theta}, \widehat{\Sigma}_{\theta}/n, \right),\]
where $\hat{\theta}$ is the minimum distance estimate and $\widehat{\Sigma}_{\theta}$ is the estimated asymptotic variance-covariance matrix of its sampling distribution. 
In order to avoid the possibility of obtaining negative draws for $\bar{\epsilon}$ and $\bar{\xi}$, we transform $\theta$ by taking $\bar{\epsilon} \mapsto \ln (\bar{\epsilon})$ and $\bar{\xi} \mapsto \ln(\bar{\xi})$, leaving the remaining components of $\theta$ unchanged, using the Delta method to obtain the implied distribution. 

Second, we need to sample jointly for {\em all} the site-level effects $\tau_1,\dots,\tau_S$. For each draw of $\theta$, we employ the same method to generate the implied value of $\tau_s$ across all sites $s=1,\dots,S$. When doing this, we can re-use the model shock draws $\epsilon_b,\xi_b$ across the sites to reduce  simulation chatter. 

Third, we face a significant data limitation that prevents us from directly estimating $F_{X,s}$, the joint distribution of individual (and site-level) characteristics by site. 
Ideally, we would have a data set that samples 
$w_{mi}$, $a_{hi}$, $y_{hi}$, and $\text{male}_i$ for migrants in each prospective site $s$. 
Unfortunately, there are no datasets representative at the district level which provide comprehensive data on households in rural areas linked to migrant members. 
We instead draw on representative data sets---which we will hereafter refer to as administrative data---to construct site-level marginal summary statistics, largely for either home households or migrants. 
In these data, we only regard the marginal distribution of gender among migrants, the marginal sample means and variances of migrant wage, home household size, and home household income conditional on migrant gender as being of sufficiently high quality for our purposes. 
We therefore need to make additional assumptions about the dependence structure of the individual-level variables in order to specify their joint distribution. 

To specify the joint distribution of individual characteristics by site, we assume that for each site, the vector $(\ln(w_{mi}), \ln(a_{hi}), \ln(y_{hi}) ) $ is jointly normally distributed conditional on migrant gender. 
We can estimate the means and the marginal variances of this distribution from the administrative data.
To specify the conditional covariance terms, we estimate the correlation matrix for the same (logged) covariates from the pilot experimental data in \citet{Lee2018}, separately by gender, and assume that the same correlation matrix holds in every prospective site $s$. 
This fully specifies $F_{X,s}$ through the probability of the migrant's gender, and the joint distribution of $(w_{mi},a_{hi},y_{hi})$ conditional on migrant's gender. 

Table \ref{table:summstat_bkash_model} demonstrates the results of our procedure for working with the administrative data.
It shows the model implied treatment effect on daily remittances sent by migrants when using the full data from \cite{Lee2018} and compares it to the model-predicted value when restricted to only using the same summary statistics we calculate in the administrative data.
The results are encouraging, with the model-implied treatment effect based on the summary statistics differing only by 8\% from the one derived using the full microdata.
\begin{table}[h]
\centering
\caption{Average Treatment Effects in Model and Data}
\label{table:summstat_bkash_model}
\begin{tabular}{lcc}
\toprule
Outcome & Model & Data \\
\cmidrule(lr){1-1}\cmidrule(lr){2-2}\cmidrule(lr){3-3}
Remittance per Day (\cite{Lee2018} Microdata) & 5.94 & 6.41\\
Remittance per Day (Summary Statistics Only) & 6.44 & 6.41\\
\bottomrule
\end{tabular}
\begin{minipage}{0.7\textwidth}
\vspace{1pt} {\footnotesize Notes: amounts are in Bangladesh taka (1  USD = 78 BDT in 2015).
The Model column gives the model-implied average treatment effect when inputting the full experimental microdata (row 1) and only summary statistics as described in Section \ref{subsec:prediction} (row 2).
The Data column gives the unrestricted average treatment computed in the data.
\par}
\end{minipage}
\end{table}

\section{Application: Designing Experiments in South Asia}
\label{sec:application}

We apply the experimental site selection methodology developed in the previous sections to determine where to locate new randomized trials in Bangladesh, India, and Pakistan.  
The experimental intervention will again be aimed at reducing the costs a migrant may face when remitting to their family in distant home districts. 
A migration corridor (site) is defined as a destination district and origin district  pair in any country.
We give site names in destination-origin format throughout the paper.
Districts are referred to by the name ``district'' in India and Pakistan, while Bangladesh refers to the same unit as a \textit{zila}.
For convenience we will refer to all such spatial units as districts in what follows.

We first describe the social welfare function and treatment effects we will target and translate costs into welfare terms, and specify the site-level data and prior distributions we will use for the analyses. 
We then apply the method developed in Section \ref{sec:experimental_design} to the problem of choosing a single experimental site in each country. 
Next, we consider site selection when it is possible to target two migration corridors in each country, which is the case we can actually take to the field. 
We finally evaluate the welfare gains and discuss the potential gains (and costs) of experimenting on multiple sites. 

\subsection{Specifying Social Welfare}
\label{sec:outcome}

The household-level outcome of interest in this application is the additive log social welfare associated with each of its home members' utility,\footnote{This is a special case of additive CES social welfare (c.f.~\citealt{deaton1997}, \citealt{alderman2019}). We focus on log welfare for transparency and compatibility with the log per-capita consumption outcomes reported in \cite{Lee2018}, but could consider alternative parameterizations. This is isomorphic to specifying $j$'s utility from consumption as log with utilitarian social welfare. We abstract from the migrant's utility gain from a decrease in the effective price of remitting via mobile money decreasing due to the training, but this could easily be incorporated. } 
\[
\sum_{j\in HH_i} \ln u_j.
\]
As in \cite{deaton1997}, we assume the utility attained by an individual $j$ in home household $i$ is household per-capita consumption which, after accounting for remittances, is
\[
u_j = \frac{1}{a_{h,i}}\left(y_{h,i}+\frac{R_i}{p_h}\right).
\]
Since all members of the household have the same per-capita consumption, the outcome of interest may be written as:
\[
Y_i(t) = a_{h,i} \ln \left[ \frac{1}{a_{h,i}}\left(y_{h,i}+\frac{R_i^t}{p_h}\right) \right], 
\]
where the dependence on the treatment status $t\in \{0,1\}$ is made explicit in the notation $R_i^t$.

For site $s$, the average treatment effect in welfare terms equals
\[
\tau_s = \mathbb{E}_\theta \left[Y_i(1)-Y_i(0)\right],
\]
where the $\theta$ subscript indicates the dependence of the distribution of remittances on the model parameters, as described in Section \ref{subsec:stochastic-spec}. The expectations operator above also aggregates over the joint distribution of $(a_{h,i}, y_{h,i}, w_{m,i})$ at site $s$, as detailed in Section \ref{subsec:prediction}.
The average treatment effect on this measure can be interpreted as approximating the home-household size-weighted average percentage change in per-capita consumption due to the training program. 

One last point to handle is that household size $a_{h,i}$ can be affected by the treatment.
In \cite{Lee2018} household size declines slightly with treatment as some additional household members migrate.
We consider these members to benefit from the increase in per-capita consumption at home, so we measure the treatment effect in a site $s$ as
\[
\mathbb{E}_\theta\left[a^{\text{baseline}}_{h,0i}\cdot \ln\left(\frac{\text{home expenditure}^{\text{endline}}_{1i}}{a_{h,1i}^{\text{endline}}}\right)\right] -
\mathbb{E}_\theta\left[a^{\text{baseline}}_{h,0i}\cdot \ln\left(\frac{\text{home expenditure}^{\text{endline}}_{0i}}{a_{h,0i}^{\text{endline}}}\right)\right],
\]
where the 0 or 1 preceding the household index $i$ denotes treatment.

A site will be recommended for treatment in the policymaking stage if the perceived welfare benefit of treatment there ($\tau_s$) exceeds the cost. 
We translate the per-migrant cost of the program, given in Table 3 of \citet{Lee2018} as 885.84 taka, into welfare terms by supposing the cost is homogeneous across sites within a country and the program is financed by a uniform consumption tax in the country.
We provide the details of our calculation in Appendix \ref{sec:welfare-cost}, which yields welfare-unit treatment costs of 0.183, 0.145, and 0.153 for Bangladesh, Pakistan, and India, respectively.

\subsection{Candidate Migration Corridors}
\label{subsec:candidate_corrs}

While data is available for a large number of migration corridors in each country (Appendix \ref{subsec:admindata} discusses the construction of our dataset of site characteristics), we focus on a smaller subset for the purposes of experimental site selection. 
We have two main objectives in doing so.
The first is to build in robustness to prior misspecification through geographic diversity: we require that the two corridors have origins in different states (India), provinces (Pakistan), and divisions (Bangladesh).
The second is survey and program administration practicality: origin locations must be sufficiently dense in households with migrants to make it feasible to sample in that location, and we can only survey in areas where our potential enumeration partners work.
We view potential program administrators as being subject to the same constraints as our enumeration partners.
Program administrators, too, must have a sufficiently dense pool of migrants to be able to fill  training sessions.
Appendix \ref{subsec:restrictions_details} provides details on the migrant density restrictions we use, and Appendix Table \ref{table:summary_characteristics} lists summary statistics on the site characteristics entering the structural model, for sites matching our criteria across the three countries.

\subsection{Description of the Priors}
\label{subsec:application_priors}

\paragraph{Mixed prior} Our preferred prior to provide the extrapolation benefits of the structural model and the robustness of the smoothing prior is the mixed prior described in Section \ref{subsec:priorspec}, with equal weight placed on the structural prior and a smoothing prior. 
The mean parameter for the prior is the vector of average treatment effects generated by the structural model. 
The covariance matrix for the smoothing prior is constructed as in Section \ref{subsubsec:smoothingprior} using two site level characteristics: the distance between destination and origin; and home household income.
These variables are important both in our fully-specified model and according to general economic reasoning.
The benefits of constant-price mobile money relative to traditional methods like migrant travel are increasing in the distance between the destination and origin points of a corridor, and the marginal value to the migrant of income sent home is higher when the home household is poorer.
We standardize both variables to have zero mean and unit standard deviation.

Migrant wage at destination would also seem important but does not vary much in Bangladesh since there are so few destinations among high-migrant-density corridors.
Furthermore, including more variables in the smoothing prior increases the sensitivity of site selections to its tuning parameters, specifically the measure of distance between site characteristics.\footnote{In Pakistan and India there are more destinations and wage at destination varies more across corridors (see Table \ref{table:summary_characteristics}). We keep the number of characteristics at two for consistency with Bangladesh and because of the aforementioned concerns about dimensionality in the smoothing prior.}
This is a key difference with the structural prior, whose parametric structure specifies the relevant notion of distance explicitly.

We set $h(\cdot)$ in the definition of the smoothing prior covariance matrix in Section \ref{subsubsec:smoothingprior} to the identity function and use Euclidean distance as the norm. 
We set the tuning parameter $c_1$ to $1$ while the parameter $c_0$ is chosen such that the scale of covariance values implied by the smoothing prior are similar to those of the structural prior.
In particular, $c_0$ is chosen to approximately minimize the maximum absolute difference in variance values as defined by the structural and smoothing priors.\footnote{Following this approach, we set $c_0 = 0.1$ for Pakistan, while for India we use $c_0 = 0.073$. In Bangladesh, data limitations require us to use predicted values from a wage regression to approximate migrant wage at destination, as described in Appendix \ref{subsec:admindata}. This makes mean migrant wages much more homogeneous in Bangladesh than in the other two countries (see Table \ref{table:summary_characteristics}), which results in a structural prior with understated uncertainty. Therefore we set $c_0$ in Bangladesh to be the same as in the pure smoothing prior.}

While most empirical results shown below are based on this mixed prior, we also consider other priors with different weights on the structural model. 

\paragraph{Structural prior} The pure structural prior we consider is a mixed prior with zero weight on the smoothing prior.

\paragraph{Smoothing prior}  We also consider a pure smoothing prior.
We intend it to represent an off-the-shelf application of the ideas behind the smoothing prior, showing an alternative that does not involve undertaking the construction of a structural model.
Instead of setting the prior mean to the structural-model-predicted vector of site treatment effects the way we do in all mixed priors, 
we set the mean parameter in the pure smoothing prior to be constant  across all sites and equal to the estimated treatment effect (in welfare terms) in the pilot bKash experiment of \cite{Lee2018}.
As in the smoothing component of the mixed prior, we construct the covariance matrix for the pure smoothing prior using destination--origin distance and home income, but we choose the parameter $c_0$ governing the overall prior scale so that the site-level variances match the sampling variance for the estimated treated effect in the pilot experiment.
This gives us $c_0\approx 0.13$, and we set $c_1$ to a benchmark value of 1. 

\paragraph{Distribution of $\hat{\tau}$ given $\tau$}
Recall that the distribution of $\hat{\tau}$ conditional on the true vector of site-level treatment effects $\tau$ is
\[
\hat{\tau}\mid \tau \sim N\left(\tau,  \Sigma_\epsilon\right).
\]
In our implementation, we set $ \Sigma_\epsilon = \text{diag}\left\{ \sigma_\epsilon^2,\dots,\sigma_\epsilon^2 \right\}$ with the sampling error $ \sigma_\epsilon^2$ defined as
\[
 \sigma_\epsilon^2 =\frac{\hat{\sigma}^2_\text{BKash}}{n_{\text{exp}}},
\]
where $\hat{\sigma}^2_\text{BKash}$ is the estimated variance of the welfare treatment effect from the \cite{Lee2018} data and $n_{\text{exp}}$ is the number of households sampled in each corridor for the new experiment.
The value of $n_{\text{exp}}$ depends on the number of corridors selected for experimentation, as described in Appendix 
\ref{sec:site_calculations_by_number}.

\begin{figure}[ht!]
    \centering
    \begin{minipage}{0.5\textwidth}
    \caption{Expected Welfare TE (Structural Prior)}
    \label{fig:welfare_te_bd_model}
    \includegraphics[width = \textwidth]{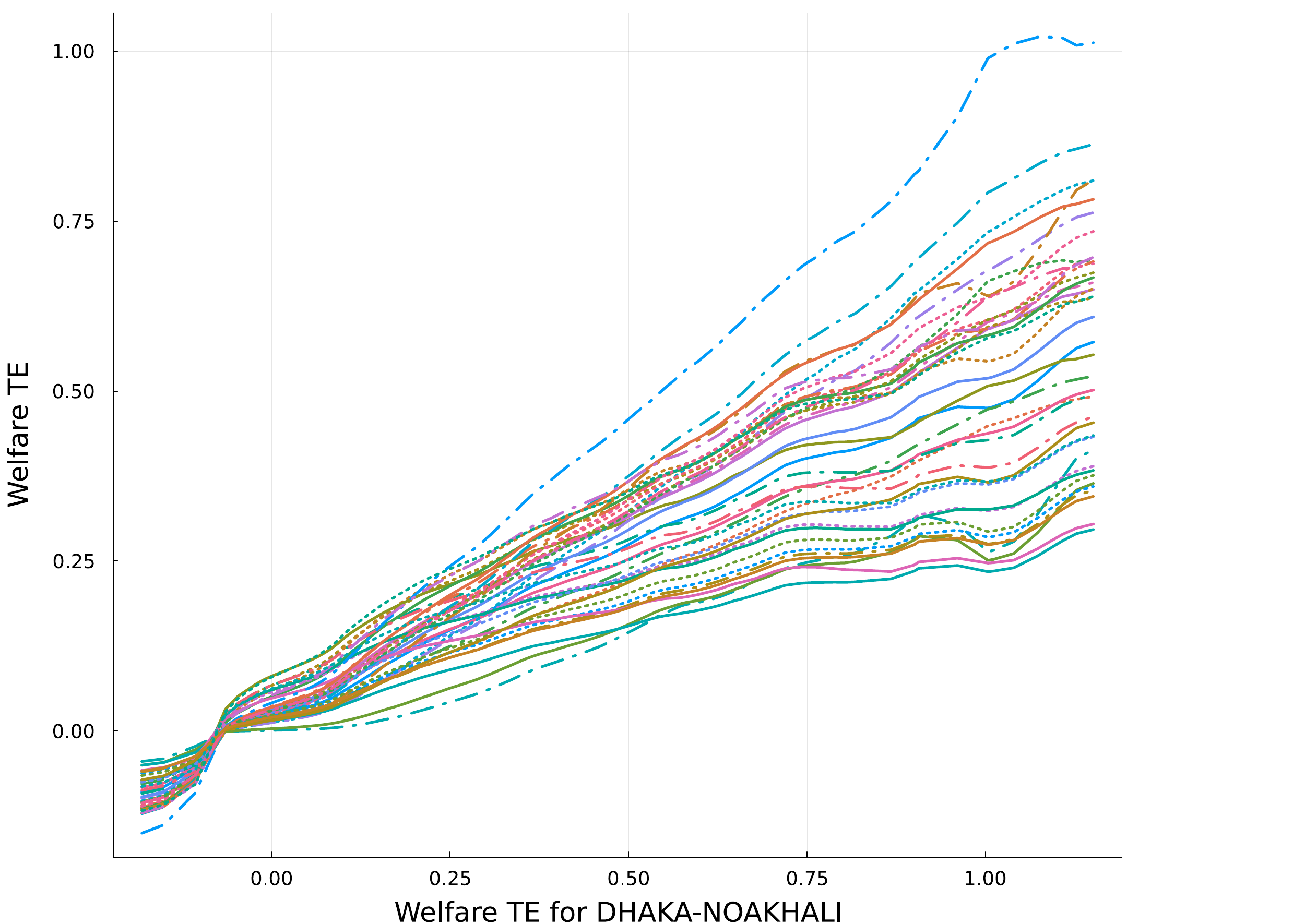}
    \end{minipage}\hfill
    \begin{minipage}{0.5\textwidth}
    \caption{Expected Welfare TE (Smoothing Prior)}
    \label{fig:welfare_te_bd_smooth}
    \includegraphics[width = \textwidth]{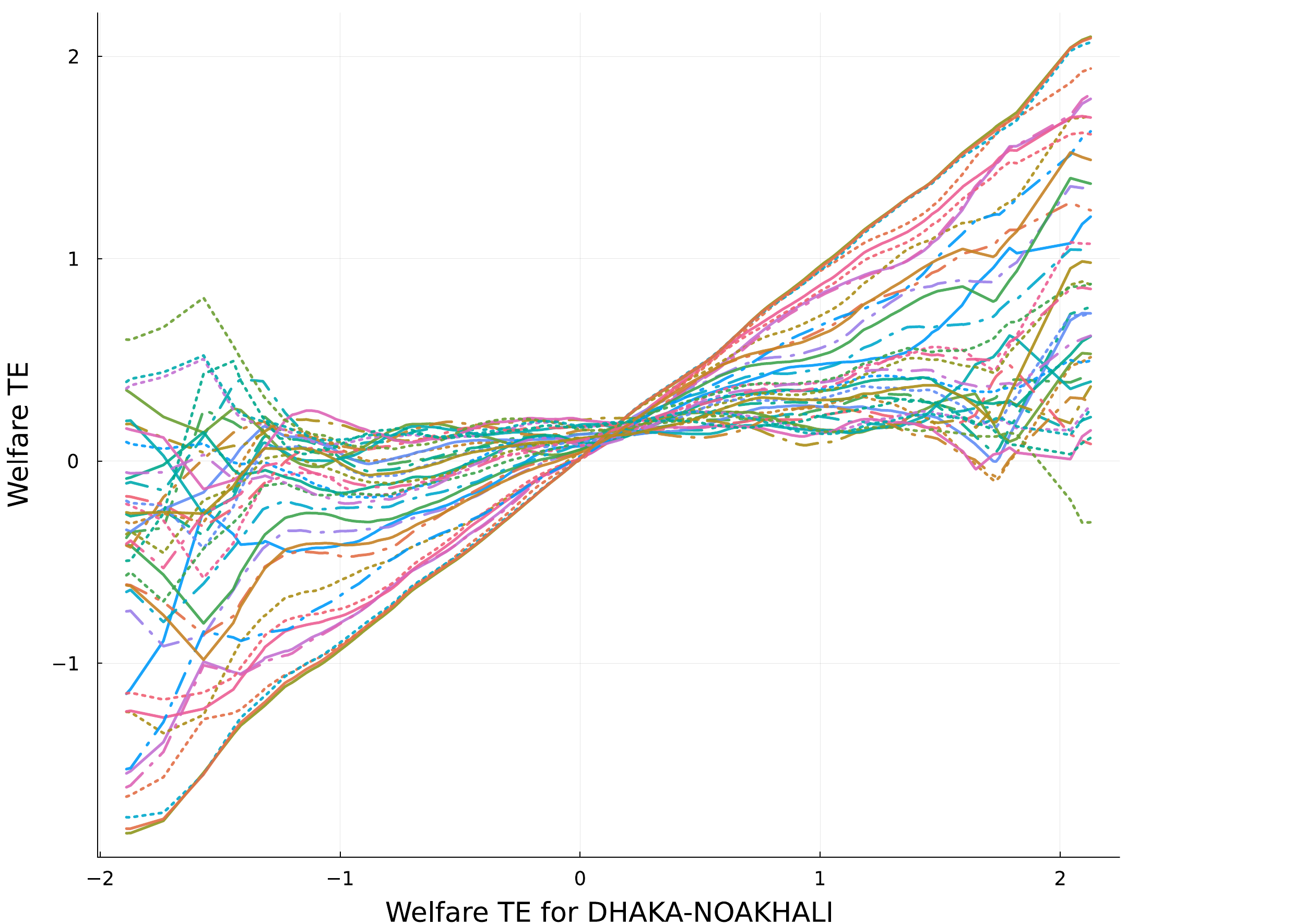}
    \end{minipage}
    \begin{minipage}{\textwidth}
\vspace{1pt} {\footnotesize Notes: Each line in the figures shows the expected welfare treatment effect in a Bangladesh corridor conditional on knowing the welfare treatment effect in the Dhaka-Noakhali corridor is equal to the value on the X-axis.
\par}
\end{minipage}
\end{figure}

\paragraph{Comparing the priors} 
As discussed above, the parametric model underlying the structural prior allows experimental results in a given site to be informative for treatment effects in sites with quite different characteristics.
We can visualize this by examining how the mean predicted treatment effect in other sites changes as we change the value of the treatment effect in a reference site. 
This amounts to plotting $\mathbb{E}[\tau_s | \tau_{s^\prime}]$ (where expectations are taken under the prior) as a function of $\tau_{s^\prime}$. 
Figure \ref{fig:welfare_te_bd_model} displays such curves for all sites $s$ under the structural prior, with the treatment effect for the reference site Dhaka--Noakhali on the horizontal axis.
Dhaka--Noakhali is a migration corridor between Noakhali on the Bay of Bengal and Dhaka, the capital of Bangladesh and the main location of garment factories.\footnote{Patterns are similar regardless of the choice of reference site.}
Most of the lines have substantial positive slopes, meaning that a large treatment effect in Dhaka--Noakhali is expected to be associated with a large and positive predicted treatment effect in most other sites. 
Figure \ref{fig:welfare_te_bd_smooth} plots the same relationships under the smoothing prior.
Here the predicted treatment effect in Dhaka--Noakhali is strongly positively correlated with the predicted effects in only a few other sites, specifically only those sites with similar average home household income and destination--origin distance.
The horizontal axes of the two plots, which are set by the range of 1,000 draws from each prior, are also of note.
It is evident that the structural prior puts very little probability on Dhaka--Noakhali having a negative welfare ATE, while the smoothing prior remains agnostic.
We provide analogous plots for Pakistan, yielding the same qualitative conclusions,  in Appendix \ref{subsec:pakistan_prior}.

\subsection{Optimal Choice of a Single Experimental Site}

\begin{figure}[ht!]
\centering
\caption{Optimal Experimental Sites Under Alternative Decision Rules, One Site Per Country}
\label{fig:optimal_sites_diff_prior_1site}
\vspace{.25cm}
\begin{minipage}{0.5\textwidth} \centering \textsf{Bangladesh}\\  
\includegraphics[width = \textwidth]{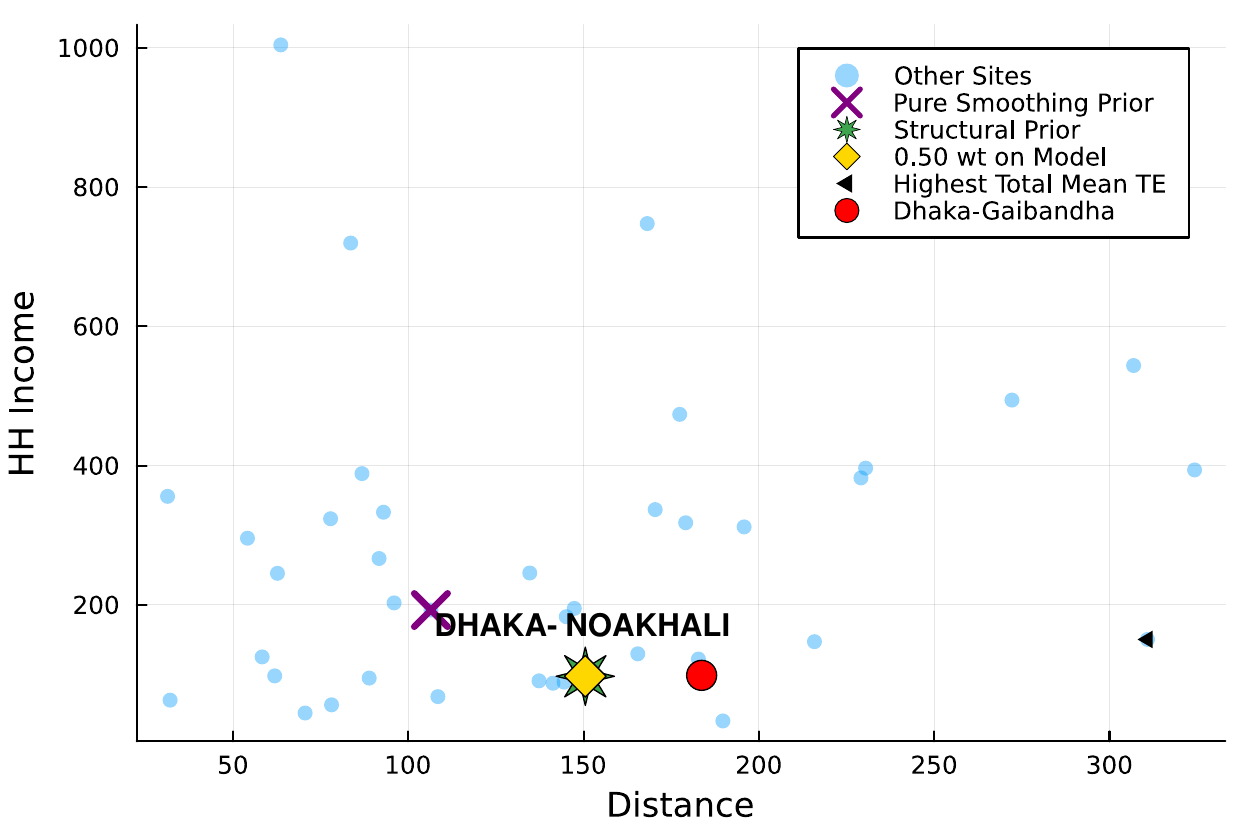}
\end{minipage}\hfill
\begin{minipage}{0.5\textwidth} \centering \textsf{Pakistan}\\
\includegraphics[width = \textwidth]{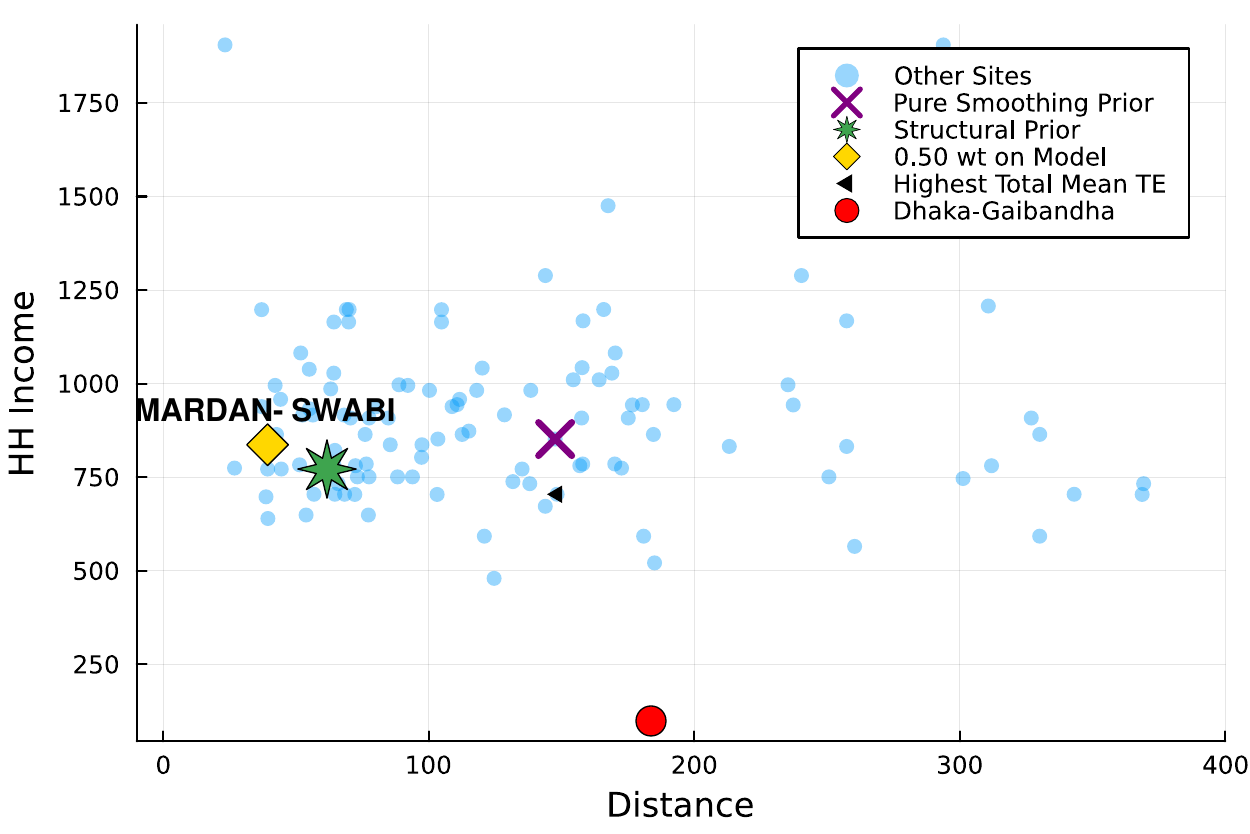}
\end{minipage}
\begin{minipage}{0.5\textwidth} \centering \textsf{India}\\
\includegraphics[width = \textwidth]{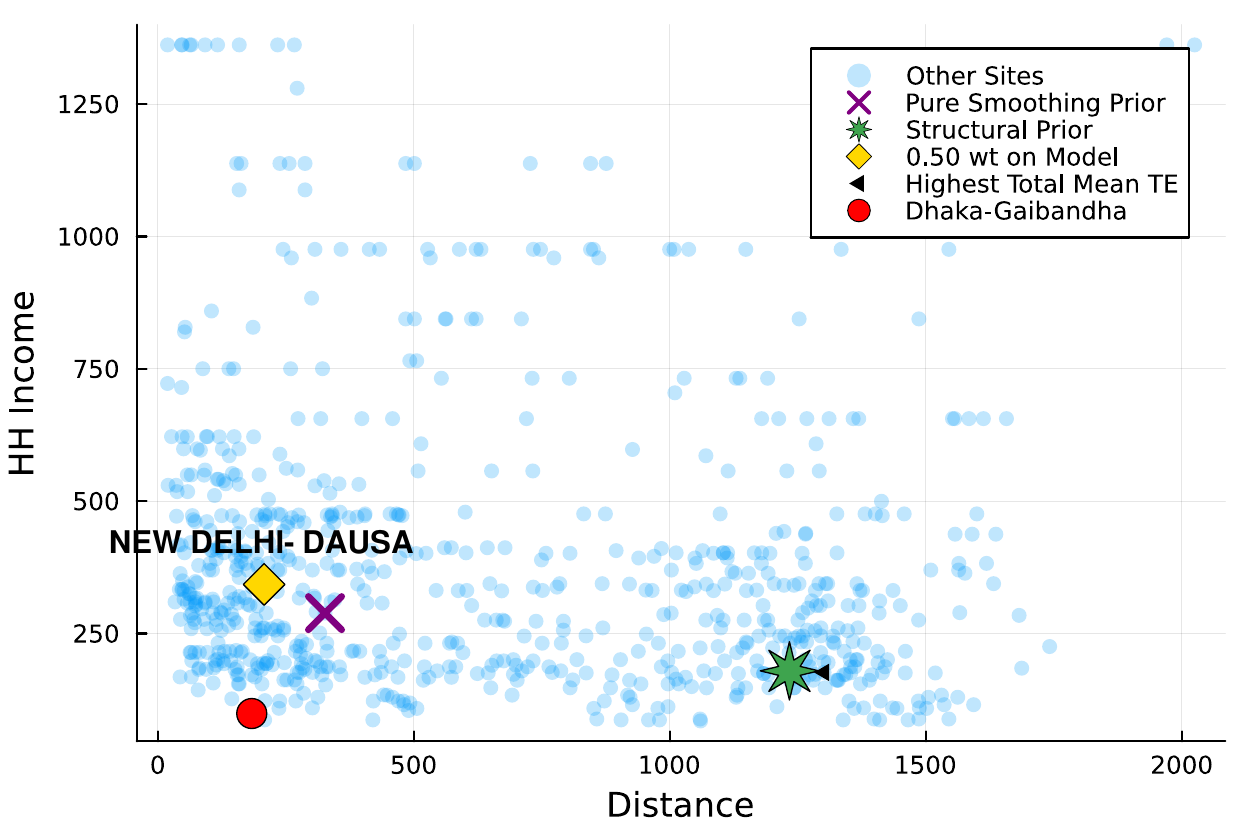} 
\end{minipage}
\begin{minipage}{0.95\textwidth}
\vspace{1pt} {\footnotesize Notes: Light dots represent candidate sites. HH Income refers to the average income in 2015 Bangladesh taka (1 2015 USD = 78 BDT), excluding remittances, of a household at migration corridor origin reporting having sent an internal migrant. Distance is km between corridor origin and destination as the crow flies. See Appendix \ref{subsec:admindata} for details. Site names are in destination-origin format and refer to the sites selected under our preferred mixed prior with 50\% weight on the structural model.
\par}
\end{minipage}
\end{figure}

We now consider the problem of choosing a single experimental site in each of the three countries, drawing on the pilot experiment to construct the prior distributions. 

The site for the original experiment in Bangladesh (the Dhaka--Gaibandha migration corridor) was chosen in part because it was a promising location for the intervention. 
\citet{Lee2018} find relatively large treatment effects, which likely reflects particular qualities of the Dhaka--Gaibandha corridor. 
First, the distance between Gaibandha and Dhaka makes it costly to travel between the two sites, and the ability to send money digitally, with a price that does not depend on distance, is thus particularly appealing. Second, the high poverty rate in Gaibandha meant that getting extra remittances could be expected to make a larger marginal impact than in better-off districts. 
Third, the infrastructure for mobile money was already in place, led by bKash, the largest provider. And, fourth, barriers existed to the adoption of bKash that could be reduced through a basic training intervention. 

The question after the experiment was whether the results in \citet{Lee2018} would generalize to other contexts. The top left panel of Figure \ref{fig:optimal_sites_diff_prior_1site} shows other migration corridors in Bangladesh. The horizontal axis gives the distance between the two ends of each corridor, and Dhaka--Gaibandha is among the longer corridors. The vertical axis is average income among households in the home district who report having sent an internal migrant, and Gaibandha is on the lower end compared to others. The red circle which locates the Dhaka--Gaibandha corridor is thus on the periphery of the main cluster of sites in the figure. 

A black triangle indicates the corridor where the highest total mean treatment effect from the intervention is expected, according to the mixed prior. It also involves a relatively poor district like Gaibandha, and it is much further from Dhaka. 
Choosing a site for additional experimentation which is similar to Dhaka--Gaibandha---or the site where the treatment effect is expected to be largest---would lead to choices that are unlike most other migration corridors in Bangladesh.

The single site in Bangladesh chosen using the pure smoothing prior depends on the distribution across sites of the average income of migrants' home households and the length of the corridor, the two axes of the figure.
The purple X in the figure shows the corridor chosen with the pure smoothing prior, and it is, by design, in the middle of the other corridors.
The figure shows that the chosen site is far more typical than the Dhaka--Gaibandha corridor in these two dimensions; it is shorter, and the populations are somewhat better off economically.\footnote{We lay out in detail the simulation algorithm underlying these choices and those in the rest of the section in Appendix \ref{sec:algorithm}.} 

As described in Section \ref{subsec:priorspec}, other dimensions can matter as well.
The microeconomic model in Section \ref{sec:model}, reflected in the pure structural prior, incorporates the migrants’ choices of labor, consumption, and remittances to predict the vector of average treatment effects across sites. 
Site selection with the structural prior thus reflects dimensions beyond just household income and corridor distance. 
The richer information yields a site selected closer to the original choice of the Dhaka--Gaibandha corridor. 
The optimal choice, marked by the green 8-pointed star, is the Dhaka--Noakhali corridor, which is longer and poorer than the site chosen with the pure smoothing prior. The mixed prior, which puts equal weight on both the smoothing and the structural prior (and which is the preferred specification), also selects the Dhaka--Noakhali corridor (shown by the yellow diamond).

The two other panels in Figure \ref{fig:optimal_sites_diff_prior_1site} show optimal single experimental site selections in Pakistan and India. 
In both contexts, choosing  migration corridors for experimentation with distances and home incomes as close as possible to the Dhaka--Ghaibandha corridor, as might be dictated by the logic of pure replication, would again lead to peripheral choices. 
So would choosing the corridor where the total mean treatment effects are expected to be highest.
Instead, our approach again optimally selects sites that are similar to other sites; when just basing choices on the pure smoothing prior, the sites are centrally located in terms of home household income and corridor distance. 
The sites selected when incorporating the structural prior again diverge from those chosen just with the smoothing prior, reflecting the additional dimensions of comparison.
But, with either prior, the figures show that choosing a single site favors selection of locations that are generally central among candidate sites. 

The bottom panel of Figure \ref{fig:optimal_sites_diff_prior_1site} shows site selection in India, and it illustrates a limitation of conducting only a single experiment.
Sites in India are dispersed widely, with two main clusters of points.
By the logic underlying the smoothing prior, a point in the center of one cluster may then not generalize easily to the other cluster.
The observation motivates the next questions: When is it valuable to experiment in more than one site? Where should sites be located when experimenting in two or more sites, and how much more informative is it for policymakers? 
In Sections \ref{subsec:choosing_2sites}--\ref{sec:margsites} we analyze the problem of selecting two migration corridors in each country, then in \ref{subsec:choosing-site-number} we quantify the welfare gains from experimenting in various numbers of sites weighed against the added costs.\footnote{Throughout our discussion of site selections in this application, we choose sites from each country independently to ease computational costs.
To ensure we are not leaving gains from joint site selection on the table, we select sites for Bangladesh and Pakistan jointly and find that the top site combinations from selecting sites independently achieve 97\% of the welfare of the top combinations from joint selection.}

\subsection{Optimally Choosing Two Sites in Each Country}
\label{subsec:choosing_2sites}

Figure \ref{fig:optimal_sites_diff_prior_1site} for India shows two clusters of sites based on distance and household income. 
When the field experiment can only be implemented in a single site, the optimal choice is relatively central to one of the clusters but far from the other cluster.  
Experimenting in two sites, instead of just one, could allow us to learn more about sites in the other cluster. 
Fortunately, in the actual experiment we are designing, we have the ability (in fact a mandate) to select two sites in each country. 
We turn to this case now.

\paragraph{Impact of Optimal Site Selection} 
Table \ref{table:two_site_best_worst_welfare} shows the range of welfare values possible under different choices of experimental sites in each country, with average welfare per migrant at the site level evaluated under our preferred mixed prior..
The impacts of optimal site selection emerge as quite substantial in Bangladesh, with the worst site combination leading to a maximized welfare which is about 13\% of the welfare in the best site combination.
In both Pakistan and India, the ratio of welfare achieved by choosing the best sites compared to the worst is smaller than in Bangladesh.
We return to this point in the next subsection.

\begin{table}[ht!]
\centering
\captionof{table}{Average Welfare from Best and Worst Site Combinations by Country}
\label{table:two_site_best_worst_welfare}
\begin{tabular}{lcc}
\toprule
Country & Highest Average Welfare & Lowest Average Welfare \\
\midrule
Bangladesh & 0.030 & 0.004 \\
Pakistan & 0.111 & 0.103 \\
India & 0.209 & 0.197 \\
\bottomrule
\end{tabular}
\begin{minipage}{0.7\textwidth}
\vspace{1pt} {\footnotesize Notes: Average Welfare refers to the average across all candidate sites of the per-migrant welfare increase from site-level implementation recommendations, predicted under our preferred mixed prior with 0.5 weight on the structural model.
\par}
\end{minipage}
\end{table}

Appendix Table \ref{table:two_site_best_worst} goes into more specific detail, showing the five best and worst migration corridor pairs in each all three country.
In Pakistan three of the worst combinations involve migration from the district containing the capital, Islamabad.
In India again corridors involving origin districts with large cities appear in many of the worst site combinations.
These corridors do have high migrant densities but the home households are outliers---they have the highest incomes among candidate sites.

Appendix Figures \ref{fig:bangladesh_bestworstsites} -- \ref{fig:india_bestworstsites} show where the best and worst site combinations for each country lie in the space of characteristics considered by the smoothing prior.
The best choices for the new experiment lie within clusters and the worst choices are always outliers in the feature space.
We will see that this is due to the influence of the smoothing prior, which, roughly speaking, tends to prefer combinations of sites such that all candidate corridors are close to at least one experimental site.\footnote{We could consider an alternative heuristic rule that minimizes the sum of the distances in characteristic space between each site and its closest experimental site. However, this would not account for the fact that for some sites, there is little uncertainty about whether the treatment will be effective there or not.}
Under the smoothing prior conducting an experiment on a migration corridor that is very different from most other sites will yield little information about how the treatment might impact other sites.
\footnote{Appendix Figures \ref{fig:pakistan_bestworstsites} and \ref{fig:india_bestworstsites} show the same pattern in Pakistan and India, respectively.}

\begin{figure}[ht!]
\centering
\caption{Optimal Experimental Sites Under Alternative Decision Rules, Two Sites Per Country}
\label{fig:optimal_sites_diff_prior}
\vspace{.25cm}
\begin{minipage}{0.5\textwidth} \centering \textsf{Bangladesh}\\  
\includegraphics[width = \textwidth]{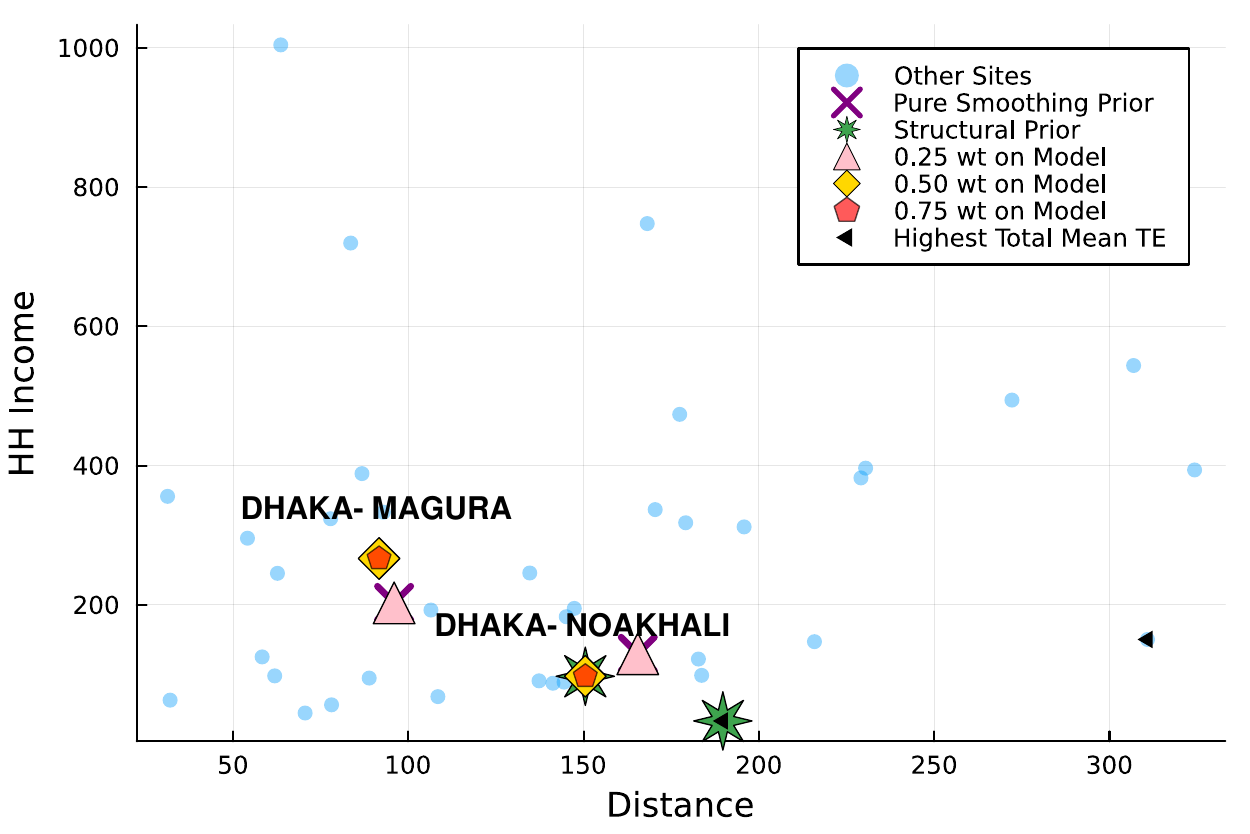}
\end{minipage}\hfill
\begin{minipage}{0.5\textwidth} \centering \textsf{Pakistan}\\
\includegraphics[width = \textwidth]{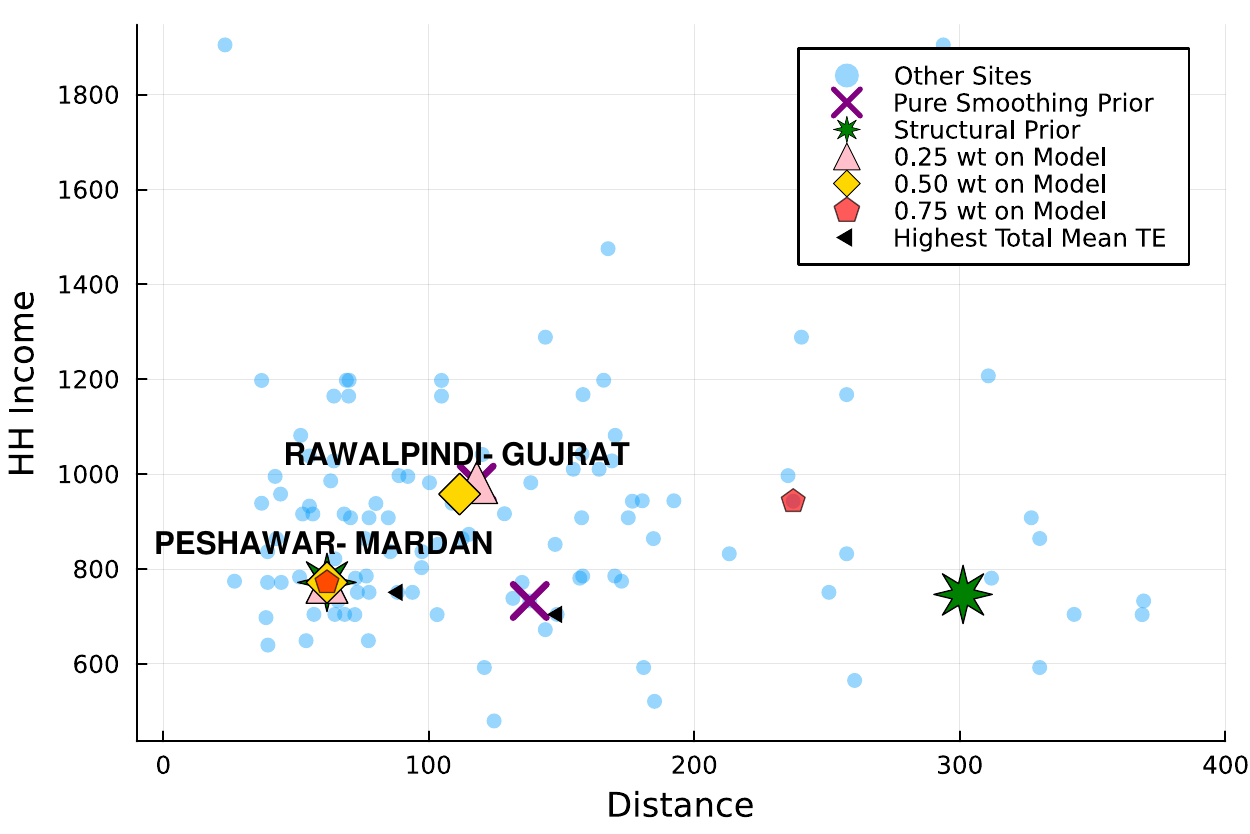}
\end{minipage}
\begin{minipage}{0.5\textwidth} \centering \textsf{India}\\
\includegraphics[width = \textwidth]{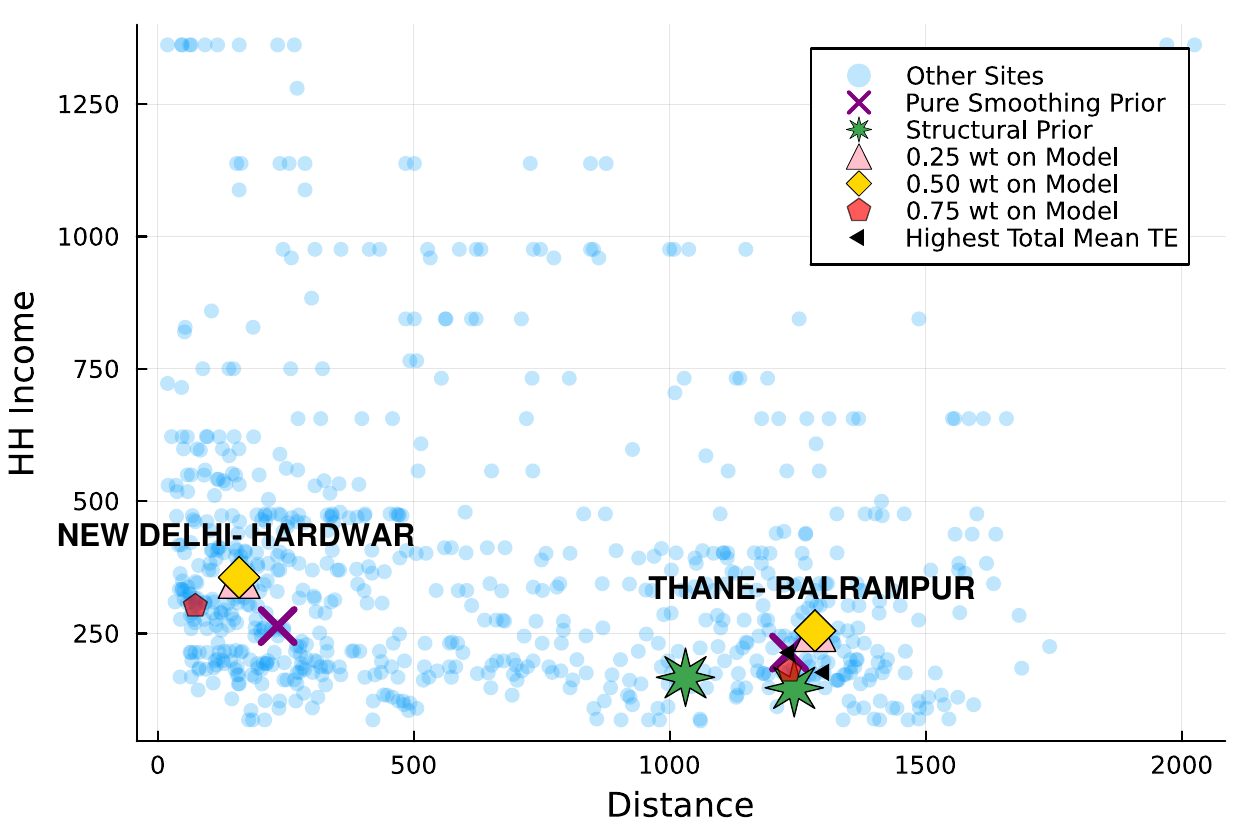} 
\end{minipage}
\begin{minipage}{0.95\textwidth}
\vspace{1pt} {\footnotesize Notes: Light dots represent candidate sites. HH Income refers to the average income in  2015 Bangladesh taka (1 2015 USD = 78 BDT), excluding remittances, of a household at origin reporting having sent an internal migrant. Distance is km between corridor origin and destination as the crow flies. See Appendix \ref{subsec:admindata} for details. Site names are in destination-origin format and refer to the sites selected under our preferred mixed prior with 50\% weight on the structural model.
\par}
\end{minipage}
\end{figure}

\paragraph{Top Site Choices by Prior} We now investigate how the choice of top site combinations varies with the prior. 
Figure \ref{fig:optimal_sites_diff_prior} shows the best destination--origin pairs under the different priors discussed in the previous subsection.
We again mark pairs of corridors selected under the pure smoothing prior, the pure structural prior (mixed prior with 0 weight on smoothing), and mixed priors with differing weights including our preferred 0.5.\footnote{Our choice of 0.5 weight in the mixing prior should not be taken to represent agnosticism between the pure smoothing and structural approaches. 
We recommend that practitioners check for robustness of selected sites and welfare to a variety of mixing weights, as we do in Appendix \ref{sec:eval_by_mixed_prior_weight}.
}
We also mark the pair of corridors with the largest sum of average predicted treatment effects in preparation for considering the effect of site selection bias, where the most promising sites in terms of treatment effects are also chosen for earlier experimentation \citep{Allcott2012}.

The pure structural prior tends to choose sites that are closer in average home household income and corridor distance than priors involving smoothing.
This is because many more site characteristics enter the structural prior, in nonlinear and interactive ways.
We show this in Appendix \ref{sec:structural_correlations} where we linearly regress correlations between site effects in the structural prior on differences in site characteristics and find (1) the relationships between characteristic differences and ATE correlations differ substantially across countries despite the common structural model and (2) the vector of simple differences between the characteristics of two sites imperfectly explains their predicted treatment effect correlation even within-country.
The pure smoothing prior, as discussed previously, tends to want all candidate sites to be close to at least one experimental site.
This feature largely carries over to the mixed priors, including the mixed prior placing only 0.25 weight on smoothing.\footnote{In Appendix \ref{sec:alternate_tes} we show that the set of selected sites is robust to variation in $\psi$, the parameter capturing the effect of the training program on the effective price of mobile money, 
which helps us be confident in our choice of sites even if the treatment effect may be moderated by unmodeled environmental changes taking place over time and space since the original \cite{Lee2018} study, such as the extent of mobile money penetration.
}

Since the pure structural prior considers more site characteristics than the pure smoothing prior, it may select sites that are more central on dimensions not targeted by the smoothing prior.
We show an example of this in Appendix Figure \ref{fig:smoothing_nontargeted_dims}, where the smoothing prior picks an outlying site in terms of both average household size at origin and the the ratio of origin prices to destination prices.
On the latter characteristic, in particular, the smoothing prior picks a quite unusual site where prices in the origin district are higher than they are at destination.

\subsection{Sites Where Policy Choices Are A Priori
Uncertain}\label{sec:margsites}

Evidence from experiments is especially valuable for decisions in sites where it is not already clear from the prior whether the treatment is likely to be effective. These ``marginal'' sites play a particularly important role in optimal site selection. 

We consider a migration corridor to be marginal if its treatment effect under a given prior has a non-trivial probability of exceeding or falling short of the cost of treatment there. 
Specifically, given a value $\kappa \in (0,1)$, a site $s$ with treatment effect $\tau_s$ and cost of treatment $\text{cost}_s$ is said to be marginal if the following conditions hold:
\begin{align*}
\pr(\tau_s \geq \text{cost}_s) \geq \kappa \quad \text{ and } \quad  \pr(\tau_s \leq \text{cost}_s) \geq \kappa,
\end{align*}
where the probabilities are calculated under the relevant prior for $\tau_s$. In our case with normal priors, these probabilities can be evaluated easily for any value of $\kappa$.

Sites that are not marginal are those which will almost always be treated, or almost never be treated, regardless of the choice of experimental sites, 
because there is little chance that the experimental data will revise beliefs enough to change the sign of $(\tau_s-\text{cost}_s)$. 
From the perspective of the policy-maker, it is more useful to focus on the potential gains in corridors for which the treatment assignment rule may be influenced in a nontrivial manner by experimentation. Welfare from sites that are always treated may also inflate the aggregate welfare values shown in Table \ref{table:two_site_best_worst_welfare}.

To understand why the percentage welfare gains from using the best site combination instead of the worst are smaller in Pakistan and India than Bangladesh, recall from Equation \eqref{eq:preposterior} that welfare from site selection strategies will differ only to the extent that they generate different recommendations for where to implement the mobile money training program.
If the mean predicted treatment effect in a given migration corridor is sufficiently high or low under the prior, or there is little uncertainty about the underlying treatment effect, it is considered very unlikely that subsequent experimentation will change the policy recommendation.
All treatment assignment rules will benefit from the treatment effect in sites receiving the training program regardless of which sites are selected for experimentation. 

Table  \ref{table:marginal_welfare} shows the per site average welfare from the same best and worst site combinations as underlie Table \ref{table:two_site_best_worst_welfare}, but averaging over marginal sites only.
In Table \ref{table:two_site_best_worst_welfare} the best pair of sites in Pakistan is predicted to achieve only 8\% more welfare than the worst pair.
Focusing on sites with at least 25\% probability of an opposite-signed net benefit of program implementation ($\kappa$ = 0.25) in Table \ref{table:marginal_welfare},
however, the best pair is predicted to achieve 41\% higher welfare than the worst pair.
For India, the corresponding numbers are 6\% without restricting to marginal sites and 33\% when doing so.
In Bangladesh all sites are marginal when $\kappa = 0.25$.
Appendix Tables \ref{table:pak_marginalwelfare_best} and \ref{table:ind_marginalwelfare_best} go into more detail, giving the identities of the specific sites in Pakistan and India, and showing how results change with $\kappa$ in Pakistan.

\begin{table}[ht!]
\centering
\caption{Average Welfare from Best and Worst Site Combinations by Country for Marginal Sites, $\kappa$ =0.25}
\label{table:marginal_welfare}
\begin{tabular}{lcc}
\toprule
Country & Highest Average Welfare & Lowest Average Welfare \\
\midrule
Pakistan & 0.048 & 0.034 \\
India & 0.101 & 0.076 \\
\bottomrule
\end{tabular}
\begin{minipage}{0.8\textwidth}
    \vspace{1pt} {\footnotesize Notes: Average Welfare refers to the average across marginal sites of the  welfare increase from site-level implementation recommendations based on the welfare-maximizing and minimizing sites from Table \ref{table:two_site_best_worst}.
    Average welfare is predicted under our preferred mixed prior with 0.5 weight on the structural model.
        Site $s$ is considered marginal if $
        \pr(\tau_s \geq \text{cost}_s) \geq \kappa \text{ and }  \pr(\tau_s \leq \text{cost}_s) \geq \kappa,$ where $\tau_s$ and $\text{cost}_s$ are the welfare benefit and cost from implementing the mobile money training program there.
        \par}
\end{minipage}
\end{table}

\subsection{Comparison to Rule-of-Thumb Procedures, Allowing Choice of Number of Experimental Sites}
\label{subsec:choosing-site-number}

In Section \ref{subsec:choosing_2sites}, we analyzed the optimal selection of two migration corridors in each country in which to run experiments. The question is not hypothetical: the funder of the research in South Asia in fact provided funding for experiments in two sites per country with a sample of 2,000 households in each site. 
It is natural to wonder whether we might improve welfare by experimenting in a different number of sites while maintaining the same budget.
Our framework allows us to explore such alternatives, by modifying the sampling error standard deviation $ \sigma_\epsilon$ to reflect the corresponding change in the sample size per site and specifying the appropriate choice set for site combinations.

\begin{figure}[ht!]
\centering
    \caption{Average Per-Migrant Welfare 
    by Number of Experimental Sites and Site Selection Method}
    \label{fig:more_less_sites}
    \vspace{.25cm}
    \begin{minipage}{0.5\textwidth}
    \centering \textsf{Bangladesh}\\ 
    \includegraphics[width = \textwidth]{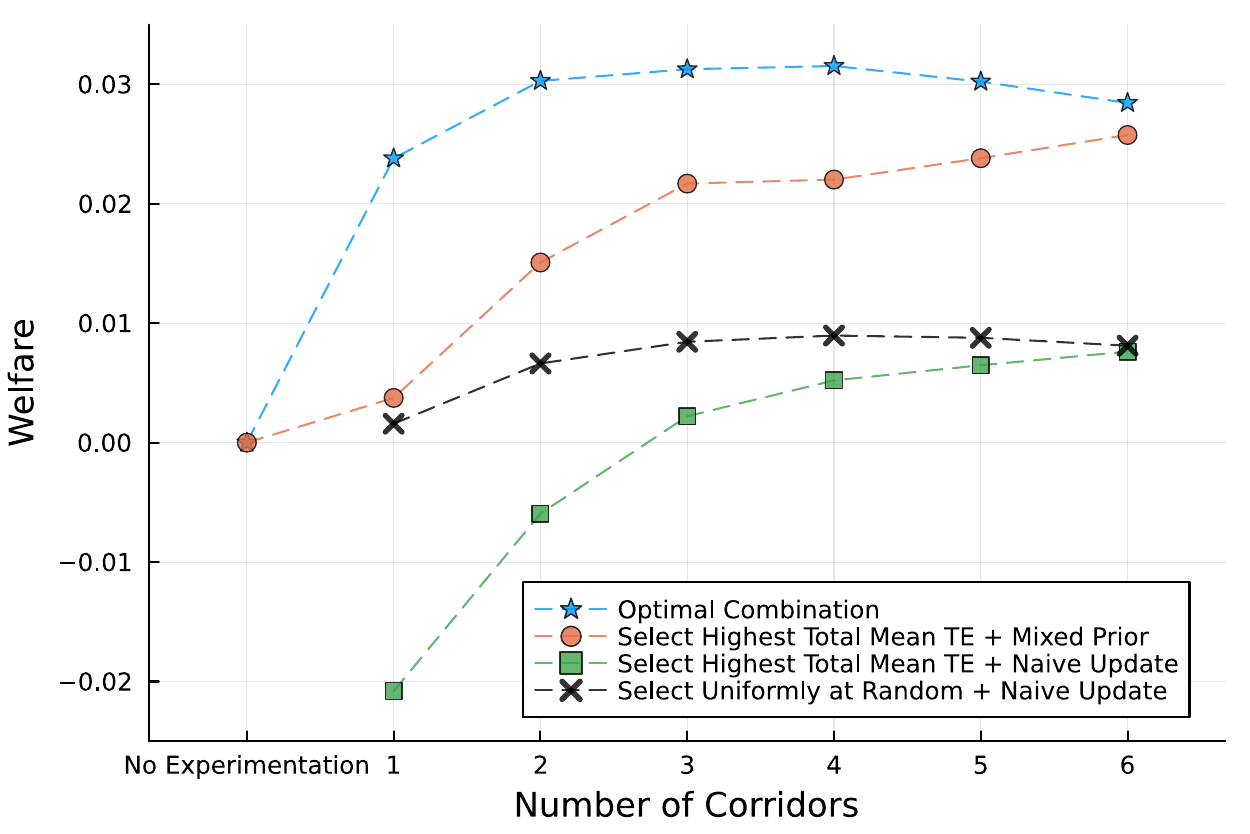}
    \end{minipage}\hfill
    \begin{minipage}{0.5\textwidth}
    \centering \textsf{Pakistan}\\ 
    \includegraphics[width = \textwidth]{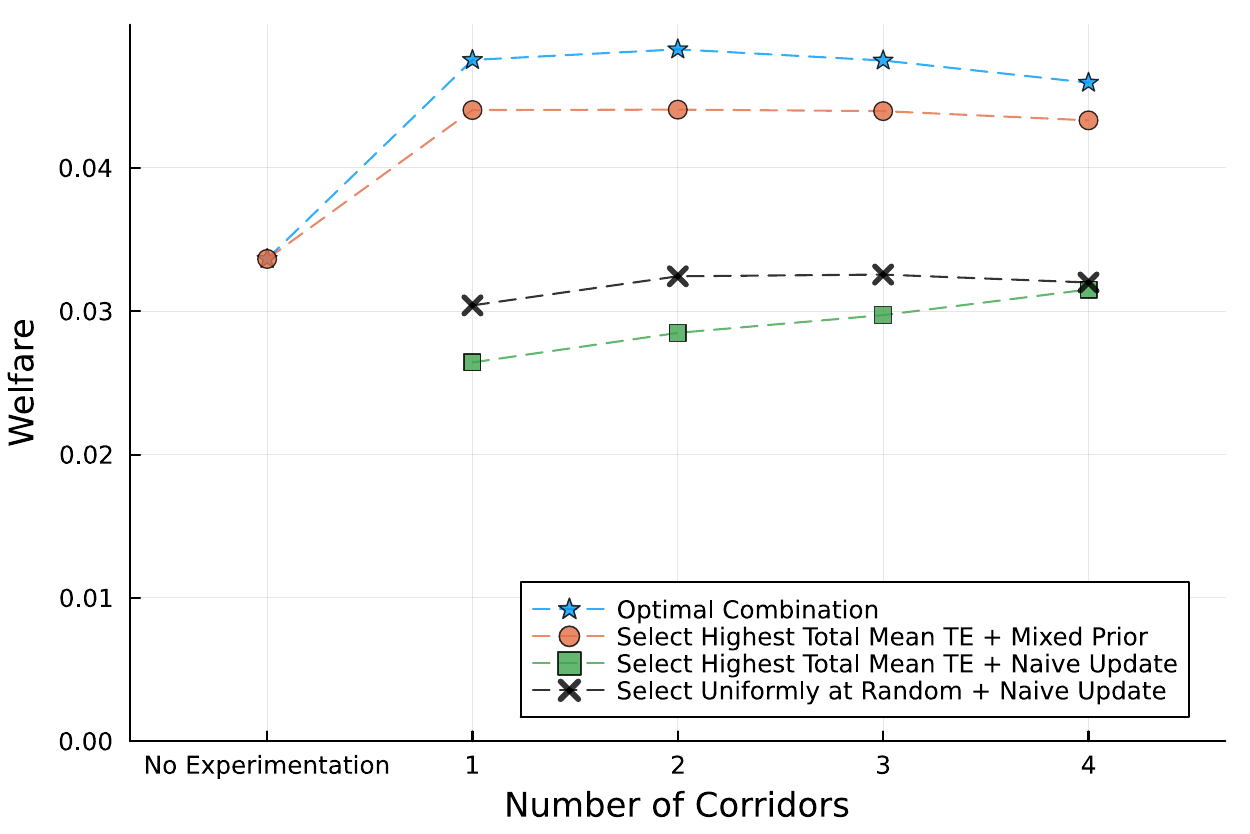}    
    \end{minipage}
    \begin{minipage}{\textwidth}
    \vspace{1pt} {\footnotesize Notes: Average Per-Migrant Welfare refers to the average across all candidate sites of the  welfare increase from site-level implementation recommendations, predicted under our preferred mixed prior with 0.5 weight on the structural model.
    Pakistan plot is restricted to marginal sites, where site $s$ is considered marginal if $
\pr(\tau_s \geq \text{cost}_s) \geq \kappa \text{ and }  \pr(\tau_s \leq \text{cost}_s) \geq \kappa,$ where $\tau_s$ and $\text{cost}_s$ are the welfare benefit and cost from implementing the mobile money training program there. Surveying budget is held fixed across different numbers of sites. See Appendix \ref{sec:site_calculations_by_number} for details.
    \par}
\end{minipage}
\end{figure}

Figure \ref{fig:more_less_sites} shows per-site average welfare in Bangladesh and Pakistan as we vary the number of experimental sites, 
based on calculations with a Bangladeshi survey firm reported in Appendix 
\ref{sec:site_calculations_by_number}.\footnote{The calculations are computationally intensive, and we illustrate ideas with results from Bangladesh and Pakistan where computation is easier due to the much smaller number of candidate sites than in India.}
For Pakistan, we report welfare values over marginal sites with $\kappa = 0.25$ to sharpen comparisons.
We maintain a constant budget based on our two-sites-per-country, 2,000-households-per-site mandate, which implies that the number of households we can sample diminishes with more sites due to the fixed costs associated with experimenting in each site. 
As a baseline, the figure also displays the welfare from no experimentation where assignment is based on the prior itself, i.e.~a site is treated if the prior mean for that site exceeds the cost of treatment. 

We predict welfare using the 0.5-weight mixed prior. 
Consequently, site selection using that prior will attain optimal welfare; 
this is represented by the line with star markers.
In Bangladesh we see large gains in welfare from the first experimental site, and diminishing welfare gains for subsequent sites  until welfare begins to decrease as experimental site effects become noisier due to smaller sample sizes with 5 or 6 sites.
In Pakistan the pattern is similar, with welfare beginning to decrease with the 3rd site.

To compare with optimal site selection, we consider two cases of site selection bias, where the sites chosen have the maximal sum of predicted ATEs according to the prior. 
In the first, indicated by circular markers, we allow the policymaker to make optimal use of the treatment effect estimates from the selected sites.
That is, the policymaker updates the predicted treatment effects for non-experimental sites according to the informativeness of the highest-treatment-effect sites for each one.
In Bangladesh, the site with the highest single predicted treatment effect is very unusual and contributes little to learning about other sites, but the site with the second-highest predicted treatment effect happens to be less outlying and is even selected by the structural prior in the two-site case.
Because some of the sites with high predicted effects are informative, the welfare loss relative to the optimal design be can moderated by a sophisticated policymaker who fully accounts for the biased site selection.

We compare these scenarios with a naive policymaker who assigns the treatment to all sites if the average of the estimated ATEs exceeds the cost of treatment, a case considered in \cite{Allcott2012}.
\cite{Hjort2019} provide evidence that 
policymakers may extrapolate experimental results naively. 
In their study, Brazilian municipal officials displayed a concern for the sample size of prototypical experiments, but not the contexts in which they were performed.
In our numerical experiment, naive policymakers fare very poorly when there is extreme site selection bias. 
Selecting sites with the highest predicted effects and applying the naive policy rule leads to the uniformly lowest expected welfare values, including negative welfare with fewer than three experimental sites in Bangladesh. 

We also consider the performance of the naive policy rule when sites are selected at random, which roughly approximates \cite{Allcott2012}'s recommendation to choose sites at random to avoid site selection bias when making uniform implementation recommendations on the basis of a set of experiments.
These are indicated in Figure \ref{fig:more_less_sites} by the points marked by an ``x.''
This strategy also fares quite poorly, in some cases little better than when there is extreme site selection.
While this may be surprising at first glance, randomization has a clear shortcoming in our setting. 
If a very small number of sites are chosen at random, there is a reasonably high probability that the chosen sites will be very unrepresentative of the entire country. 

\section{Conclusion}

We have developed a general framework for experimental site selection that incorporates external validity as its key aim. 
Our approach is flexible, allowing for various specifications of the decision-maker's social welfare function, various constraints on the choice set of feasible combinations of sites, and a range of prior specifications that incorporate different sources of information about the heterogeneity of effects across sites. 

In our application, we developed a structural model of individual-level responses to the intervention, which allowed us to leverage the rich data available in a pilot experiment to build an informative prior for the treatment effects at other sites. 
Then, to allow the possibility that the structural model may not capture all the underlying causal mechanisms and sources of heterogeneity, we mixed the structural prior with a smoothing prior based on a set of aggregate site-level characteristics. 
Working with quasi-posteriors associated with standard estimation methods makes it convenient to incorporate standard structural estimation strategies without needing to invoke the full machinery of Bayesian posterior inference on the structural model.  

This approach to incorporating structural modeling in experimental analyses while allowing for departures from the model may be useful in other problems. 
In our application to an ongoing experimental design in Bangladesh, Pakistan, and India, our method is computationally feasible and leads to substantial prospective gains in expected welfare relative to {\em ad hoc} approaches that approximate common practice. 

Further extensions of our approach are possible. 
Our two-stage adaptive design setting could be extended to multiple rounds, where at each round, the available information can be used to inform future experimental design choices. 
Additional dimensions of choice beyond the selection of sites, such as the joint choice of sites and sample sizes, or more sophisticated choices of conditional treatment randomization probabilities, could also be incorporated into the approach. 

\setlength{\bibsep}{1pt}
{\setstretch{1}
\footnotesize
\bibliography{library}}

\clearpage
\appendix
\newgeometry{left=0.7in, right=0.7in, top=1in, bottom=1in}
\setcounter{page}{1}
\renewcommand{\thepage}{OA\arabic{page}}
\setcounter{table}{0}
\setcounter{figure}{0}
\renewcommand{\thetable}{A\arabic{table}}
\renewcommand{\thefigure}{A\arabic{figure}}
\section*{Appendices, for Online Publication Only}

\section{Additional Tables Referenced in the Main Text}
\begin{table}[h]
\centering
\caption{Summary Statistics for Estimation Sample from the \cite{Lee2018} Experiment by Migrant Gender}
\label{table:bkash_summary_stats}
\scalebox{0.8}{\begin{tabular}{lrrrrrrrr}
\toprule																	
	&	\multicolumn{4}{c}{Treated}	&	\multicolumn{4}{c}{Control}													\\
	\cmidrule(lr){2-5}	\cmidrule(lr){6-9}															
	&	\multicolumn{2}{c}{Male}	&			\multicolumn{2}{c}{Female}			&	\multicolumn{2}{c}{Male}			&	\multicolumn{2}{c}{Female}			\\
		\cmidrule(lr){2-3}				\cmidrule(lr){4-5}				\cmidrule(lr){6-7}				\cmidrule(lr){8-9}			
Variable	&	Mean	&	Std. Dev.	&	Mean 	&	Std. Dev	&	Mean	&	Std. Dev.	&	Mean 	&	Std. Dev	\\
\cmidrule(lr){1-1}		\cmidrule(lr){2-3}				\cmidrule(lr){4-5}				\cmidrule(lr){6-7}				\cmidrule(lr){8-9}			
Average Remittance Amount per Day	&	 92.49 	&	 81.00 	&	 49.16 	&	 48.83 	&	 83.12 	&	 68.87 	&	 51.12 	&	 56.52 	\\
Mobile Money Usage (cond. on remitting) 	&	 0.54 	&	 0.45 	&	 0.46 	&	 0.47 	&	 0.43 	&	 0.46 	&	 0.34 	&	 0.46 	\\
Migrant Hourly Wage	&	 32.77 	&	 9.07 	&	 25.49 	&	 5.68 	&	 32.01 	&	 8.89 	&	 25.68 	&	 5.14 	\\
Home Income Per Day	&	 362.02 	&	 260.36 	&	 317.47 	&	 216.01 	&	 369.06 	&	 247.75 	&	 352.18 	&	 239.63 	\\
Home Household Size	&	 4.09 	&	 1.57 	&	 4.10 	&	 1.41 	&	 4.27 	&	 1.64 	&	 4.18 	&	 1.51 	\\
$N$	&	\multicolumn{2}{c}{241}			&	\multicolumn{2}{c}{83}			&	\multicolumn{2}{c}{233}			&	\multicolumn{2}{c}{100}			\\
\bottomrule					
\end{tabular}
}
\begin{minipage}{0.97\textwidth}
\vspace{1pt} {\footnotesize Notes: currency amounts are in Bangladesh taka (1  USD = 78 BDT in 2015). Mobile money usage conditional on remitting is the fraction of remittance transactions reported to have been executed via mobile money.
\par}
\end{minipage}
\end{table}

\begin{table}[h!]
\centering
\caption{Site Characteristics, Summary Statistics by Country}
\label{table:summary_characteristics}
\makebox[\textwidth][c]{
    \scalebox{0.8}{
        \begin{tabular}{llcccc}
        \toprule
        \multicolumn{2}{c}{Characteristic} & \multicolumn{1}{c}{Bangladesh} & \multicolumn{1}{c}{Pakistan} & \multicolumn{1}{c}{India} \\
        \cmidrule(lr){1-2} \cmidrule(lr){3-3} \cmidrule(lr){4-4} \cmidrule(lr){5-5}
        \multirow{2}{*}{Mean Household Income (Daily)}     & Mean & $267.05$ & $897.51$ & $378.94$ \\
                                                           & SD   & $211.38$ & $262.3$  & $261.15$ \\
        \cmidrule(lr){1-5}
        \multirow{2}{*}{SD Household Income (Daily)}       & Mean & $239.57$   & $836.32$   & $581.02$    \\
                                                           & SD   & $230.49$   & $276.15$   & $390.39$   \\
        \cmidrule(lr){1-5}
        \multirow{2}{*}{Mean Household Size}               & Mean & $4.03$   & $5.31$   & $6.37$    \\
                                                           & SD   & $0.89$   & $1.03$   & $1.04$   \\
        \cmidrule(lr){1-5}
        \multirow{2}{*}{SD Household Size}                 & Mean & $2.35$   & $2.73$   & $2.97$    \\
                                                           & SD   & $1.83$   & $0.72$   & $0.75$   \\
        \cmidrule(lr){1-5}
        \multirow{2}{*}{Mean Male Migrant Wage (Hourly)}   & Mean & $55.55$  & $125.49$ & $77.03$  \\
                                                           & SD   & $5.53$   & $61.54$  & $29.16$   \\
        \cmidrule(lr){1-5}
        \multirow{2}{*}{SD Male Migrant Wage (Hourly)}     & Mean & $72.96$   & $75.37$   & $65.19$    \\
                                                           & SD   & $8.26$   & $63.54$   & $34.87$   \\
        \cmidrule(lr){1-5}
        \multirow{2}{*}{Mean Female Migrant Wage (Hourly)} & Mean & $33.07$  & $44.17$  & $50.40$  \\
                                                           & SD   & $25.52$  & $53.52$  & $93.10$ \\
        \cmidrule(lr){1-5}
        \multirow{2}{*}{SD Female Migrant Wage (Hourly)}   & Mean & $43.03$   & $26.04$   & $27.53$    \\
                                                           & SD   & $33.67$   & $54.21$   & $83.47$   \\
        \cmidrule(lr){1-5}
        \multirow{2}{*}{Mean Remittance (Daily)}           & Mean & $178.08$ & $459.41$ & $227.46$ \\
                                                           & SD   & $71.05$  & $152.03$ & $215.361$ \\
        \cmidrule(lr){1-5}
        \multirow{2}{*}{Operator Density}                  & Mean & $0.07$   & $0.04$   & $0.05$    \\
                                                           & SD   & $0.02$   & $0.01$   & $0.05$   \\
        \cmidrule(lr){1-5}
        \multirow{1}{*}{Missing Operator Density (District Level)} & Percent  & $41.5$   & $48.3$   & $48.8$    \\
        \cmidrule(lr){1-5}
        \multirow{2}{*}{Price Index}                       & Mean & $0.88$   & $0.99$   & $0.81$    \\
                                                           & SD   & $0.145$   & $0.02$   & $0.17$   \\
        \cmidrule(lr){1-5}
        \multirow{2}{*}{Distance between Sites in Corridor}& Mean & $142.63$   & $227.83$   & $652.70$    \\
                                                           & SD   & $75.85$   & $315.12$   & $502.16$   \\
        \cmidrule(lr){1-5}
        \multirow{2}{*}{Migrant Density}                   & Mean & $0.03$   & $0.01$   & $0.17$    \\
                                                           & SD   & $0.022$   & $0.004$   & $0.22$   \\
        \cmidrule(lr){1-5}
        \multirow{1}{*}{Number of Sites} & Count & $41$   & $60$   & $740$    \\
        \bottomrule
        \end{tabular}
    }
}
\begin{minipage}{0.8\textwidth}
\vspace{1pt} {\footnotesize Notes: Site characteristics are themselves means and standard deviations across households within each site. The means and standard deviations reported in this table are the mean of means and standard deviations across sites in each country, and the standard deviations of site-level means and standard deviations.
Wages and incomes are in Bangladesh taka (1 USD = 78 BDT as of December 2015). Sites missing operator density have operator density replaced with the mean operator density at the next-highest administrative level (e.g.~the state mean in India).
Sites come from a set restricted to ensure feasibility of treatment delivery and robustness to prior misspecification. See Appendix \ref{subsec:admindata} for variable construction and site restriction details.
\par}
\end{minipage}
\end{table}

\begin{table}[ht!]
\centering
\captionof{table}{Welfare-Maximizing and Minimizing Experimental Site Combinations}
\label{table:two_site_best_worst}
\makebox[\textwidth][c]{
\scalebox{0.75}{
\begin{tabular}{llcllc}
\toprule
\multicolumn{6}{c}{Bangladesh} \\
\midrule
\multicolumn{3}{c}{Best Combinations}			&	\multicolumn{3}{c}{Worst Combinations}							\\					
\cmidrule(lr){1-3}	\cmidrule(lr){4-6}										
Corridor 1	&	Corridor 2	&	Avg.~Welfare	&	Corridor 1	&	Corridor 2	&	Avg.~Welfare	\\
\cmidrule(lr){1-1}	\cmidrule(lr){2-2}	\cmidrule(lr){3-3}				\cmidrule(lr){4-4}	\cmidrule(lr){5-5}	\cmidrule(lr){6-6}			
Dhaka-Magura	&	Dhaka-Noakhali	&	 0.030 	&		Dhaka-Madaripur	&	Dhaka-Panchagarh	&	 0.004 	\\
Dhaka-Kishoregonj	&	Dhaka-Noakhali	&	 0.030 	&		Dhaka-Madaripur	&	Dhaka-Thakurgaon	&	 0.004 	\\
Dhaka-Barguna	&	Dhaka-Kishoregonj	&	 0.030 	&		Dhaka-Gopalganj	&	Dhaka-Thakurgaon	&	 0.004 	\\
Dhaka-Bagerhat	&	Dhaka-Kishoregonj	&	 0.030 	&		Dhaka-Bhola	&	Dhaka-Thakurgaon	&	 0.004 	\\
Dhaka-Feni	&	Dhaka-Narsingdi	&	 0.029 	&		Dhaka-Bhola	&	Dhaka-Panchagarh	&	 0.004 	\\
\midrule
\multicolumn{6}{c}{Pakistan} \\
\midrule
\multicolumn{3}{c}{Best Combinations}			&	\multicolumn{3}{c}{Worst Combinations}							\\					
\cmidrule(lr){1-3}	\cmidrule(lr){4-6}										
Corridor 1	&	Corridor 2	&	Avg.~Welfare	&	Corridor 1	&	Corridor 2	&	Avg.~Welfare	\\
\cmidrule(lr){1-1}	\cmidrule(lr){2-2}	\cmidrule(lr){3-3}				\cmidrule(lr){4-4}	\cmidrule(lr){5-5}	\cmidrule(lr){6-6}	
Rawalpindi-Gujrat	&	Peshawar-Mardan	&	 0.111 	&	Karachi-Mianwali	&	Karachi-Sukkur	&	 0.103 	\\
Rawalpindi-Mandi Bahauddin	&	Peshawar-Mardan	&	 0.111 	&	Karachi-Bannu	&	Rawalpindi-Islamabad	&	 0.103 	\\
Rawalpindi-Gujrat	&	Mardan-Swabi	&	 0.111 	&	Karachi-Bannu	&	Chitral-Islamabad	&	 0.103 	\\
Rawalpindi-Gujrat	&	Hangu-Kohat	&	 0.111 	&	Rawalpindi-Islamabad	&	Karachi-Sukkur	&	 0.103 	\\
Faisalabad-Bahawalnagar	&	Peshawar-Mardan	&	 0.111 	&	Chitral-Islamabad	&	Karachi-Sukkur	&	 0.103 	\\
\midrule
\multicolumn{6}{c}{India} \\
\midrule
\multicolumn{3}{c}{Best Combinations}			&	\multicolumn{3}{c}{Worst Combinations}							\\					
\cmidrule(lr){1-3}	\cmidrule(lr){4-6}										
Corridor 1	&	Corridor 2	&	Avg.~Welfare	&	Corridor 1	&	Corridor 2	&	Avg.~Welfare	\\
\cmidrule(lr){1-1}	\cmidrule(lr){2-2}	\cmidrule(lr){3-3}				\cmidrule(lr){4-4}	\cmidrule(lr){5-5}	\cmidrule(lr){6-6}			
Thane-Balrampur	&	New Delhi-Hardwar	&	 0.209 		&	Kancheepuram-Chandigarh	&	Sitamarhi-Mumbai	&	 0.197 	\\
New Delhi-Hardwar	&	Mumbai-Sultanpur	&	 0.209 		&	Kancheepuram-Chandigarh	&	New Delhi-Chennai	&	 0.197 	\\
New Delhi-Hardwar	&	Thane-Sultanpur	&	 0.209 		&	Bangalore-Chandigarh	&	New Delhi-Chennai	&	 0.197 	\\
New Delhi-Hardwar	&	Thane-Sant Kabir Nagar	&	 0.209 		&	Bangalore-Chandigarh	&	Sitamarhi-Mumbai	&	 0.197 	\\
Thane-Basti	&	New Delhi-Hardwar	&	 0.209 		&	New Delhi-Bangalore	&	Kancheepuram-Chandigarh	&	 0.197 	\\
\bottomrule
\end{tabular}
}
}
\begin{minipage}{\textwidth}
\vspace{1pt} {\footnotesize Notes: Avg. Welfare refers to the average across all candidate sites of the per-migrant welfare increase from site-level implementation recommendations, predicted under our preferred mixed prior with 0.5 weight on the structural model. Site names are in destination-origin format.
\par}
\end{minipage}
\end{table}

\begin{table}[ht!]
	\centering
	\captionof{table}{Contributions of Marginal Sites to Average Per-Migrant Welfare in Pakistan}
	\label{table:pak_marginalwelfare_best}
	\scalebox{0.7}{
		\begin{tabular}{llcccllccc}
			\toprule
			\multicolumn{2}{c}{Best Combinations}			&	\multicolumn{3}{c}{$\kappa$}					&	\multicolumn{2}{c}{Worst Combinations}			&	\multicolumn{3}{c}{$\kappa$}					\\
			\cmidrule(lr){1-2}	\cmidrule(lr){3-5}	\cmidrule(lr){6-7} \cmidrule(lr){8-10}
			Corridor 1	&	Corridor 2	&	0.05	&	0.15	&	0.25	&	Corridor 1	&	Corridor 2	&	0.05	&	0.15	&	0.25	\\
			\cmidrule(lr){1-1}	\cmidrule(lr){2-2}	\cmidrule(lr){3-5}	\cmidrule(lr){6-6}	\cmidrule(lr){7-7}	\cmidrule(lr){8-10}							
			Rawalpindi-Gujrat	&	Peshawar-Mardan	&	0.109	&	0.077	&	0.048	&	Karachi-Mianwali	&	Karachi-Sukkur	&	0.100	&	0.066	&	0.035	\\
			Rawalpindi-Mandi Bahauddin	&	Peshawar-Mardan	&	0.109	&	0.077	&	0.048	&	Karachi-Bannu	&	Rawalpindi-Islamabad	&	0.100	&	0.066	&	0.034	\\
			Rawalpindi-Gujrat	&	Mardan-Swabi	&	0.109	&	0.077	&	0.048	&	Karachi-Bannu	&	Chitral-Islamabad	&	0.100	&	0.066	&	0.034	\\
			Rawalpindi-Gujrat	&	Hangu-Kohat	&	0.109	&	0.076	&	0.048	&	Rawalpindi-Islamabad	&	Karachi-Sukkur	&	0.100	&	0.066	&	0.034	\\
			Faisalabad-Bahawalnagar	&	Peshawar-Mardan	&	0.109	&	0.076	&	0.048	&	Chitral-Islamabad	&	Karachi-Sukkur	&	0.100	&	0.066	&	0.034	\\
			\bottomrule																			
		\end{tabular}
	}
	\captionof{table}{Contributions of Marginal Sites to Average Per-Migrant Welfare in India}
	\label{table:ind_marginalwelfare_best}
	\scalebox{0.7}{
		\begin{tabular}{llcllc}
			\toprule
			\multicolumn{2}{c}{Best Combinations}			&	$\kappa$	&	\multicolumn{2}{c}{Worst Combinations}			&	$\kappa$	\\
			\cmidrule(lr){1-2}	\cmidrule(lr){3-3}	\cmidrule(lr){4-5}		\cmidrule(lr){6-6}							
			Corridor 1	&	Corridor 2	&	0.25	&	Corridor 1	&	Corridor 2	&	0.25	\\
			\cmidrule(lr){1-1}	\cmidrule(lr){2-2}	\cmidrule(lr){3-3}				\cmidrule(lr){4-4}	\cmidrule(lr){5-5}	\cmidrule(lr){6-6}			
			Thane-Balrampur	&	New Delhi-Hardwar	&	 0.100 	&	New Delhi-Bangalore	&	Kancheepuram-Chandigarh	&	 0.078 	\\
			New Delhi-Hardwar	&	Mumbai-Sultanpur	&	 0.100 	&	Bangalore-Chandigarh	&	Sitamarhi-Mumbai	&	 0.077 	\\
			New Delhi-Hardwar	&	Thane-Sultanpur	&	 0.101	&	Bangalore-Chandigarh	&	New Delhi-Chennai	&	 0.077 	\\
			New Delhi-Hardwar	&	Thane-Sant Kabir Nagar	&	 0.101	&	Kancheepuram-Chandigarh	&	New Delhi-Chennai	&	 0.076 	\\
			Thane-Basti	&	New Delhi-Hardwar	&	 0.099 	&	Kancheepuram-Chandigarh	&	Sitamarhi-Mumbai	&	 0.077 	\\
			\bottomrule											
		\end{tabular}
	}
	\begin{minipage}{\textwidth}
		\vspace{1pt} {\footnotesize Notes: Average Per-Migrant Welfare refers to the average across all candidate sites of the  welfare increase from site-level implementation recommendations, predicted under our preferred mixed prior with 0.5 weight on the structural model.
			Site $s$ is considered marginal if $
			\pr(\tau_s \geq \text{cost}_s) \geq \kappa \text{ and }  \pr(\tau_s \leq \text{cost}_s) \geq \kappa,$ where $\tau_s$ and $\text{cost}_s$ are the welfare benefit and cost from implementing the mobile money training program there. Site names are in destination-origin format.
			\par}
	\end{minipage}
\end{table}

\clearpage
\section{Additional Figures Referenced in the Main Text}

\subsection{Prior Implied Conditional Mean Welfare Treatment Effects for Pakistan}
\label{subsec:pakistan_prior}

\begin{figure}[ht!]
    \centering
    \begin{minipage}{0.50\textwidth}
    \caption{Expected Welfare TE (Structural Prior)}
    \label{fig:welfare_te_pk_model}
    \includegraphics[width = \textwidth]{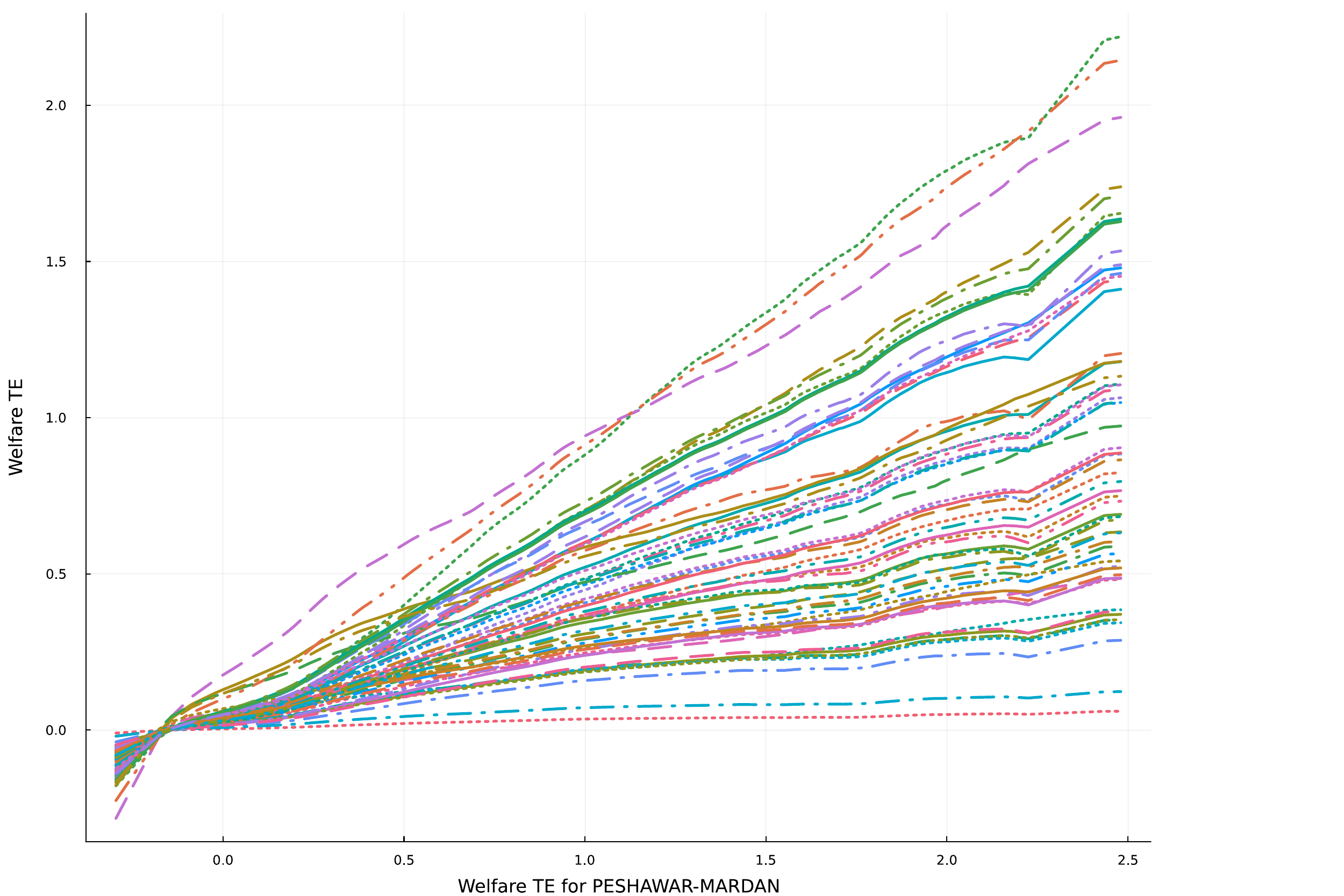}
    \end{minipage}\hfill
    \begin{minipage}{0.5\textwidth}
    \caption{Expected Welfare TE (Smoothing Prior)}
    \label{fig:welfare_te_pk_smooth}
    \includegraphics[width = \textwidth]{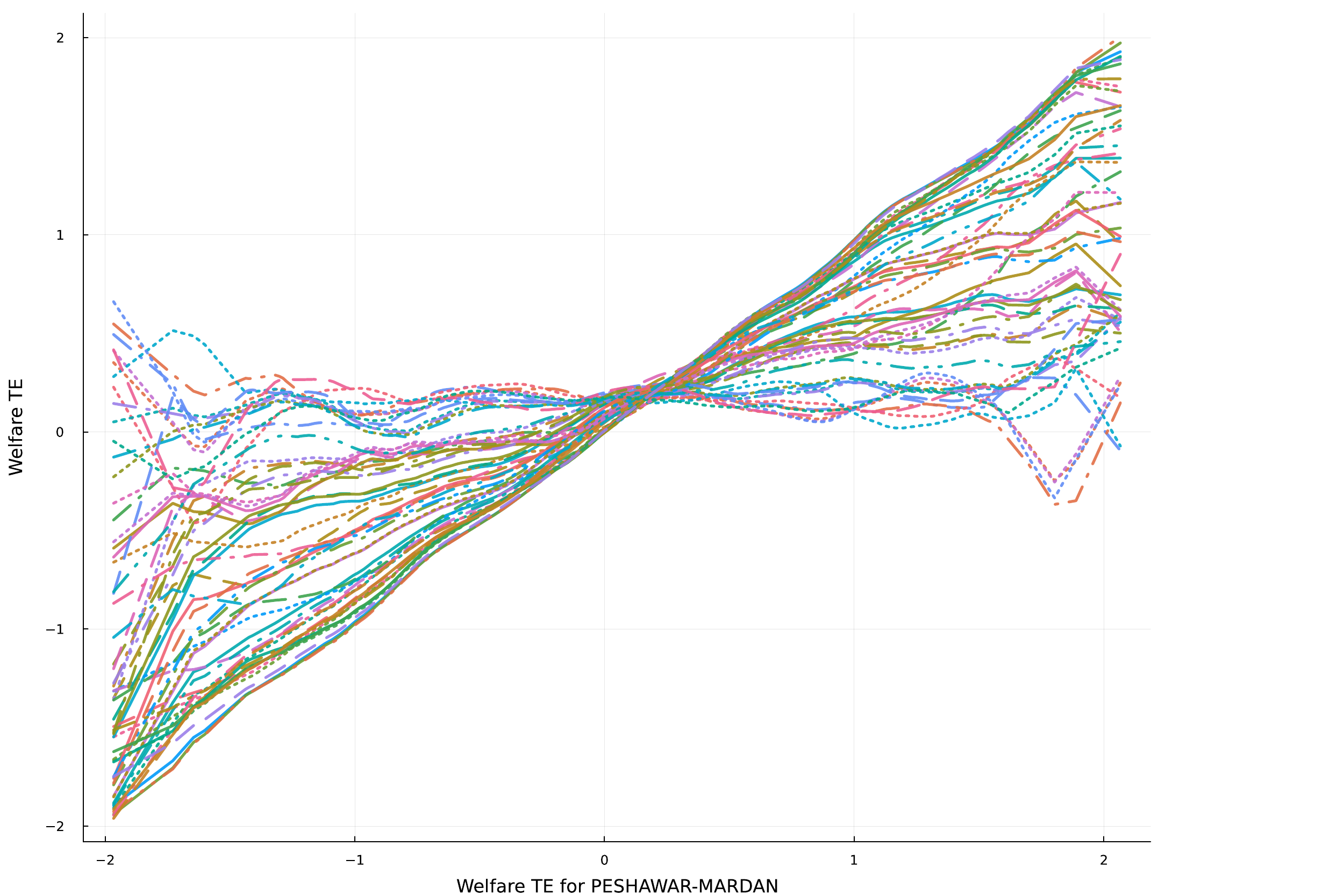}
    \end{minipage}
\begin{minipage}{\textwidth}
\vspace{1pt} {\footnotesize Notes: Each line in the figures shows the expected welfare treatment effect in a Pakistani migration corridor conditional on knowing the welfare treatment effect in the Peshawar-Mardan corridor is equal to the value on the X-axis.
\par}
\end{minipage}
\end{figure}

 Figures \ref{fig:welfare_te_pk_model} and \ref{fig:welfare_te_pk_smooth} replicate Figures \ref{fig:welfare_te_bd_model} and \ref{fig:welfare_te_bd_smooth} for potential experimental sites in Pakistan. Mean welfare treatment effects are shown, conditional on the welfare treatment effect in the corridor originating in Mardan and going to Peshawar. 

\subsection{Top 5 Best Corridor Combinations in Characteristic Space}
\label{subsec:}

\begin{figure}[ht!]
    \centering
    \caption{Welfare Maximizing (Left) and Minimizing (Right) Experimental Site Combinations in Bangladesh} 
    \label{fig:bangladesh_bestworstsites}
    \begin{minipage}{0.5\textwidth}
    \includegraphics[width = \textwidth]{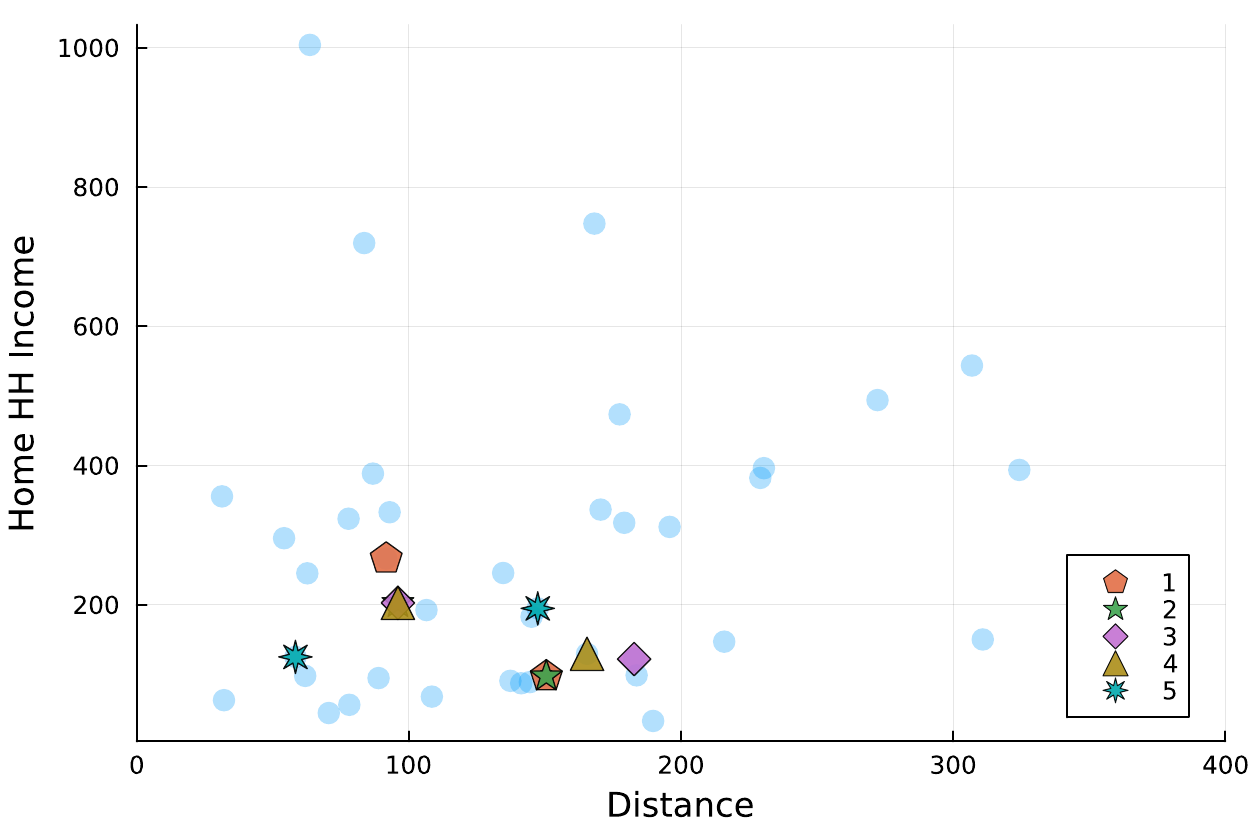}
    \end{minipage}\hfill
    \begin{minipage}{0.5\textwidth}
    \includegraphics[width = \textwidth]{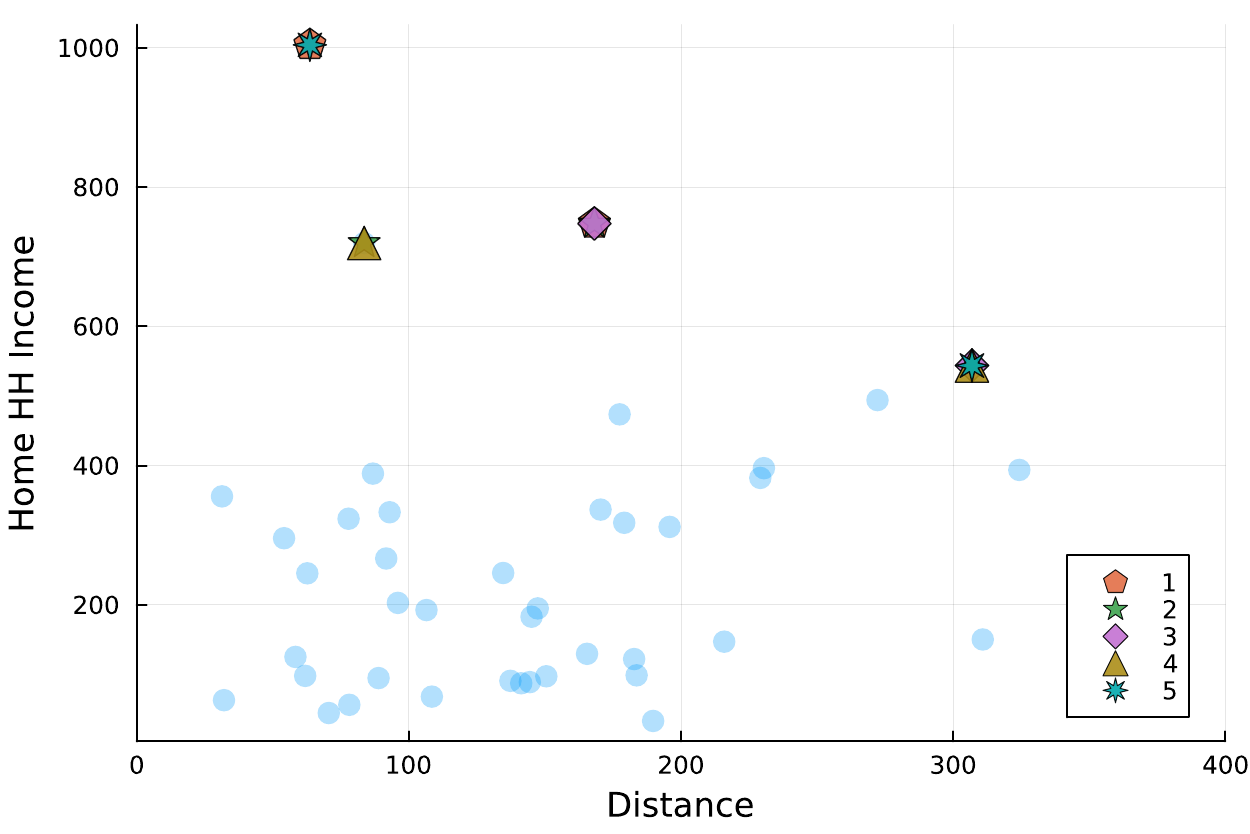}    
    \end{minipage}
    \begin{minipage}{0.95\textwidth}
        \vspace{1pt} {\footnotesize Notes: Light dots represent candidate sites.
        Numbers in the legend refer to rank among the top or bottom sites. E.g. 1 refers to the best site pair in the left panel and the worst pair in the right panel.
        Ranks are computed using our preferred mixed prior, with 0.5 weight on the structural model.
        \par}
    \end{minipage}
\end{figure}

\begin{figure}[ht!]
    \centering
    \caption{Welfare Maximizing (Left) and Minimizing (Right) Experimental Site Combinations in Pakistan} 
    \label{fig:pakistan_bestworstsites}
    \begin{minipage}{0.5\textwidth}
    \includegraphics[width = \textwidth]{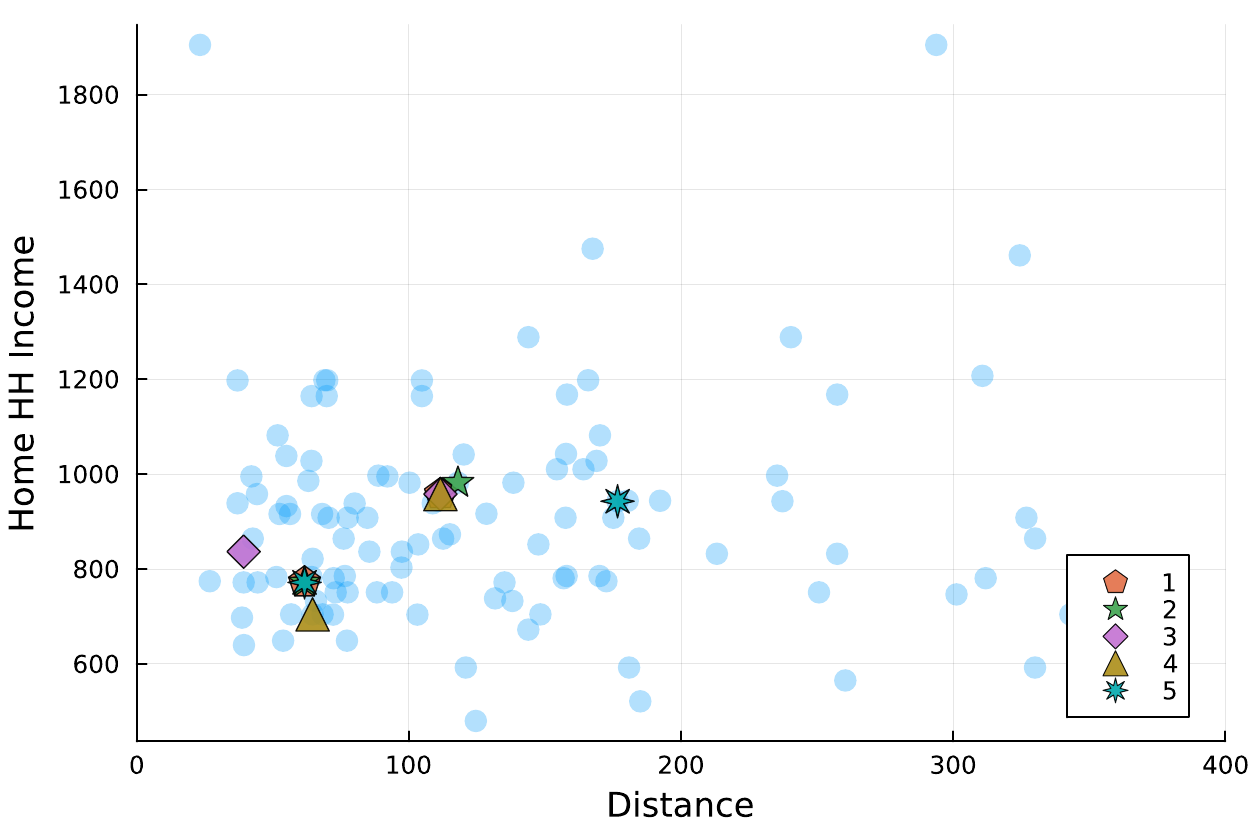}    
    \end{minipage}\hfill
    \begin{minipage}{0.5\textwidth}
    \includegraphics[width = \textwidth]{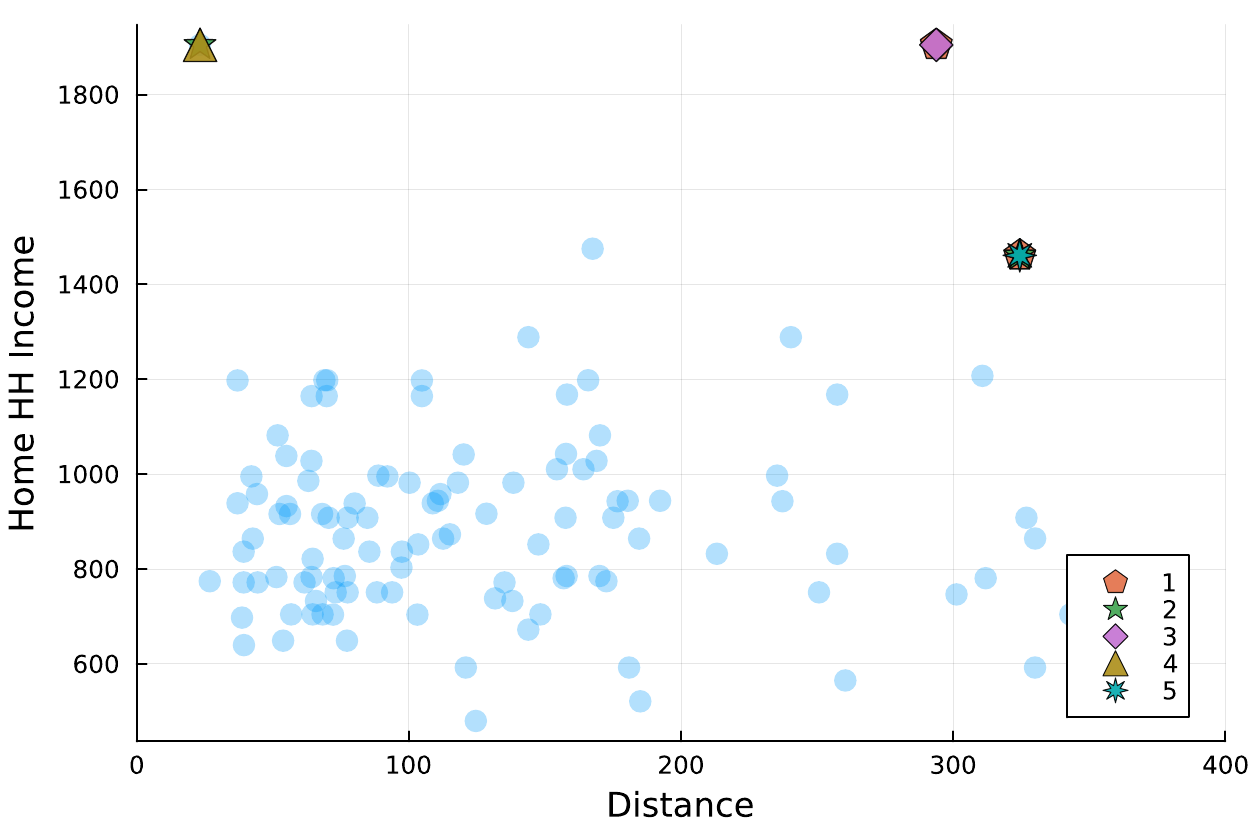}    
    \end{minipage}
    \begin{minipage}{0.95\textwidth}
        \vspace{1pt} {\footnotesize Notes: Light dots represent candidate sites.
        Numbers in the legend refer to rank among the top or bottom sites. E.g. 1 refers to the best site pair in the left panel and the worst pair in the right panel.
        Ranks are computed using our preferred mixed prior, with 0.5 weight on the structural model.
        \par}
    \end{minipage}
\end{figure}

\begin{figure}[ht!]
    \centering
    \caption{Welfare Maximizing (Left) and Minimizing (Right) Experimental Site Combinations in India} 
    \label{fig:india_bestworstsites}
    \begin{minipage}{0.5\textwidth}
    \includegraphics[width = \textwidth]{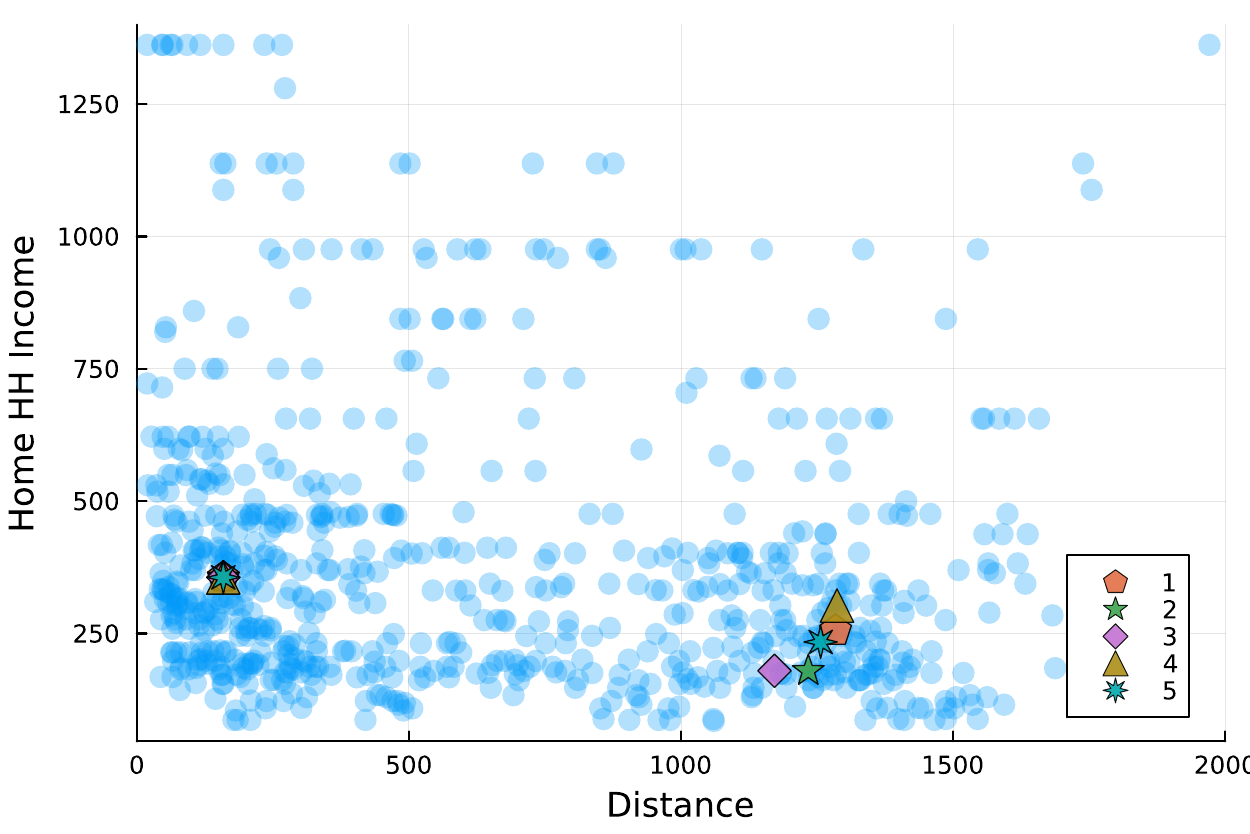}
    \end{minipage}\hfill
    \begin{minipage}{0.5\textwidth}
    \includegraphics[width = \textwidth]{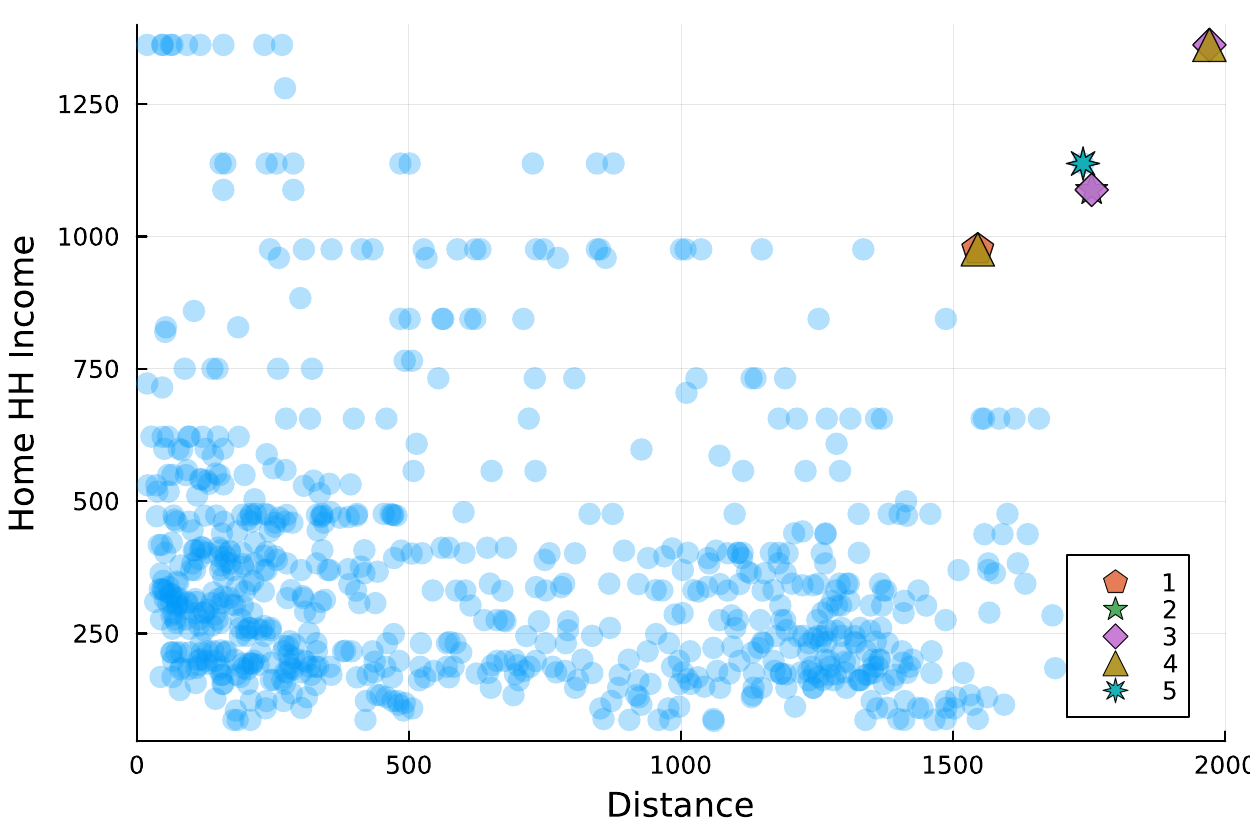}    
    \end{minipage}
    \begin{minipage}{0.95\textwidth}
        \vspace{1pt} {\footnotesize Notes: Light dots represent candidate sites.
        Numbers in the legend refer to rank among the top or bottom sites. E.g. 1 refers to the best site pair in the left panel and the worst pair in the right panel.
        Ranks are computed using our preferred mixed prior, with 0.5 weight on the structural model.
        \par}
    \end{minipage}
\end{figure}

Figures \ref{fig:bangladesh_bestworstsites} - \ref{fig:india_bestworstsites} show where the best and worst combinations of two sites according to our preferred mixed prior lie in the space of characteristics considered by the smoothing prior.
The best site combinations are located in the middle of the distribution of characteristics while the worst site combinations are outliers.

\begin{figure}[h!]
    \caption{Sites Selected by the Smoothing Prior Need Not Be Central on Non-Targeted Dimensions}
\label{fig:smoothing_nontargeted_dims}\vspace{.25cm}
    \centering \textsf{Bangladesh}\\
    \includegraphics[width = 0.5\textwidth]{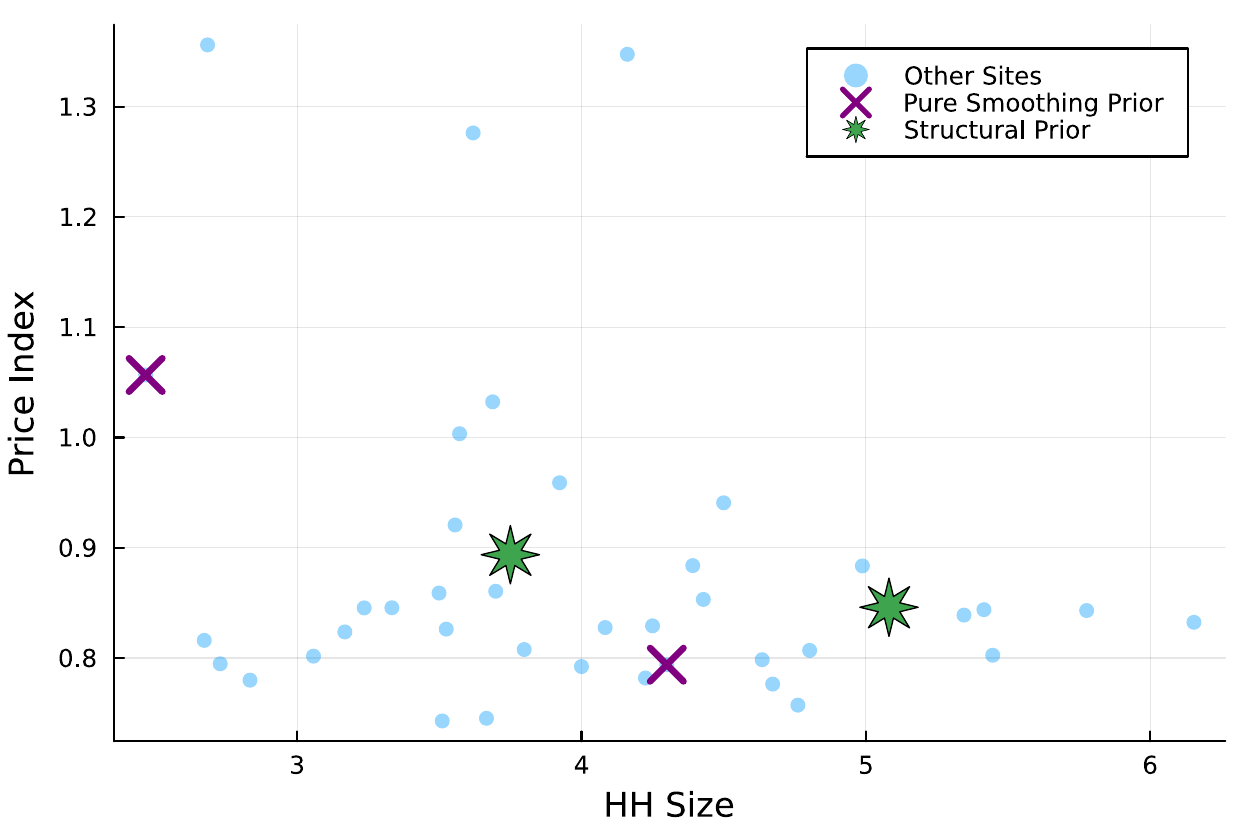}    \begin{minipage}{0.98\textwidth}
        \vspace{1pt} {\footnotesize Notes: Light dots represent candidate sites.
        Y-axis is the ratio of price level in the origin district to that in the destination district, X-axis is the average household size in the origin district.
        \par}
    \end{minipage}
\end{figure}

\clearpage

\section{Posterior Updating Rule}
\label{sec:posteriorexample}
Let the prior for $\tau$ be $\tau \sim N(\mu_{\tau},\Sigma_{\tau})$. 
Since each $\hat{\tau}_s$ is distributed $N(\tau_s,\sigma_{\epsilon,s}^2)$ conditional on $\tau$, 
we can write 
the full potential vector of estimates $\hat{\tau}=(\hat{\tau}_1,\dots,\hat{\tau}_S)$ as 
\[ \hat{\tau} \mid \tau \sim N\left(\tau,  \Sigma_\epsilon \right),\]
where $ \Sigma_\epsilon = \text{diag}\left\{ \sigma_{\epsilon,1}^2,\dots,\sigma_{\epsilon,S}^2 \right\}$ is the diagonal matrix of sampling variances. 
Combining these, we have 
\[ \begin{pmatrix}
        \tau \\ \hat{\tau} 
    \end{pmatrix} 
    \sim N_{2S}\left( 
        \begin{pmatrix}
        \mu_\tau \\ \mu_\tau 
        \end{pmatrix}, 
        \begin{pmatrix}
            \Sigma_\tau & \Sigma_\tau \\ 
            \Sigma_\tau' & \Sigma_\tau +  \Sigma_\epsilon
        \end{pmatrix}
    \right).
\] 
However, given a site-selection subset $\mathcal{S}$, we will only observe $\hat{\tau}_s$ for $s\in \mathcal{S}$. 
Let $\hat{\tau}[\mathcal{S}]$ denote the subvector of $\hat{\tau}$ selected by $\mathcal{S}$, and define $\hat{\mu}_{\tau}[\mathcal{S}]$ analogously. For any 
matrix $\Sigma$, let $\Sigma[:,\mathcal{S}]$ denote the submatrix with columns selected by $\mathcal{S}$ and all rows, and let $\Sigma[\mathcal{S},\mathcal{S}]$ denote the square submatrix with rows and columns selected by $\mathcal{S}$. 
Then the marginal distribution of $(\tau,\hat{\tau}[\mathcal{S}])$ is given by: 
\[ \begin{pmatrix}
        \tau \\ \hat{\tau}[\mathcal{S}]
    \end{pmatrix} 
    \sim N\left( 
        \begin{pmatrix}
        \mu_\tau \\ \mu_\tau[\mathcal{S}]
        \end{pmatrix}, 
        \begin{pmatrix}
            \Sigma_\tau & \Sigma_\tau[:, \mathcal{S} ] \\ 
            \Sigma_\tau[:, \mathcal{S} ]' & (\Sigma_\tau +  \Sigma_\epsilon)[\mathcal{S},\mathcal{S}]
        \end{pmatrix}
    \right).
\] 

By standard calculations, the conditional distribution of $\tau$ given $\hat{\tau}[\mathcal{S}]$ is multivariate normal with mean
\[ \mathbb{E}\left[\tau \mid \hat{\tau}[\mathcal{S}]\right] = 
\mu_\tau + \Sigma_\tau[:, \mathcal{S}] \left\{ (\Sigma_\tau +  \Sigma_\epsilon)[\mathcal{S},\mathcal{S}] \right\}^{-1}
(\hat{\tau}[\mathcal{S}] - \mu_\tau[\mathcal{S}]),
\] 
and variance 
\[ \mathbb{V}\left[\tau \mid \hat{\tau}[\mathcal{S}]\right] =
    \Sigma_{\tau} -  \Sigma_\tau[ :, \mathcal{S}] \left\{ (\Sigma_\tau +  \Sigma_\epsilon)[\mathcal{S},\mathcal{S}] \right\}^{-1} \Sigma_\tau[:,\mathcal{S}]'.
\] 
With these formulas in hand, the integrals required to calculate the preposterior welfare for any choice of $\mathcal{S}$ can be approximated efficiently by numerical simulation. 

To illustrate the posterior updating rule under multivariate normal priors, suppose there are two potential sites $s=1,2$, and under the prior their treatment effects $\tau = (\tau_1,\tau_2)^\prime$ have distribution
\[ 
	\tau = \begin{pmatrix}
		\tau_1 \\ \tau_2
	\end{pmatrix} 
	\sim 
	N\left( \begin{pmatrix}
		\mu_1 \\ \mu_2
	\end{pmatrix}, \begin{pmatrix}
		\sigma_1^2 & \sigma_{12} \\
		\sigma_{12} & \sigma_2^2
	\end{pmatrix}\right).
\] 

Suppose we are considering experimenting only on site $s=1$, so that the site selection vector is $\mathcal{S} = [1]$.
As in the main text, suppose that the estimation error if an experiment is conducted in a site is independent of the site, and determined by the sample size of the prospective experiment so that the variance matrix for the estimation errors is $ \Sigma_\epsilon =  \sigma_\epsilon^2 I$.  
Then we will observe $\hat{\tau}[\mathcal{S}] = \hat{\tau}_1 = \tau_1 + \epsilon_s$, which has conditional distribution 
\[ \hat{\tau}_1 \mid \tau \sim N\left(\tau_1,  \sigma_\epsilon^2\right). \] 

If we choose to experiment in site 1 and observe $\hat{\tau}_1$, we will be able to update our beliefs about $\tau = (\tau_1,\tau_2)$ to its posterior distribution given $\hat{\tau}_1$.
We apply the formulas above with 
\[ \hat{\tau}[\mathcal{S}] = \hat{\tau}_1, \quad \mu[\mathcal{S}] = \mu_1, 
\quad 
\Sigma_{\tau}[:,\mathcal{S}] = \begin{bmatrix}
	\sigma_1^2 \\ \sigma_{12}
\end{bmatrix},
\quad 
    (\Sigma_{\tau} +  \Sigma_\epsilon)[\mathcal{S},\mathcal{S}] = \left[ \sigma_1^2 +  \sigma_\epsilon^2\right]. 
\] 
Then, the posterior mean for $\tau$ given $\hat{\tau}_1$ is 
\begin{align*}
	\mathbb{E}\left[\tau\mid \hat{\tau}_1\right] 
	&= \mu_\tau + \Sigma_\tau[:, \mathcal{S}] \left\{ (\Sigma_\tau +  \Sigma_\epsilon)[\mathcal{S},\mathcal{S}] \right\}^{-1}
	(\hat{\tau}[\mathcal{S}] - \mu_\tau[\mathcal{S}]) \\
	&= \begin{pmatrix}
		\mu_1 \\ \mu_2 
	\end{pmatrix} + \begin{bmatrix}
		\sigma_1^2 \\ \sigma_{12}
	\end{bmatrix} \frac{1}{\sigma_1^2 +  \sigma_\epsilon^2} \left(
		\hat{\tau}_1 - \mu_1 
	\right) \\
	&= \begin{pmatrix}
		\mu_1 + \frac{\sigma_1^2}{\sigma_1^2 +  \sigma_\epsilon^2} (\hat{\tau}_1 - \mu_1) \\
		\mu_2 + \frac{\sigma_{12}}{\sigma_1^2 +  \sigma_\epsilon^2} (\hat{\tau}_1 - \mu_1)
	\end{pmatrix}.
\end{align*}
The posterior variance is 
\begin{align*}
	\mathbb{V}\left[\tau \mid \hat{\tau}_1\right] 
	&= \Sigma_{\tau} -  \Sigma_\tau[ :, \mathcal{S}] \left\{ (\Sigma_\tau +  \Sigma_\epsilon)[\mathcal{S},\mathcal{S}] \right\}^{-1} \Sigma_\tau[:,\mathcal{S}]' \\
	&= \begin{bmatrix}
		\sigma_1^2 & \sigma_{12} \\
		\sigma_{12} & \sigma_2^2
	\end{bmatrix} - \begin{bmatrix}
		\sigma_1^2 \\ \sigma_{12}
	\end{bmatrix} \frac{1}{\sigma_1^2 +  \sigma_\epsilon^2} \begin{bmatrix}
		\sigma_1^2 & \sigma_{12}
	\end{bmatrix} \\
	&= \begin{bmatrix}
		\sigma_1^2\left( 1- \frac{\sigma_1^2}{\sigma_1^2+ \sigma_\epsilon^2}\right) & \sigma_{12}\left( 1- \frac{\sigma_1^2}{\sigma_1^2+ \sigma_\epsilon^2}\right) \\
		\sigma_{12}\left( 1- \frac{\sigma_1^2}{\sigma_1^2+ \sigma_\epsilon^2}\right) & \sigma_2^2 - \frac{\sigma_{12}^2}{\sigma_1^2 +  \sigma_\epsilon^2}
	\end{bmatrix}.
\end{align*}

To interpret the posterior distribution, first consider the posterior mean of $\tau_1$. This is 
\[ \mathbb{E}[\tau_1\mid \hat{\tau}_1] = \mu_1 + \frac{\sigma_1^2}{\sigma_1^2 +  \sigma_\epsilon^2} \left(\hat{\tau}_1 - \mu_1 \right) .\] 
The second term on the right gives the incremental update to the prior mean towards the estimate $\hat{\tau}_1$. 
If we observe $\hat{\tau}_1 > \mu_1$, then the posterior mean will be greater than the prior mean $\mu_1$. We can also rearrange the expression to write 
\[ 
	\mathbb{E}[\tau_1 \mid \hat{\tau}_1] = \frac{ \sigma_\epsilon^2}{\sigma_1^2 +  \sigma_\epsilon^2}\cdot  \mu_1 + \frac{\sigma_1^2}{\sigma_1^2 +  \sigma_\epsilon^2}\cdot  \hat{\tau}_1. 
\]	
The posterior mean of $\tau_1$ will be a weighted average of the prior mean $\mu_1$ and the experimental estimate $\hat{\tau}_1$, with weights depending on the relative uncertainty in the prior (measured by $\sigma_1^2$) and the noisiness of the estimate (through $ \sigma_\epsilon^2$). For example, if $ \sigma_\epsilon^2$ is close to zero, so that the estimate is known to be very precise, then the posterior mean will be close to the estimate $\hat{\tau}_1$. 

Next, consider the posterior mean of $\tau_2$ given $\hat{\tau}_1$: 
\[ 
	\mathbb{E}[\tau_2 \mid \hat{\tau}_1] = \mu_2 + \frac{\sigma_{12}}{\sigma_1^2 +  \sigma_\epsilon^2} \left(\hat{\tau}_1 - \mu_1\right).
\] 
If $\sigma_{12} > 0$, so that according to the prior $\tau_1$ and $\tau_2$ are positively correlated because we consider them similar in some way, then observing $\hat{\tau}_1 > \mu_1$ implies that the posterior mean for $\tau_2$ will be greater than its prior mean. On the other hand, if $\sigma_{12}<0$, then the mean for $\tau_2$ will be updated in the opposite direction to $\tau_1$. 

To interpret the posterior variances, note that 
\[ \mathbb{V}\left[\tau_1 \mid \hat{\tau}_1 \right] = \sigma_1^2\left( 1- \frac{\sigma_1^2}{\sigma_1^2+ \sigma_\epsilon^2}\right).
\] 
The marginal posterior variance of $\tau_1$ will be smaller than the prior variance, unless $\sigma_1^2=0$ to begin with (meaning there was no prior uncertainty about $\tau_1$), or $ \sigma_\epsilon^2 = \infty$ (which implies that the estimator $\hat{\tau}_1$ is completely uninformative). 

To interpret the posterior variance of $\tau_2$, let $\rho = \sigma_{12}/(\sigma_1\sigma_2)$ be the  correlation of $(\tau_1,\tau_2)$ under the prior. The posterior variance can be written as
\begin{align*}
\mathbb{V}\left[ \tau_2 \mid \hat{\tau}_1 \right] 
 &= \sigma_2^2 \left( 1- \frac{\rho^2 \sigma_1^2}{\sigma_1^2 +  \sigma_\epsilon^2}\right).
\end{align*}
If $\sigma_1^2>0$ and $ \sigma_\epsilon^2 < \infty$, then the posterior variance will be smaller than the prior variance ($\sigma_2^2$) by a factor that depends on the correlation between $\tau_1$ and $\tau_2$. If $\rho=0$ then there is no information about $\tau_2$ from $\hat{\tau}_1$, and the posterior variance will equal the prior variance. 

\section{Model Solutions}
\label{sec:model-solution}

\subsection{Optimal Choices Given Remittance Price}
\label{subsec:model-solution-given-pr}

The  model laid out in Section \ref{subsec:model-solution-given-pr} can be solved to obtain the optimal remittance amount $R^*$in closed form. Depending on the values of the parameters, optimal remittances may be zero, or strictly positive as indicated in the following equation:
\begin{align}
	\label{opt_remittances}
	R^*=
	\begin{cases}
		0\hspace{4.6cm}\text{if}\hspace{3mm} &p_r\geq \frac{24\lambda a_h \eta w_m}{p_h y_h}\\
		p_h\left\{\frac{\lambda a_h\eta\left(24 w_m + p_r p_h y_h\right)}{p_h p_r \left(1+ \lambda a_h\eta\right)}-y_h\right\}\hspace{4mm}\text{if}\hspace{3mm} &p_r <\frac{24\lambda a_h \eta w_m}{p_h y_h}
	\end{cases}.
\end{align}
The mapping $\lambda \mapsto R^*$ is monotonic and continuous at the boundary points. 
The optimal values of migrant consumption and leisure are:
\[
C_m^*=\frac{(24 w_m + p_r p_h y_h)\cdot\alpha}{1+\lambda a_h\eta}; \quad 
L_m^*=\frac{(24 w_m + p_r p_h y_h)\cdot(1-\alpha)}{w_m \left(1+\lambda a_h\eta\right)}.
\]

\subsection{Optimal Distribution of Remittances with Stochastic Remittance Price}
\label{subsec:model-solution-stochastic}

Let $X$ denote the vector of household characteristics, including treatment status $t$:
\[  X = 	(p_{h}, a_{h}, w_{m}, y_{h}, \text{male},o, d, t). \] 
Let the outcome variables be 
\[
	Y = (R^*, M^*\cdot {\bf 1}(R^* > 0)).\]  
We first derive 
the distribution of the price to remit $p_r$. 
Given the distributions of $\epsilon, \xi$, we have:
\begin{align*}
	\pr\left(p_r \geq \tilde{p} \mid X \right)= &\exp\left(\min\left\{-\frac{\tilde{p}-1}{\overline{\epsilon}\cdot d}, 0\right\}\right)\cdot \exp\left(\min\left\{-\frac{\tilde{p}-1-\gamma}{\overline{\xi}\cdot e^{\psi\cdot t+\delta \cdot o}}, 0\right\}\right)\\
	=&\begin{cases}
		1\hspace{4cm}\text{ if }\hspace{3mm} \tilde{p}<1,\\
		 \exp\left(-\frac{\tilde{p}-1}{\overline{\epsilon}\cdot d}\right)\hspace{2.4cm}\text{if} \hspace{3mm} 1\leq \tilde{p}< 1+\gamma,\\
		 \exp\left(-\frac{\tilde{p}-1}{\overline{\epsilon}\cdot d}-\frac{\tilde{p}-1-\gamma}{\overline{\xi}\cdot e^{\psi\cdot t+\delta \cdot o}}\right)\hspace{9mm}\text{if} \hspace{3mm} 1+\gamma\leq \tilde{p}.
	\end{cases}
\end{align*}
Given the expression for optimal remittances, for any value $r\geq 0$ we have  
\begin{align}
	\label{eq:optimal_remit_dist}
	\pr(R^*\leq r\mid X ) &=\pr\left(p_r\geq \left.\frac{24\lambda a_h w_m \eta}{r(1+\lambda a_h \eta)+p_hy_h} \;\right\lvert\; X \right).
\end{align}
The optimal distribution of remittances now follows from the expression for $\Pr(p_r\geq \tilde{p} \mid X )$ for any $\tilde{p} \in \mathbb{R}_+$.

\subsection{Probability of Remitting Using Mobile Money}
\label{subsec:model-solution-m}

  Another quantity of interest is the probability that a migrant will remit a positive amount through mobile money. This probability, conditional on migrant characteristics, is
\begin{align}
	\label{eq:prob_mm}
	\pr(M^* =1, R^* > 0 | X )
	&= \exp\left(\frac{-\gamma}{\overline{\epsilon}d}\right)\cdot\frac{\overline{\epsilon}d}{\overline{\epsilon}d+\overline{\xi}e^{\psi t +\delta o}} \nonumber\\
	&\times {\bf 1}\left\{w_m \geq \frac{(1+\gamma)\cdot p_h y_h}{24\lambda a_h \eta}\right\}\cdot\left(1-\exp\left(-\left(\frac{1}{\overline{\xi}e^{\psi\cdot t+\delta o}}+\frac{1}{\overline{\epsilon}   d}\right)\cdot \left(\frac{24 \lambda a_h \eta w_m }{p_h y_h}-1-\gamma\right)\right)\right).
\end{align}
The indicator variable in Equation (\ref{eq:prob_mm}) reflects the necessary condition for mobile money to be the preferred mode of remittance.

\section{Asymptotic Distribution of the Structural Parameter Estimator}
\label{sec:asymptotics_appendix}

The minimum distance estimator $\hat{\theta}$ solves: 
\[ \min_{\theta} \left( \hat{\pi} - q(\theta) \right)^{\prime} \hat{W}_n 
\left( \hat{\pi} - q(\theta) \right),\]
where $\hat{W}_n \cvgp W$ for some $W$. 
We assume there is a unique solution $\theta_0$ to the population minimum distance problem 
\[ \min_\theta (\pi_0 - q(\theta))^{\prime} W (\pi_0 - q(\theta)). \] 
Under suitable conditions, we will have 
\[ 
	\sqrt{n}(\hat{\theta}-\theta_0) \cvgd N(0,\Sigma_\theta),
\]  
where 
\[ \Sigma_\theta = (G'W G)^{-1} G' W \Sigma_\pi W G (G'W G)^{-1},\]
and
$G = \frac{\partial}{\partial \theta} q(\theta_{0})$.
In order to estimate the sampling variance of $\hat{\theta}$, we estimate $G$ by evaluating the Jacobian at  $\theta = \hat{\theta}$, taking the 
derivative analytically (see Appendix \ref{subsec:moment_derivs} for the specific expressions). 
The matrix $\Sigma_\pi$ is the asymptotic variance matrix of $\hat{\pi}$, which can also be estimated using the expressions in Appendix \ref{subsec:hat_pi_asymptotics}. 

\subsection{Asymptotic Distribution of Reduced Form Estimator}
\label{subsec:hat_pi_asymptotics}

The vector $\pi$ in the minimum distance estimator consists of elements of the form 
\[ \pr\left(R\leq r\mid X\in A\right), \quad  \pr\left(M=1\mid R>0, X\in A\right),\]
for suitable choices of $r,X,A$. In addition, $\pi$ contains one component equal to the expression for $\alpha$ in Equation (\ref{eq:alpha_def}). The reduced form estimator $\hat{\pi}$ consists of sample analogs of the population quantities in $\pi$. 
Under i.i.d.~sampling and ruling out degenerate cases, we will have 
\[ \sqrt{n} (\hat{\pi}-\pi) = 
\frac{1}{\sqrt{n}}\sum_{i=1}^n \zeta_i  + o_P(1) \xrightarrow{d}N(0, \Sigma_{\pi}),
\] 
where $\Sigma_\pi = \mathbb{E}\left[\zeta_i \zeta_i ^\prime\right]$. 
The asymptotic variance can be estimated by the bootstrap, or by a sample analog 
\[ \hat{\Sigma}_{\pi} = \frac{1}{n}\sum_{i=1}^n \hat{\zeta}_i  \hat{\zeta}_i,\]
where the $\hat{\zeta}_i$ replace unknown population quantities in the formulas for $\zeta_i$ with sample estimates. 
For completeness, we give specific expressions for the influence function $\zeta_i$ next. 

For components of $\pi$ containing quantities of the form $\pr\left(R\leq r\mid X\in A\right)$, the corresponding component of $\zeta_i$ is 
\[ \frac{{\bf 1}\left\{R_i\leq r, X_i\in A\right\}-\pr(R \leq r, X \in A)}{\pr(X\in A)}-
	\frac{\pr(R\leq r\mid X\in A)}{\pr(X\in A)}\Big({\bf 1}\left\{X_i\in A\right\}-\pr\left(X\in A\right)\Big). 
\]
For components of $\pi$ containing quantities of the form $\pr\left(M=1\mid R>0, X\in A\right)$, the corresponding component of $\zeta_i$ is 
\[ 
\frac{{\bf 1}\left\{M_i=1, R_i>0, X_i\in A\right\}-\pr(M=1, R>0, X \in A)}{\pr(R>0, X\in A)} -
	\frac{\pr(M=1\mid R>0, X\in A)}{\pr(R>0, X\in A)}\Big({\bf 1}\left\{R_i>0, X_i\in A\right\}-\pr\left(R>0, X\in A\right)\Big)
\]
Finally, for the element of $\pi$ equal to $\alpha$, we have the following expression for the corresponding element of $\zeta_i$:
\[\frac{\mathbb{E}[E_m]^2}{\left(\mathbb{E}[E_m]+\mathbb{E}[W_m]\cdot\mathbb{E}[L_m]\right)^2}\Bigg\{\frac{\mathbb{E}[W_m]\cdot\mathbb{E}[L_m]}{\mathbb{E}[E_m]^2}\cdot \left(E_{m,i}-\mathbb{E}[E_m]\right)-\frac{\mathbb{E}[W_m]}{\mathbb{E}[E_m]}\left(L_{m,i}-\mathbb{E}[L_m]\right)
	-\frac{\mathbb{E}[L_m]}{\mathbb{E}[E_m]}\left(W_{m,i}-\mathbb{E}[W_m]\right)\Bigg\}. \]

\subsection{Derivatives of Estimation Moments}
\label{subsec:moment_derivs}
Consider the estimation moments,
\[\pr(R^*(\theta)\leq r \mid T=t, \text{male}=m).\]
Fixing $t$ and $m$, and denoting $\lambda=\exp(\phi_0+\phi_1 \cdot m)$, the model (refer to Equation \eqref{eq:optimal_remit_dist}) implies the following value for the joint probability $\pr(R^*(\theta)\leq r, T=t, \text{male}=m)$ :
\begin{align*}
	&\sum_a \int\limits_0^\infty \int\limits_0^{\frac{r(1+\lambda a \eta)+p_h y}{24\lambda a \eta}} f_X(w, y, a) \hspace{0.5mm}dw dy\\[2ex]
	&+ \sum_a \int\limits_0^\infty \int\limits_{\frac{r(1+\lambda a \eta)+p_h y}{24\lambda a \eta}}^{\frac{(1+\gamma)\cdot (r(1+\lambda a \eta)+p_h y)}{24\lambda a \eta}} \exp\left(-\frac{1}{\overline{\epsilon}d}\left(\frac{24\lambda a w \eta}{r(1+\lambda a \eta)+p_h y}-1\right)\right)\hspace{1mm}f_X(w, y, a)\hspace{0.5mm}dw dy\\[2ex]
	&+ \sum_a \int\limits_0^\infty \int\limits_{\frac{(1+\gamma)\cdot( r(1+\lambda a \eta)+p_h y)}{24\lambda a \eta}}^{\infty} \exp\left(-\frac{1}{\overline{\epsilon}d}\left(\frac{24\lambda a w \eta}{r(1+\lambda a \eta)+p_h y}-1\right)-\frac{1}{\overline{\xi}e^{\psi t+ \delta o}}\left(\frac{24\lambda a w \eta}{r(1+\lambda a \eta)+p_h y}-1-\gamma\right)\right)\hspace{1mm}f_X(w, y, a)\hspace{0.5mm}dw dy,
\end{align*}
where $f_X$ denotes the joint density function for $(w_m, y_h, a_h)$. The derivatives of this object with respect to $(\psi, \lambda, \overline{\epsilon}, \overline{\xi})$ are displayed below.
\begin{align*}
	&\frac{\partial \pr(R^*(\theta)\leq r, T=t, \text{male}=m)}{\partial \lambda}=\\[2ex]
	&-\sum_a \int\limits_0^\infty \int\limits_{\frac{r(1+\lambda a \eta)+p_h y}{24\lambda a \eta}}^{\frac{(1+\gamma)\cdot( r(1+\lambda a \eta)+p_h y)}{24\lambda a \eta}} \exp\left(-\frac{1}{\overline{\epsilon}  d}\left(\frac{24\lambda a w \eta}{r(1+\lambda a \eta)+p_h y}-1\right)\right)\cdot \left(\frac{(r+ p_h y)24 a \eta w}{(r(1+\lambda a \eta)+p_h y)^2}\cdot \frac{1}{\overline{\epsilon}   d}\right)\hspace{1mm}f_X(w, y, a)\hspace{0.5mm}dw dy\\[2ex]
	&- \sum_a \int\limits_0^\infty \int\limits_{\frac{(1+\gamma)\cdot (r(1+\lambda a \eta)+p_h y)}{24\lambda a \eta}}^{\infty} \exp\left(-\frac{1}{\overline{\epsilon}  d}\left(\frac{24\lambda a w \eta}{r(1+\lambda a \eta)+p_h y}-1\right)-\frac{1}{\overline{\xi}e^{\psi t+ \delta o}}\left(\frac{24\lambda a w \eta}{r(1+\lambda a \eta)+p_h y}-1-\gamma\right)\right)  \\[2ex]
	&\hspace{7cm}\times \left(\frac{(r+ p_h y)24 a \eta w}{(r(1+\lambda a \eta)+p_h y)^2}\cdot \left(\frac{1}{\overline{\epsilon}   d}+\frac{1}{\overline{\xi}e^{\psi t+\delta o}}\right)\right)\hspace{1mm}f_X(w, y, a)\hspace{0.5mm}dw dy.
\end{align*}
\begin{align*}
	&\frac{\partial \pr(R^*(\theta)\leq r, T=t, \text{male}=m)}{\partial \overline{\epsilon}}=\\[2ex]
	&\sum_a \int\limits_0^\infty \int\limits_{\frac{r(1+\lambda a \eta)+p_h y}{24\lambda a \eta}}^{\frac{(1+\gamma)\cdot (r(1+\lambda a \eta)+p_h y)}{24\lambda a \eta}} \exp\left(-\frac{1}{\overline{\epsilon}  d}\left(\frac{24\lambda a w \eta}{r(1+\lambda a \eta)+p_h y}-1\right)\right)\cdot \left(\frac{24\lambda a w \eta}{r(1+\lambda a \eta)+p_h y}-1\right)\cdot \frac{1}{\overline{\epsilon}^2  d}\hspace{1mm}f_X(w, y, a)\hspace{0.5mm}dw dy\\[2ex]
	&+ \sum_a \int\limits_0^\infty \int\limits_{\frac{(1+\gamma)\cdot (r(1+\lambda a \eta)+p_h y)}{24\lambda a \eta}}^{\infty} \exp\left(-\frac{1}{\overline{\epsilon}  d}\left(\frac{24\lambda a w \eta}{r(1+\lambda a \eta)+p_h y}-1\right)-\frac{1}{\overline{\xi}e^{\psi t+ \delta o}}\left(\frac{24\lambda a w \eta}{r(1+\lambda a \eta)+p_h y}-1-\gamma\right)\right)  \\[2ex]
	&\hspace{7cm}\times \left(\frac{24\lambda a w \eta}{r(1+\lambda a \eta)+p_h y}-1\right)\cdot \frac{1}{\overline{\epsilon}^2  d}\hspace{1mm}f_X(w, y, a)\hspace{0.5mm}dw  dy.
\end{align*}

\begin{align*}
	&\frac{\partial \pr(R^*(\theta)\leq r, T=t, \text{male}=m)}{\partial \overline{\xi}}=\\[2ex]
	&\sum_a \int\limits_0^\infty \int\limits_{\frac{(1+\gamma)\cdot (r(1+\lambda a \eta)+p_h y)}{24\lambda a \eta}}^{\infty} \exp\left(-\frac{1}{\overline{\epsilon}  d}\left(\frac{24\lambda a w \eta}{r(1+\lambda a \eta)+p_h y}-1\right)-\frac{1}{\overline{\xi}e^{\psi t+ \delta o}}\left(\frac{24\lambda a w \eta}{r(1+\lambda a \eta)+p_h y}-1-\gamma\right)\right) \\[2ex]
	&\hspace{6cm}\times \left(\frac{24\lambda a w \eta}{r(1+\lambda a \eta)+p_h y}-1-\gamma\right)\cdot \frac{1}{\overline{\xi}^2e^{\psi t+ \delta o}}\hspace{1mm}f_X(w, y, a)\hspace{0.5mm} dw dy.
\end{align*}

\begin{align*}
	&\frac{\partial \pr(R^*(\theta)\leq r, T=t, \text{male}=m)}{\partial \psi}=\\[2ex]
	&\sum_a \int\limits_0^\infty \int\limits_{\frac{(1+\gamma)\cdot (r(1+\lambda a \eta)+p_h y)}{24\lambda a \eta}}^{\infty} \exp\left(-\frac{1}{\overline{\epsilon}  d}\left(\frac{24\lambda a w \eta}{r(1+\lambda a \eta)+p_h y}-1\right)-\frac{1}{\overline{\xi}e^{\psi t+ \delta o}}\left(\frac{24\lambda a w \eta}{r(1+\lambda a \eta)+p_h y}-1-\gamma\right)\right)  \\[2ex]
	&\hspace{6cm}\times \left(\frac{24\lambda a w \eta}{r(1+\lambda a \eta)+p_h y}-1-\gamma\right)\cdot \frac{t}{\overline{\xi}e^{\psi t+ \delta o}}\hspace{1mm}f_X(w, y, a)\hspace{0.5mm} dw dy.
\end{align*}
Estimation also involves the probability of using mobile money, conditional on remitting a positive amount. Given characteristics $X$, this is the probability assigned to the event:
\[\gamma+\xi e^{\psi\cdot t+\delta \cdot o}\leq \epsilon   \cdot d \quad \text{and}\quad p_r\leq \frac{24 \lambda a_h\eta w_m}{p_h\cdot y_h}.\]
The above event is non-empty only if, 
\[w_m \geq (1+\gamma)\cdot \frac{p_h \cdot y_h}{24 \lambda a_h \eta}.\]
Once again, fixing $t, m$ and denoting $\lambda=\exp(\phi_0+\phi_1 \cdot m)$, recall that the probability $\pr(M^*=1, R^*>0 \mid T=t, \text{male}=m)$ equals
\begin{align*}
	&\exp\left(\frac{-\gamma}{\overline{\epsilon}  d}\right)\cdot\frac{\overline{\epsilon}  d}{\overline{\epsilon}  d+\overline{\xi}e^{\psi t +\delta o}}\\
	&\hspace{2cm} \times \sum\limits_a\int\limits_0^\infty\int\limits_{\frac{(1+\gamma)\cdot p_h y}{24\lambda a \eta}}^\infty\left(1-\exp\left(-\left(\frac{1}{\overline{\xi}e^{\psi\cdot t+\delta o}}+\frac{1}{\overline{\epsilon}  d}\right)\cdot \left(\frac{24 \lambda a \eta w}{p_h y}-1-\gamma\right)\right)\right)f_{X}(w, y, a)\hspace{1mm} dw dy.
\end{align*}
Denote the integral shown above as $I(\theta)$, and the function multiplying it $C(\theta)$. Partial derivatives are displayed below.
\begin{align*}
	&\frac{\partial \pr(M^*=1, R^*>0, T=t, \text{male}=m)}{\partial \lambda}=\\[2ex]
	&C(\theta)\cdot \sum_a \int\limits_0^\infty \int\limits_{\frac{(1+\gamma)\cdot p_h y}{24\lambda a \eta}}^{\infty} \exp\left(-\left(\frac{1}{\overline{\xi}e^{\psi\cdot t+\delta o}}+\frac{1}{\overline{\epsilon}  d}\right)\cdot \left(\frac{24 \lambda a \eta w}{p_h y}-1-\gamma\right)\right)\left(\frac{1}{\overline{\xi}e^{\psi\cdot t+\delta o}}+\frac{1}{\overline{\epsilon}  d}\right)\cdot \frac{24 \eta w a}{p_h \cdot y}\hspace{1mm}f_X(w, y, a)\hspace{0.5mm} dw dy.
\end{align*}
\begin{align*}
	&\frac{\partial \pr(M^*=1, R^*>0, T=t, \text{male}=m)}{\partial \overline{\epsilon}}=\\[2ex]
	&-C(\theta)\cdot \sum_a \int\limits_0^\infty \int\limits_{\frac{(1+\gamma)\cdot p_h y}{24\lambda a \eta}}^{\infty} \exp\left(-\left(\frac{1}{\overline{\xi}e^{\psi\cdot t+\delta o}}+\frac{1}{\overline{\epsilon}  d}\right)\cdot \left(\frac{24 \lambda a_h \eta w_m}{p_h y_h}-1-\gamma\right)\right)\left(\frac{24 \lambda a_h \eta w_m}{p_h y_h}-1-\gamma\right)\frac{1}{\overline{\epsilon}^2  d}\hspace{1mm}f_X(w, y, a)\hspace{0.5mm} dw dy\\
	&\hspace{2cm}+\exp\left(\frac{-\gamma}{\overline{\epsilon}  d}\right)\cdot\left[\frac{\gamma}{\overline{\epsilon}\cdot\left(\overline{\epsilon}  d+\overline{\xi}e^{\psi t+\delta o}\right)}+\frac{\overline{\xi}e^{\psi t+\delta o}  d}{\left(\overline{\epsilon}  d+\overline{\xi}e^{\psi t+\delta o}\right)^2}\right]\cdot I(\theta).
\end{align*}
\begin{align*}
	&\frac{\partial \pr(M^*=1, R^*>0, T=t, \text{male}=m)}{\partial \overline{\xi}}=\\[2ex]
	&-C(\theta)\cdot \sum_a \int\limits_0^\infty \int\limits_{\frac{(1+\gamma)\cdot p_h y}{24\lambda a \eta}}^{\infty} \exp\left(-\left(\frac{1}{\overline{\xi}e^{\psi\cdot t+\delta o}}+\frac{1}{\overline{\epsilon}  d}\right)\cdot \left(\frac{24 \lambda a_h \eta w_m}{p_h y_h}-1-\gamma\right)\right)\left(\frac{24 \lambda a_h \eta w_m}{p_h y_h}-1-\gamma\right)\frac{1}{\overline{\xi}^2e^{\psi t +\delta o}}\hspace{1mm}f_X(w, y, a)\hspace{0.5mm} dw dy\\
	&\hspace{2cm}-\exp\left(\frac{-\gamma}{\overline{\epsilon}  d}\right)\cdot \left[\frac{\overline{\epsilon}  de^{\psi t+\delta o}}{\left(\overline{\epsilon}  d+\overline{\xi}e^{\psi t+\delta o}\right)^2}\right]\cdot I(\theta).
\end{align*}
\begin{align*}
	&\frac{\partial \pr(M^*=1, R^*>0, T=t, \text{male}=m)}{\partial \psi}=\\[2ex]
	&-C(\theta)\cdot \sum_a \int\limits_0^\infty \int\limits_{\frac{(1+\gamma)\cdot p_h y}{24\lambda a \eta}}^{\infty} \exp\left(-\left(\frac{1}{\overline{\xi}e^{\psi\cdot t+\delta o}}+\frac{1}{\overline{\epsilon}  d}\right)\cdot \left(\frac{24 \lambda a_h \eta w_m}{p_h y_h}-1-\gamma\right)\right)\left(\frac{24 \lambda a_h \eta w_m}{p_h y_h}-1-\gamma\right)\frac{t}{\overline{\xi}e^{\psi t +\delta o}}\hspace{1mm}f_X(w, y, a)\hspace{0.5mm}dw dy\\
	&\hspace{2cm}-\exp\left(\frac{\gamma}{\overline{\epsilon}  d}\right)\cdot \frac{t}{\overline{\xi}\cdot e^{\psi\cdot t+\delta o}}\left(\frac{2}{\overline{\xi}e^{\psi t+\delta o}}+\frac{1}{\overline{\epsilon}  d}\right)\cdot I(\theta).
\end{align*}

\section{Parameters Normalized and Estimated Separately}
\label{sec:normalized_params}
The parameter $\gamma$, reflecting the price of remittance transfer via mobile money, is obtained directly from the actual price schedules as described in Section \ref{subsec:admindata}.
The parameter $\eta$ will be normalized to equal one, because only the product of $\eta$ and $\lambda$ appear in the expression for the distribution of optimal remittances in Equation (\ref{eq:optimal_remit_dist}).    

A similar issue arises for inferring $\delta$ and $\overline{\xi}$. 
The experimental data has no variation in operator density since all observations came from a single migration corridor. 
For estimation purposes we set $\delta=0$ and estimate the parameter $\overline{\xi}$. 
In Section \ref{subsec:prediction} we describe how the model is used to make predictions about treatment effects in sites not covered in the study of \citet{Lee2018}. These sites have different values for operator density, and we can incorporate this information by calibrating a value for $\delta$ given our estimate of $\overline{\xi}$. In particular, let $o_g$ denote the operator density in Gaibandha, and let $\hat{\overline{\xi}}$ denote the estimated value of $\overline{\xi}$. We then set:
\[
\hat{\delta} = \frac{\ln \hat{\overline{\xi}}}{o_g}.
\]

\section{Model Fit to the Pilot Experiment}
\label{subsec:modelfit}

Table \ref{table:modelfit_moments} displays model implied values for the treatment effect on daily remittances sent and the probability of remitting a positive amount via mobile money (bKash).

\begin{table}[h]
\caption{Model-Implied and Estimated Average Treatment Effects}
\label{table:modelfit_moments}
\centering
\begin{tabular}{lcc}
    \toprule
    Calculated Using&Daily Remittances& $\pr (MM=1, R>0)$\\
    \cmidrule(lr){1-1}\cmidrule(lr){2-2}\cmidrule(lr){3-3}
    \cite{Lee2018}  Data & 6.41 & 0.10\\
    Structural Model & 5.94 & 0.11\\
    \bottomrule
\end{tabular}
\end{table}

\begin{figure}[ht!]
    \centering
        \caption{Average Daily Remittances for Female Migrants}
    \label{fig:modelfit_female}
    \vspace{.25cm}
    \begin{minipage}{0.5\textwidth} 
    \centering \textsf{Control}\\
    \includegraphics[width = \textwidth]{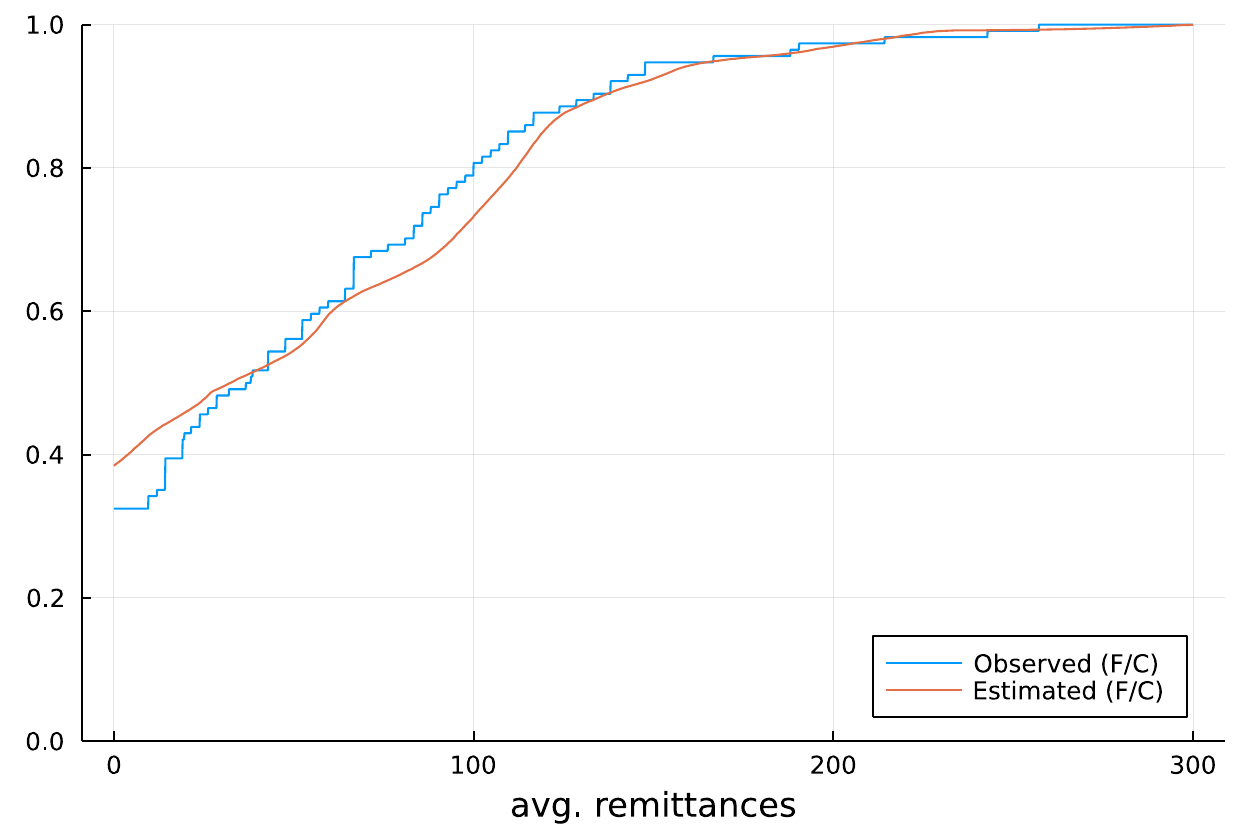}
    \end{minipage}\hfill
    \begin{minipage}{0.5\textwidth}
    \centering \textsf{Treated}\\ 
    \includegraphics[width = \textwidth]{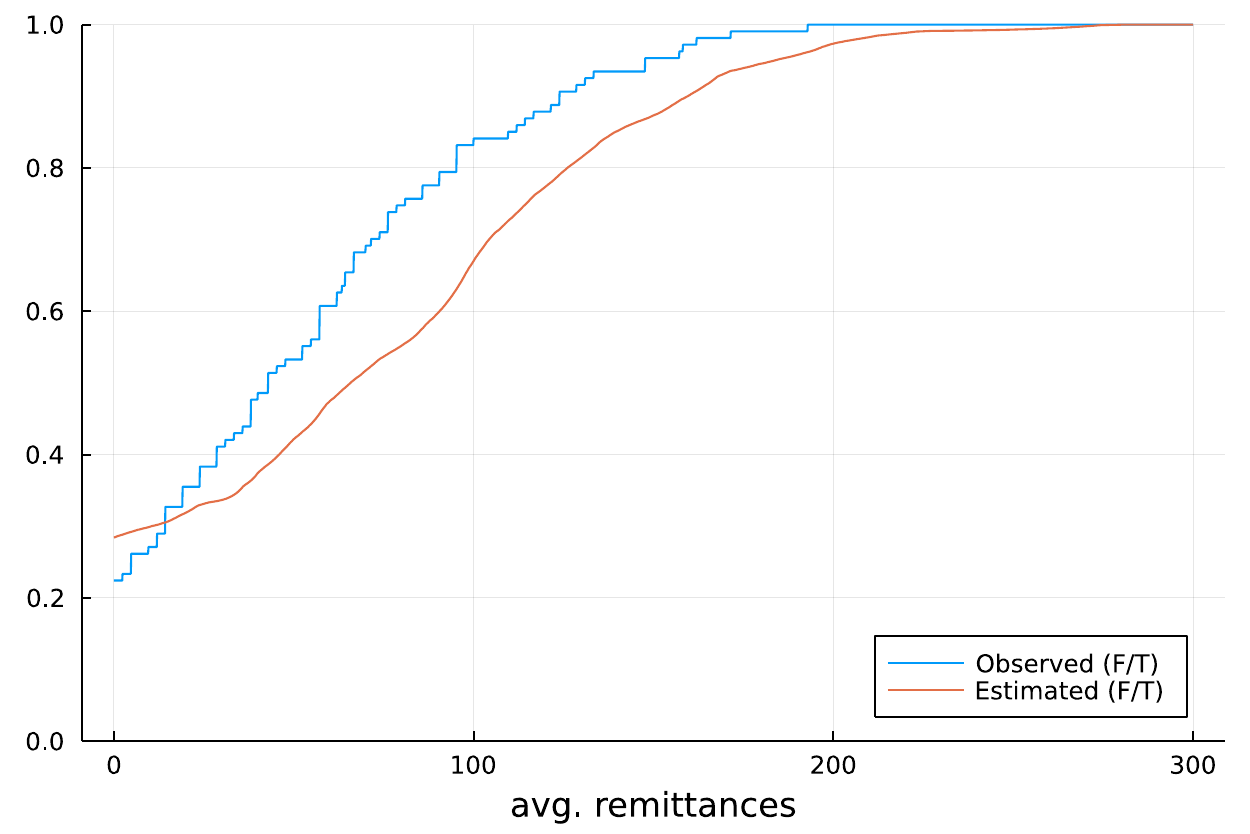}
    \end{minipage}
\end{figure}

\begin{figure}[ht!]
    \centering
        \caption{Average Daily Remittances for Male Migrants}
    \label{fig:modelfit_male}
    \vspace{.25cm}
    \begin{minipage}{0.5\textwidth}
    \centering \textsf{Control}\\
    \includegraphics[width = \textwidth]{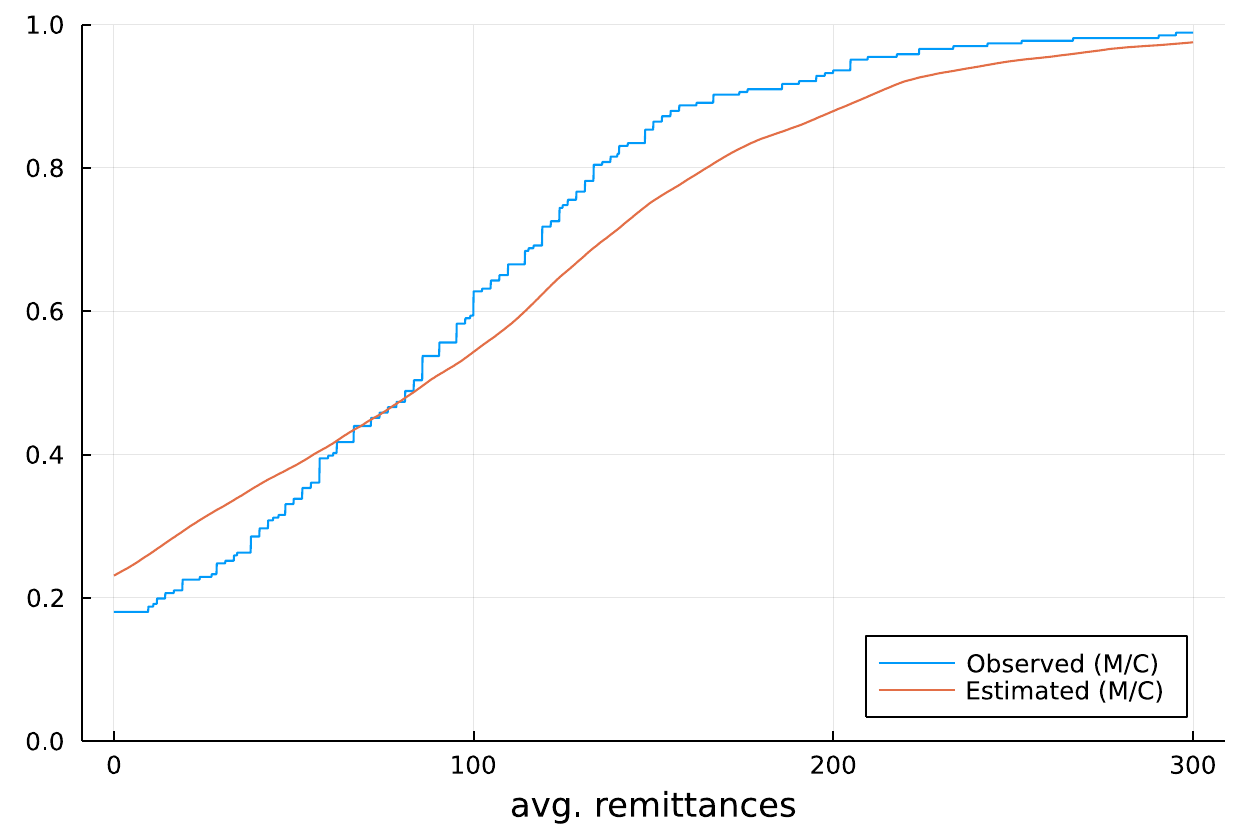}
    \end{minipage}\hfill
    \begin{minipage}{0.5\textwidth}
    \centering \textsf{Treated}\\
    \includegraphics[width = \textwidth]{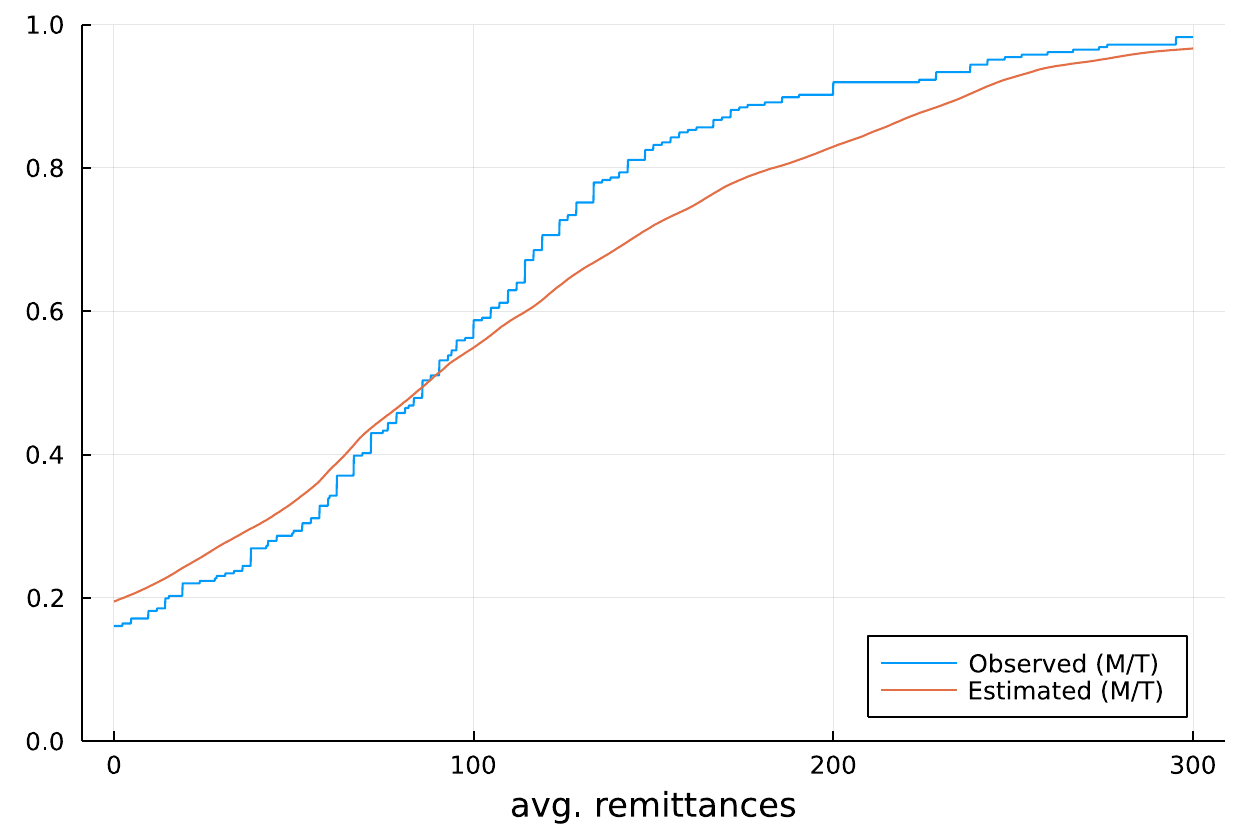}
    \end{minipage}
    \begin{minipage}{\textwidth}
    \vspace{1pt} {\footnotesize Notes: observed refers to the empirical CDF from the \cite{Lee2018} data, estimated to the structural model-implied CDF at the estimated parameters.
    \par}
\end{minipage}
\end{figure}

Figures \ref{fig:modelfit_female} and \ref{fig:modelfit_male} display the observed and model implied cumulative distribution functions of average daily remittances when the model of Section \ref{sec:model} is fit to the experimental data of \cite{Lee2018}. 
We use the point estimates in Table \ref{table:estimates} to generate the model-implied distributions of daily remittances, and show the empirical and model-implied CDFs separately by gender and treatment status.
The parsimonious model largely fits the data well, with the exception of overpredicting the treatment effect on remittances for female migrants, which consitute a small share of the total sample, as shown in Table \ref{table:bkash_summary_stats}.

\section{Deriving the Welfare Cost of Treating a Site}
\label{sec:welfare-cost}

Following the standard benchmark in the public finance literature (c.f.~\citealt{Finkelstein2020}), we assume taxpayers incur a 30\% deadweight loss, so that the cost of the program from their perspective is 1152 taka.\footnote{This may differ in developing countries --- for example because of lower overall tax burdens, a lower labor supply elasticity and the mode of taxes (e.g.~VAT vs.~labor tax).  These factors may result in a different, potentially lower deadweight loss.}
Consider the consumption tax rate $\omega$ needed to pay for the program on a per-migrant basis.
As above, $u_j$ for taxpayer $j$ is per-capita expenditure in $j$'s household.
To pay for the program with a uniform consumption tax at rate $\omega$, $u_j$ must go down by
\[
  \omega u_j = \omega \cdot \frac{\text{household expenditure}_j}{\text{hh size}_j}.
\]

We take a first-order approximation to the impact of this change in $u_j$ on social welfare, multiplying the change in $u_j$ by the social marginal utility associated with $u_j$.
Because of the log specification for the social value of $u_j$, the social marginal utility of a change in $u_j$ is
\[
  \frac{\text{hh size}_j}{\text{household expenditure}_j}.
\]
All members $j$ within the same household have the same social marginal utility and change in $u_j$ so we can sum over households $h$ to get the aggregate effect of the the tax,
\begin{equation}\label{eq:tax_welfare_loss}
 \sum_h \text{hh size}_h \cdot \omega.
\end{equation}
It remains to derive the size of $\omega$ needed to pay for the program on a per migrant basis.
This solves:
\begin{align}
    \label{eq:tax_size}
    1152 = \omega \cdot  \text{total expenditure in the country}.
\end{align}
Combining \eqref{eq:tax_welfare_loss} and \eqref{eq:tax_size}, we can rewrite the treatment cost in welfare terms as:
$$
  1152 \cdot \frac{1}{\text{national per-capita expenditure}}.
$$

To compute national per-capita expenditure in the three countries, we use the ``Final consumption expenditure'' series from the World Bank's World Development Indicators database in constant local currency units and divide it by the contemporaneous value of total population, also from World Development Indicators.
For compatibility with the per-capita expenditures from \citet{Lee2018}, we convert ``Final consumption expenditure'' to the monthly level and use 2015 values.
For Pakistan and India, we convert rupee values to Bangladesh taka using the April 1, 2015 market exchange rate from xe.com.

\section{Migration Corridor Data and Restrictions}

\subsection{Data on Corridor Characteristics}
\label{subsec:admindata}

As discussed in Section \ref{subsec:prediction}, to make predictions for future experimental sites or migration corridors, we need site-level data such as the migrant's wage rate and their family's income. This information is taken from multiple sources for each country. Monetary values are adjusted for inflation to ensure that all values accord with the original experiment of \cite{Lee2018}. 
Home households are defined as those who report having sent a domestic migrant or, when this information is not available, receiving remittances from a domestic source.

For all three countries, we measure mobile money operator density in the origin district information from the latest available waves of InterMedia's Financial Inclusion Insights surveys, conducted in collaboration with the Bill and Melinda Gates Foundation. For Bangladesh and India, we use data from 2018, while for Pakistan we use data from 2020. 
Our operator density measure is based on the time it takes an individual to reach their nearest mobile money agent. A lower average time corresponds to higher operator density in a district.
To ease interpretation of this measure as operator density, we invert it so that it is 1 divided by the time taken (in minutes) to the nearest mobile money operator.
We measure the length of a corridor as the distance in kilometers as the crow flies between the centroid of the origin district and the destination district.

The explicit per-unit monetary cost of remitting via mobile money, parameterized by $\gamma$, is calculated separately for each country. For Bangladesh we use the per-unit cost to remit using mobile money as reported in \cite{Lee2018}. For Pakistan, we use the cost schedule from a popular mobile money service (Telenor Microfinance Bank) to construct the parameter. For India, we construct $\gamma$ using the cost of remitting though common digital transfer methods available to those who have bank accounts at post offices. Specifically, we use the per-unit cost of transferring money from one bank account to another through the National Electronic Funds Transfer (NEFT) protocol.\footnote{In Appendix \ref{sec:gamma-def} we describe how we approximate the nonlinear price schedules in Pakistan and India with a single price per currency unit of remittance sent in each country.}

\paragraph{Bangladesh} For Bangladesh, we use the Bangladesh Household Income and Expenditure Survey 2016 (BHIES) for home household incomes and size. We only use data for households which report having sent a migrant at any point  in the past, and for which the migrant in question had been working within the country at the time of the survey. 
A household surveyed in the BHIES reports if any of their members have migrated in the past, where they migrated to, and their age and sex, but does not record their wages which we need for our analysis. 
To resolve this missing data problem we turn to another data source. 
The Bangladesh Integrated Household Survey 2015 (BIHS) contains wage and demographic information for migrants, but is a much smaller survey than the BHIES. To leverage the BHIES, we estimate a wage model using data from the BIHS. The model specifies that 
\[
\ln w_{\text{dist},l} = \beta_0 + \beta_1 \cdot \text{age}_l + \beta_2 \cdot \text{age}^2_l + \beta_3\cdot \text{male}_l + \beta_4\cdot \text{migrant}_l + \alpha_{\text{dist}} +\epsilon_l,
\]
where the district-level effects $\alpha_{\text{dist}}$ are treated as i.i.d.~normal random effects with mean zero and a constant variance. We also specify $\epsilon_{l}$ as i.i.d.~normal with mean zero and constant variance, independent of the covariates and random effects. 
We estimate the $\beta$ coefficients and the variance parameters by maximum likelihood, and plug in the estimates to obtain empirical Bayes-type estimates of the district effects $\alpha_{\text{dist}}$. 
This fitted model is used to simulate the distribution of migrant wages in each corridor according to BHIES, and obtain values for the mean and variance of log migrant wage used in the construction of predicted site-level effects outlined in Section \ref{subsec:prediction}.
Price indices are constructed from the 2016--17 Household Income and Expenditure Survey for Bangladesh.

\paragraph{India }For India, we use the 10th schedule of the 64th (2007-08) round of the National Sample Survey (NSS) for both home household and migrant data, as well as price indices. We define a home household as one which sent a migrant at any point of time in the past and reports having lived in its current district for at least a year. The 2011 Census provides additional information for long term migrant flows along a corridor. 

\paragraph{Pakistan} For Pakistan, we obtain data for migrants from the 2014--15 Labour Force Survey (LFS) and use data from the 2019--20 Pakistan Social and Living Standards Measurement Survey (PSLM) for home households. We construct home household summary statistics using only data for households which report having received remittances from a domestic source during the prior year. Price indices are constructed using the Pakistan Household Integrated Economic Survey (PHIES). 

\subsection{Restrictions on the Set of Candidate Corridors}
\label{subsec:restrictions_details}

In the rest of this subsection we describe the density and other  restrictions placed on the set of candidate sites in each country, in addition to the requirement that the two corridors have origins
in different states (India), provinces (Pakistan), and divisions (Bangladesh), which we mentioned in the main text.

\paragraph{Bangladesh} For Bangladesh, we require each site to possess a minimal migrant density of $0.01$ in the origin district of that corridor. Migrant density in a district is defined as the number of migrants originating there as a fraction of the total households from that district for which we have data.
The density restriction yields 41 possible migration corridors, and 619 possible two-site combinations.

\paragraph{Pakistan} Migration flows in Pakistan are much lower in magnitude that in Bangladesh so we require a lower minimal migrant density of $0.003$.
Our restrictions leave us with 129 possible corridors and 4,763 possible combinations.

\paragraph{India} Following \cite{imbert:papp:2019}, for India, we first combine data from the NSS and the 2011 Census to derive a measure of migrant density.
This is because while the 10th schedule of the 64th round of the NSS does record how many migrants are sent by a household in a district, it does not record where those migrants move to.
The 2011 Census records the number of migrants along each migration corridor but these correspond to long term migrants, while short term migrants are targeted for the training intervention.
The migrant density measure we use is defined as,
\[
\frac{\text{Migrants along a corridor (Census 2011)}}{\text{Total migrants from an origin district (Census 2011)}}\times
\frac{\text{Total migrants from an origin district (NSS 07-08)}}{\text{Total HHs in an origin (NSS 07-08)}}.
\] 
The above product is meant to approximate the ratio of the total migrants along a corridor to the total households in an origin district. 
This approach uses long term migration flows to allocate the short term migrants from an origin district to multiple destination districts. 
This relies on the assumption that conditional on migrating, short term and long term migrants choose similar destinations. 
\cite{imbert:papp:2019} provide evidence in favor of this assumption using the REDS survey which measures both short and long term migration.

For India, we impose more stringent restrictions so that the number of site combinations remains computationally tractable.
In addition to a minimal migrant density of $0.01$ at the origin, as in Bangladesh, we require any site to have at least $0.75$ long term migrants (as measured in the 2011 Census) per unit area of the origin. 
While in Bangladesh and Pakistan there are only a handful of destinations within high-density migration corridors, in India there are many more destinations so we additionally require  site combinations to have destinations in different states.
These criteria give us a feasible set of 740 corridors and 223,2000 site-pair combinations in India.

\section{Simulation Algorithm}
\label{sec:algorithm}

This section describes how we use the mixed prior to arrive at the welfare estimates given in e.g.~Table \ref{table:two_site_best_worst}, and how these estimates determine the sites we select.

\subsection{Setup}

As described in Section \ref{subsubsec:mixed_prior_intro} we compose the mixed prior from a combination of the structural and smoothing priors.
The mean predicted site effects, $\mu_\tau$,
and the model-based variance matrix $\Sigma_{\text{model}}$ are obtained from the structural model as discussed in 
Section \ref{subsec:prediction}. 
The mixed prior with weight $w \in [0,1]$ is normal with mean $\mu_{\tau}$, and variance matrix 
$$
    \Sigma_{\text{mixed}} = w \cdot \Sigma_{\text{model}} + (1 - w) \cdot \Sigma_{\text{smooth}}.
$$
Sampling uncertainty is independent and identically distributed across sites, with variance $\sigma^2_\epsilon$ computed as described in Section \ref{subsec:application_priors}, so the matrix $\Sigma_\epsilon$ is the diagonal matrix $\Sigma_{\epsilon} = \sigma_{\epsilon}^2 I_S$.

\subsection{Simulations}

Our goal is to approximate by simulation the double integral in Equation \eqref{eq:preposterior} for each set of possible experimental sites $\mathcal{S}$.
For $b=1,\dots,B = 10,000$, we draw a vector of true and (potential) estimated treatment effects for all sites from the joint distribution given in Appendix \ref{sec:posteriorexample}: 
\[ 
\begin{pmatrix}
    \tau \\ \hat{\tau}
\end{pmatrix} \sim \mathcal{N}\left(\begin{bmatrix} \mu_{\tau} \\ \mu_{\tau}  \end{bmatrix},  \begin{bmatrix}\Sigma_{\text{mixed}} & \Sigma_{\text{mixed}} \\ \Sigma_{\text{mixed}} & \Sigma_{\text{mixed}} + \Sigma_{\epsilon} \end{bmatrix} \right).
\]

To evaluate alternative choices for $\mathcal{S}$, we take these simulation draws and estimate the welfare that would obtain from selecting each feasible subset of sites.  
Given $\mathcal{S}$, we keep sub-vector $\hat{\tau}[\mathcal{S}]$, the estimated treatment effects in the experimental sites,
$$
    \hat{\tau}[\mathcal{S}] = \left\{ \hat{\tau}[s] \text{ for } s \in \mathcal{S} \right\},
$$     
and compute $\mathbb{E}[\tau \mid \hat{\tau}[\mathcal{S}]]$, the posterior mean of the full vector of true site effects $\tau$ given $\hat{\tau}[\mathcal{S}]$ using the formula in  Appendix \ref{sec:posteriorexample}.
We then compute the optimal site-level treatment vector $T^*$, a vector of indicators for whether each site's posterior mean treatment effect exceeds the cost of implementation:
$$
    T^* = \mathbbm{1}\{\mathbb{E}[\tau \mid \hat{\tau}[\mathcal{S}]] \geq c\}.
$$ 
Finally, we calculate welfare under $T^*$ for simulation $b$.
For site $s$, welfare is 0 if not treated and $\tau_s - c$ otherwise, giving aggregate welfare
$$
    W_b[\mathcal{S}] = \sum_{s=1}^S (\tau_s - c)T_s^*,
$$ 
where $\tau_s$ is an element of the vector we drew at the beginning of simulation $b$.

We then compute the average welfare over all the simulation draws
$$
    \overline{W}[\mathcal{S}] = \frac{1}{B}\sum_{b=1}^B W_b[\mathcal{S}],
$$ 
which gives us an entry in e.g.~Table  \ref{table:two_site_best_worst} for experimental site combination $\mathcal{S}$.
To select sites, we choose the $\mathcal{S}$ with the highest $\overline{W}[\mathcal{S}]$.

\section{Site Selection Rule Performance Under Alternative Priors}
\label{sec:eval_by_mixed_prior_weight}

\begin{figure}[ht!]
\centering
\caption{Average Per-Migrant Welfare of Experimental Sites by Site Selection Method and Evaluating Mixed Prior}
\label{fig:robustness_plot}
\vspace{.25cm}
\begin{minipage}{0.49\textwidth}
\centering \textsf{Bangladesh}
\includegraphics[width = \textwidth]{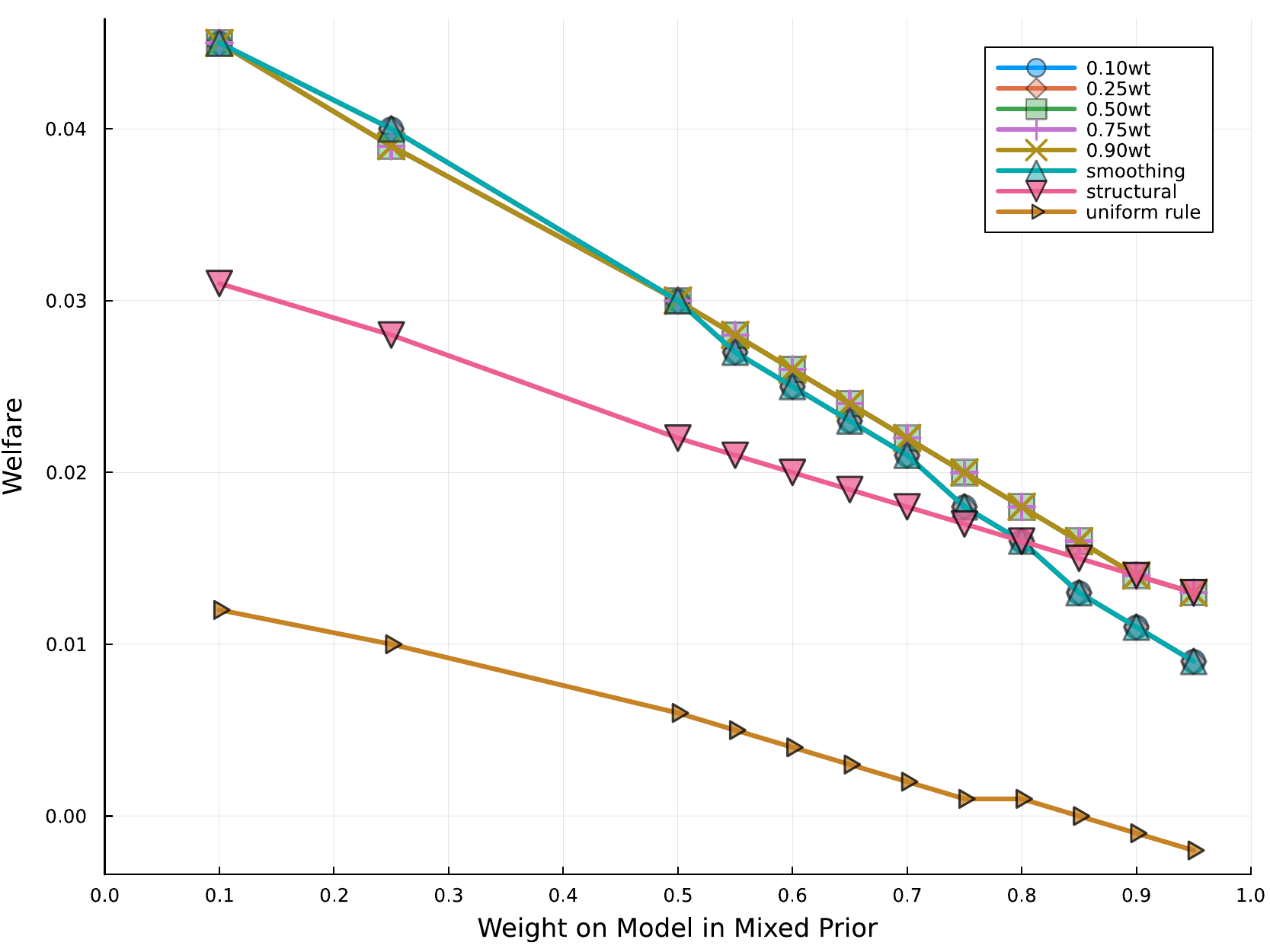}
\end{minipage}
\hfill
\begin{minipage}{0.49\textwidth}
\centering \textsf{Pakistan}
\includegraphics[width = \textwidth]{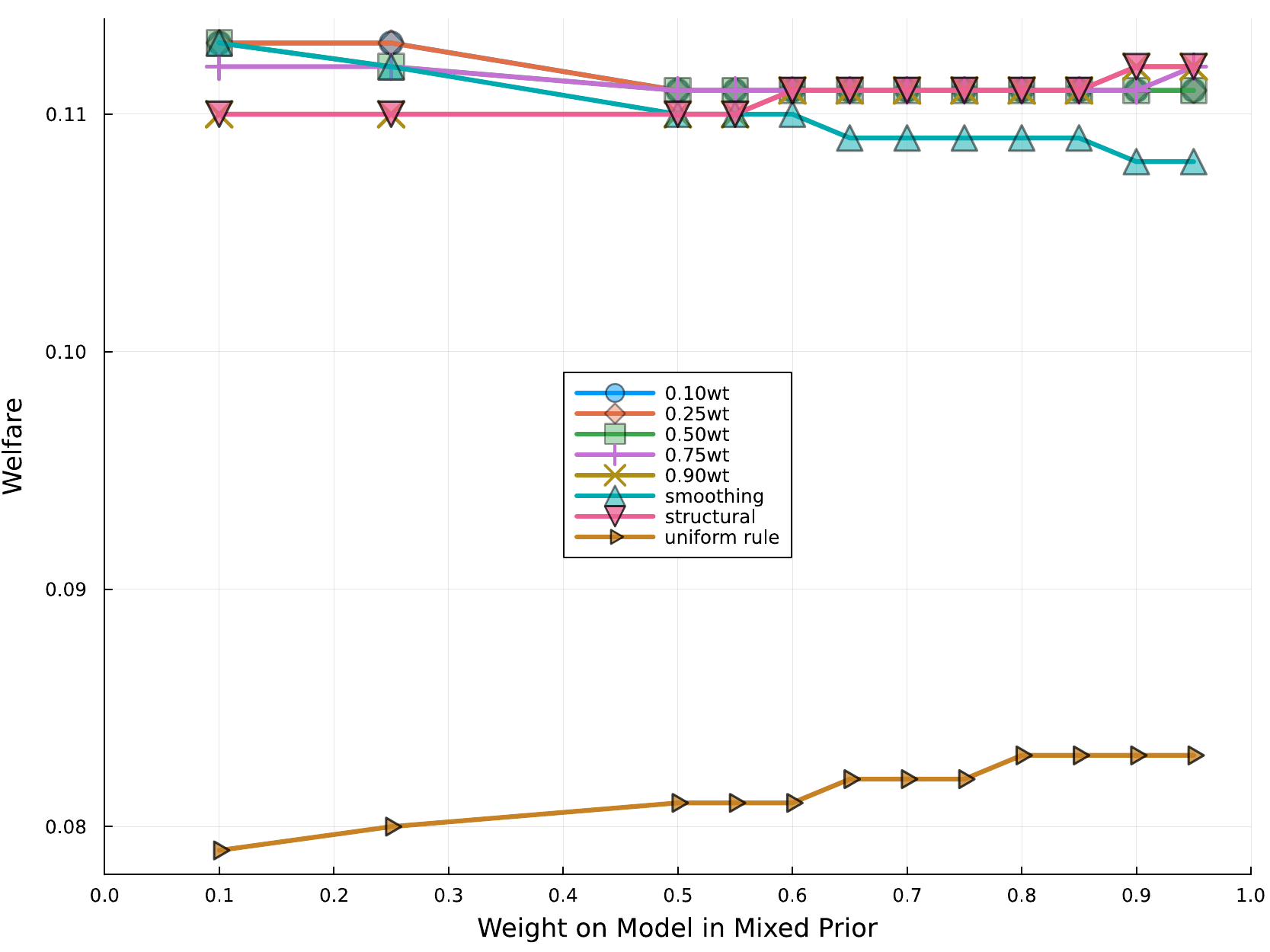}
\end{minipage}
\begin{minipage}{0.5\textwidth}
\centering \textsf{India}
\includegraphics[width = \textwidth]{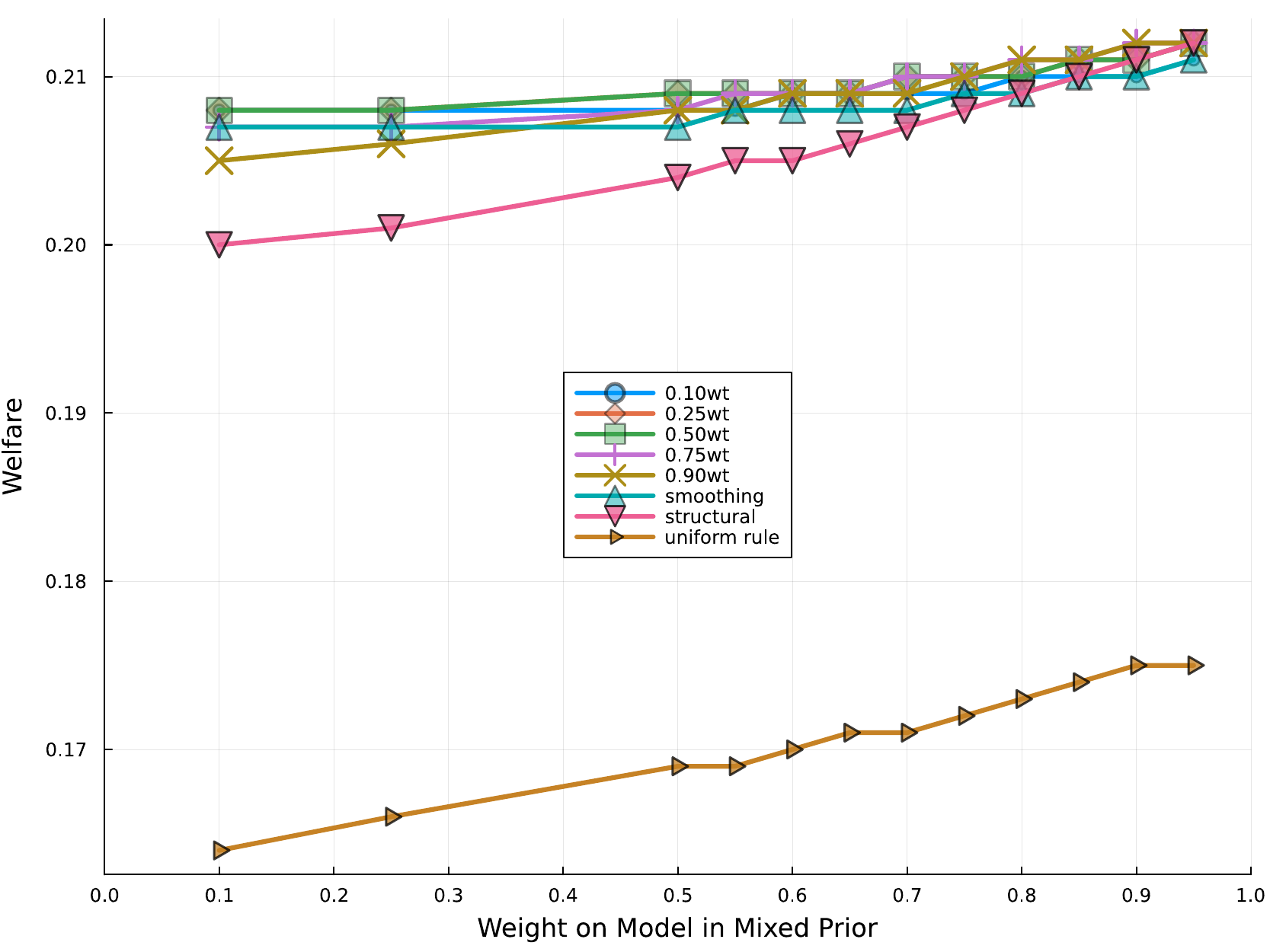}
\end{minipage}
\begin{minipage}{\textwidth}
\vspace{1pt} {\footnotesize Notes: Welfare refers to the average across all candidate sites of the per-migrant welfare increase from site-level implementation recommendations, 
predicted under the mixed prior with model weight given by the X-coordinate.
Lines index different site selection methods, with $p$wt indicating a mixed prior with $p$ weight on the structural model.
\par}
\end{minipage}
\end{figure}

We now examine the welfare predicted under specific mixed priors when choosing sites using different priors. 
Figure \ref{fig:robustness_plot} shows the results.
Each line in the graph gives the welfare anticipated under the mixed prior specified on the X-axis when sites are selected according the prior represented by a given line.
In other words, sites are selected using the prior indicated by the line but evaluated under the prior on the X-axis.
For instance, the line marked with upside-down triangles shows the welfare predicted when selecting sites using the pure structural prior, according to different mixed priors.

It is never optimal to use the pure structural prior to select sites since Figure \ref{fig:robustness_plot} only predicts welfare using mixed priors placing between 0.1 and 0.95 weight on the structural prior, but it may be close.
It \emph{is} optimal to use the 0.5 mixed prior when also predicting welfare from the site combination chosen using the same prior.
Matching the mixed prior used to select sites and to predict welfare yields the upper envelope for expected performance.

Looking at the figure, we see that performance of selecting sites using any mixed prior is quite robust to the mixing weight used when predicting welfare.
This is in part because the mixed priors often select the same sites.
We suggest all practitioners using our method perform a similar robustness check since it is difficult to determine the appropriate weight to place on the structural model \emph{a priori}.
One might believe that a 0.5 weight reflects complete ambivalence between the structural and smoothing priors but - in all countries - selecting sites using the pure smoothing prior leads to better predicted performance than selection using the pure structural prior when predictions are computed using the 0.5-weight prior.
Selecting sites using the model only begins to outperform selecting sites using the pure smoothing prior when the weight on the model in the prior used to generate predicted welfare exceeds $0.8$, $0.6$, and $0.85$ in Bangladesh, Pakistan, and India respectively.
This also shows the importance of allowing for deviations from the structural model when selecting sites, since the performance of site selections using only the structural model degrades when considering such deviations.

Nevertheless, site selection using the pure structural prior always outperforms by a wide margin the naive approach of selecting sites randomly and recommending the program be implemented in all sites if the average of experimental site ATEs exceeds the implementation cost.
We consider this approach in Section \ref{subsec:choosing-site-number} and refer to it as the uniform rule.

\section{Choosing Different Numbers of Experimental Sites}

\subsection{Best/Worst Single Experimental Site Choices}
\label{appendix:more_less_sites}
 Table \ref{table:bangladesh_single_site} shows the five best and worst choices when a single corridor is chosen to be experimented on in Bangladesh.
 Table \ref{table:pakistan_single_site} shows corresponding results for Pakistan, with the welfare values are based on the full set of admissible corridors instead of just marginal corridors.

\begin{table}[ht!]
    \centering
    \caption{Welfare-Maximizing and Minimizing Experimental Sites, Single Site Case, Bangladesh}
    \label{table:bangladesh_single_site}
    \makebox[\textwidth][c]{
    \scalebox{0.8}{
    \begin{tabular}{lclc}
    \toprule
    \multicolumn{2}{c}{Best Choice}	&	\multicolumn{2}{c}{Worst Choice} \\
    \cmidrule(lr){1-2}	\cmidrule(lr){3-4}						
    Corridor	&	Avg.~Welfare	&	Corridor	&	Avg.~Welfare	\\
    \cmidrule(lr){1-1}	\cmidrule(lr){2-2}			\cmidrule(lr){3-3}	\cmidrule(lr){4-4}
Dhaka-Noakhali	&	0.024	&	Dhaka-Thakurgaon	&	0.004	\\
Dhaka-Feni	&	0.024	&	Gazipur-Panchagarh	&	0.003	\\
Dhaka-Kishoregonj	&	0.023	&	Dhaka-Gopalganj	&	0.001	\\
Dhaka-Bogra	&	0.023	&	Dhaka-Madaripur	&	0.001	\\
Chittagong-Comilla	&	0.023	&	Dhaka-Bhola	&	0.001	\\
    \bottomrule							
    \end{tabular}
    }
    }
\begin{minipage}{0.6\textwidth}
\vspace{1pt} {\footnotesize Notes: Avg. Welfare refers to the average across all candidate sites of the per-migrant welfare increase from site-level implementation recommendations, 
predicted under our preferred mixed prior with 0.5 weight on the structural model.
\par}
\end{minipage}
\end{table}

\begin{table}[ht!]
    \centering
    \caption{Welfare-Maximizing and Minimizing Experimental Sites, Single Site Case, Pakistan}
    \label{table:pakistan_single_site}
    \makebox[\textwidth][c]{
    \scalebox{0.8}{
    \begin{tabular}{lclc}
    \multicolumn{2}{c}{Best Choice}	&	\multicolumn{2}{c}{Worst Choice}					\\
    \cmidrule(lr){1-2}	\cmidrule(lr){3-4}						
    Corridor	&	Avg.~Welfare	&	Corridor	&	Avg.~Welfare	\\
    \cmidrule(lr){1-1}	\cmidrule(lr){2-2}			\cmidrule(lr){3-3}	\cmidrule(lr){4-4}		
    Mardan-Swabi	&	0.111	&	Karachi-Bannu	&	0.103	\\
Peshawar-Swabi	&	0.111	&	Karachi-Mianwali	&	0.103	\\
Peshawar-Mardan	&	0.111	&	Karachi-Sukkur	&	0.103	\\
Kohat-Hangu	&	0.111	&	Rawalpindi-Islamabad	&	0.103	\\
Swat-Shangla	&	0.111	&	Chitral-Islamabad	&	0.103	\\				
    \bottomrule							
    \end{tabular}
    }}
    \begin{minipage}{0.6\textwidth}
\vspace{1pt} {\footnotesize Notes: Avg. Welfare refers to the average across all candidate sites of the per-migrant welfare increase from site-level implementation recommendations, 
predicted under our preferred mixed prior with 0.5 weight on the structural model.
\par}
\end{minipage}
\end{table}

\begin{table}[ht!]
    \centering
    \caption{Welfare-Maximizing and Minimizing Experimental Sites, Single Site Case, India}
    \label{table:india_single_site}
    \makebox[\textwidth][c]{
    \scalebox{0.8}{
    \begin{tabular}{lclc}
    \multicolumn{2}{c}{Best Choice}	&	\multicolumn{2}{c}{Worst Choice}					\\
    \cmidrule(lr){1-2}	\cmidrule(lr){3-4}						
    Corridor	&	Avg.~Welfare	&	Corridor	&	Avg.~Welfare	\\
    \cmidrule(lr){1-1}	\cmidrule(lr){2-2}			\cmidrule(lr){3-3}	\cmidrule(lr){4-4}		
    New Delhi-Dausa	&	0.207	&	Patiala-Chandigarh	&	0.197	\\
New Delhi-Hardwar	&	0.207	&	New Delhi-Chennai	&	0.197	\\
New Delhi-Jaipur	&	0.207	&	Sitamarhi-Mumbai	&	0.197	\\
New Delhi-Agra	&	0.206	&	Bangalore-Chandigarh	&	0.197	\\
New Delhi-Pilibhit	&	0.206	&	Kancheepuram-Chandigarh	&	0.197	\\	
    \bottomrule							
    \end{tabular}
    }}
\begin{minipage}{0.7\textwidth}
\vspace{1pt} {\footnotesize Notes: Avg. Welfare refers to the average across all candidate sites of the per-migrant welfare increase from site-level implementation recommendations, 
predicted under our preferred mixed prior with 0.5 weight on the structural model.
\par}
\end{minipage}
\end{table}

\subsection{Calculating the Number of Observations with Differing Numbers of Sites}
\label{sec:site_calculations_by_number}

To calculate the cost of collecting data in different numbers of corridors, we started with the scenario above where 8000 surveys would be collected in 2 corridors, one for each migrant and one for their household at home.
We then asked the survey firm that conducted the surveys in \citet{Lee2018} to create a hypothetical budget for survey collection in ``typical'' sites in Bangladesh. 

The firm's work led to an estimate of the total cost of the 8000 surveys in the 2 corridors of \$175,415 or \$21.93 per survey (converted to dollars from taka on July 18, 2023). We then took that budget as fixed and calculated how many surveys would be possible with 1, 3, 4, 5, and 6 corridors, using the survey firm's assumptions about fixed and variable costs. Variable costs are not linear in the number of surveys since some costs increase with the number of sites and some increase with the number of surveys. 

Before collecting surveys, the firm completes a process of ``listing'' which involves collecting basic data on the households. The firm assumes that there would be 15 enumerators for this part of the work, each conducting 20 listing surveys per day (with buffer days added). They collect data for 110\% of the eventual surveys to be completed. Other staff members include 3 supervisors, 2 research assistants (who assist with training, data collection, and data cleaning), and one research manager (who monitors and supervises the complete process).  
For the main household survey, the budget covers 30 enumerators who each can collect 3 surveys per day (with buffer days added). Again there will also be 2 research assistants and a research manager. This phase of the work includes additional staff to ensure data quality: 2 ``scutinizers'' and 6 ``back-checkers.''

The costs also include subscriptions of survey software and the costs of training the enumerators (including space rental, food, and transportation). Travel costs for visiting the sites by supervisors and research assistants are included. 

The costs were constructed for ``typical''  locations several hours from Dhaka. Survey costs in Dhaka itself would be higher in general, but there would be savings on travel costs and related expenses.

The cost calculations allow us to determine how many surveys could be collected in each corridor if we kept the budget capped at \$175,415. As noted, with the 2 corridor benchmark, 8,000 surveys could be collected.
Table \ref{table:Max_number_of_surveys_for_fixed_budget} shows that with only 1 corridor, it is possible to increase the number to 8,673 surveys in total. With 6 corridors, however, only 5,305 can be collected with the same budget.

\begin{table}[ht!]
    \centering
    \caption{Maximum Number of Surveys Possible for a Fixed Budget, By Number of Corridors (Costs in USD)}
    \label{table:Max_number_of_surveys_for_fixed_budget}
    \makebox[\textwidth][c]{
    \scalebox{0.8}{
    \begin{tabular}{lcccccc}
    \toprule									
     &  1	&	 2	&	3	&	4	&	5  &  6	\\
    	\cmidrule(lr){2-2}	\cmidrule(lr){3-3}		\cmidrule(lr){4-4}		\cmidrule(lr){5-5}	\cmidrule(lr){6-6} \cmidrule(lr){7-7}			
    Fixed cost	&	18.308	&	 18,308 	&	18,308 &  18,308 & 	18,308 &  18,308 \\
    Total variable costs	&	134,219 &	134,227 & 134,220 &  134,213 & 134,220 & 134,213 \\
    Indirect costs	&	  22,879 & 22,880 & 22,879 & 22,878 & 22,879 & 22,878 \\
    	\cmidrule(lr){2-2}	\cmidrule(lr){3-3}		\cmidrule(lr){4-4}		\cmidrule(lr){5-5}	\cmidrule(lr){6-6} \cmidrule(lr){7-7}		
    Total cost	&	175,406 & 175,415 & 175,407 & 175,399 & 175,407 & 175,400 \\
   \midrule
    Fixed cost per survey	&	2.11	&	 2.29 	&	2.50 &  2.75    &	3.06 &  3.45 \\
    Variable cost per survey	&	15.48	&	 16.78 	&	18.32 &  20.18 
    &	22.45 & 23.50 \\
    Indirect cost per survey & 2.64 & 2.86 & 3.12 & 3.44 & 3.83 & 4.31 \\
    	\cmidrule(lr){2-2}	\cmidrule(lr){3-3}		\cmidrule(lr){4-4}		\cmidrule(lr){5-5}	\cmidrule(lr){6-6} \cmidrule(lr){7-7}		
    Total cost per survey	&	20.22 & 21.93 & 23.94 & 26.37 & 29.34 & 33.06 \\
    \midrule
    Maximum survey number	&	 8,673 & 8,000 & 7,326 & 6,652 & 5,979 & 5,305\\
    Surveys per corridor &	 8,673 & 4,000 & 2,442 & 1,663 & 1,196 & 884 \\  
    \bottomrule									
    \end{tabular}
    }
    }
\end{table}

\section{Choosing Sites Jointly Over Multiple Countries}
\label{sec:joint_site_selection}

Instead of choosing a pair of experimental sites in each country separately, we could choose migration corridors jointly across some or all of the countries.  
Table \ref{table:bd_pak_joint_best} shows the best choices for four migration corridors with two each from Bangladesh and Pakistan, respectively, and Table \ref{table:bd_pak_joint_worst} shows the worst performing sets. 
Sites in India are not considered for reasons of computational tractability. As before, we assume a total of 4000 sampled migrant-household pairs in each country.

\begin{table}[ht!]
    \centering
    \caption{Best Site Choices when Choosing Jointly across Bangladesh and Pakistan}
    \label{table:bd_pak_joint_best}
    \scalebox{0.75}{
    \begin{tabular}{llllc}
        \toprule									
        \multicolumn{2}{c}{Bangladesh}	&			\multicolumn{2}{c}{Pakistan}					\\
        \cmidrule(lr){1-2}		\cmidrule(lr){3-4}		\cmidrule(lr){5-5}					
        Corridor 1	&	Corridor 2	&	Corridor 1	&	Corridor 2	&	Avg.~Welfare	\\
        \cmidrule(lr){1-1}	\cmidrule(lr){2-2}	\cmidrule(lr){3-3}		\cmidrule(lr){4-4}		\cmidrule(lr){5-5}			

Dhaka-Barguna	&	Dhaka-Kishoregonj	&	Multan-Khanewal	&	Peshawar-Swat	&	 0.1064 	\\
Dhaka-Kishoregonj	&	Dhaka-Noakhali	&	Multan-Muzaffar Garh	&	Peshawar-Swat	&	 0.1063 	\\
Dhaka-Comilla	&	Dhaka-Pirojpur	&	Rawalpindi-Gujrat	&	Peshawar-Swat	&	 0.1063 	\\
Dhaka-Comilla	&	Dhaka-Pirojpur	&	Gujranwala-Gujrat	&	Peshawar-Swat	&	 0.1063 	\\
Dhaka-Comilla	&	Dhaka-Pirojpur	&	Rawalpindi-Gujrat	&	Rawalpindi-Peshawar	&	 0.1062 	\\

        \bottomrule									
    \end{tabular}
    }
\end{table}

\begin{table}[ht!]
    \centering
    \caption{Worst Site Choices when Choosing Jointly across Bangladesh and Pakistan}
    \label{table:bd_pak_joint_worst}
    \scalebox{0.75}{
    \begin{tabular}{llllc}
        \toprule									
        \multicolumn{2}{c}{Bangladesh}	&			\multicolumn{2}{c}{Pakistan}					\\
        \cmidrule(lr){1-2}		\cmidrule(lr){3-4}		\cmidrule(lr){5-5}					
        Corridor 1	&	Corridor 2	&	Corridor 1	&	Corridor 2	&	Avg.~Welfare	\\
        \cmidrule(lr){1-1}	\cmidrule(lr){2-2}	\cmidrule(lr){3-3}		\cmidrule(lr){4-4}		\cmidrule(lr){5-5}			

Dhaka-Bhola	&	Dhaka-Madaripur	&	Chitral-Islamabad	&	Peshawar-Lakki Marwat	&	 0.0708 	\\
Dhaka-Bhola	&	Dhaka-Madaripur	&	Rawalpindi-Islamabad	&	Peshawar-Lakki Marwat	&	 0.0708 	\\
Dhaka-Bhola	&	Dhaka-Madaripur	&	Peshawar-Lakki Marwat	&	Karachi-Sukkur	&	 0.0707 	\\
Dhaka-Bhola	&	Dhaka-Madaripur	&	Chitral-Islamabad	&	Karachi-Sukkur	&	 0.0702 	\\
Dhaka-Bhola	&	Dhaka-Madaripur	&	Rawalpindi-Islamabad	&	Karachi-Sukkur	&	 0.0701 	\\

        \bottomrule									
    \end{tabular}
    }
\end{table}

\section{Structural Correlations and Site Characteristics}
\label{sec:structural_correlations}

We investigate how the structural model generates correlations across site-level treatment effects in the structural prior. 
We fit a linear regression model that relates the structural prior correlations to the absolute difference in site characteristics: 
\begin{align}
\rho_{s,s^\prime} = \beta_0 + \lvert X_{s} - X_{s^\prime}\rvert^\prime \beta + \varepsilon_{s, s^\prime}, \nonumber
\end{align}
where $\rho_{s, s^\prime}$ denotes the correlation in outcomes between sites $s$ and $s^\prime$ in the structural prior, and $X_s$ denotes characteristics for site $s$. 
All differences in site level characteristics are standardised to have mean zero and unit variance. 

The regression estimates are reported in Tables \ref{table:struc_corr_bd}--\ref{table:struc_corr_ind}.
In all the estimated regressions, the constant term is large.   
This shows that the structural prior implies high correlations in ATEs across sites, which reaffirms the message from Figure \ref{fig:welfare_te_bd_model} that the parametric model allows quite different sites to be informative for one another. 
Some of the site-level characteristics have explanatory power for the prior correlations, but these patterns differ substantially across the three countries. 
This suggests that the true relationship between site-level characteristics and prior correlations may be highly nonlinear or involve interactions that our simple linear approximation does not capture. 

\begin{table}[ht!]
\centering
\caption{Regressing Structural Correlations on Differences in Site Characteristics in Bangladesh}
\label{table:struc_corr_bd}
\scalebox{0.8}{
\begin{tabular}{lcccc}
\toprule                                  
Variable    &   Coef.   &   Std. Error  &   Lower 95\%   &   Upper 95\%   \\
\cmidrule(lr){1-1}      \cmidrule(lr){2-2}      \cmidrule(lr){3-3}      \cmidrule(lr){4-4}      \cmidrule(lr){5-5}    
Intercept   &   0.9908  &   0.0002  &   0.9903  &   0.9913  \\
Distance    &   -0.0024 &   0.0002  &   -0.0029 &   -0.0019 \\
Home Income &   -0.0011 &   0.0004  &   -0.0020 &   -0.0003 \\
Migrant Wage &   0.0016  &   0.0005  &   0.0006  &   0.0026 \\
Home HH Size &   0.0005  &   0.0003  &   -0.0001 &   0.0010 \\
Operator Density &   -0.0102 &   0.0003  &   -0.0108 &   -0.0096 \\
Home Income SD &   0.0020  &   0.0004  &   0.0012  &   0.0028 \\
Migrant Wage SD &   -0.0021 &   0.0005  &   -0.0031 &   -0.0011 \\
Home HH Size SD &   0.0026  &   0.0003  &   0.0019  &   0.0032 \\
\cmidrule(lr){1-5}                                  
Number of Corrs.    &   820 &       &   F   &   200.7426    \\
SSR &   0.0395  &       &   (Prob. >F)  &   0.0000    \\
R$^2$   &   0.6645  &       &       &       \\
\bottomrule                                  
\end{tabular}
}
\begin{minipage}{0.55\textwidth}
\vspace{1pt} {\footnotesize Notes: results from regressing correlations in welfare ATEs according to the structural prior across all site pairs on the vector of differences in site characteristics for the corresponding pair. Numbers are rounded to the fourth decimal place.
\par}
\end{minipage}
\end{table}

\begin{table}[ht!]
\centering
\caption{Regressing Structural Correlations on Differences in Site Characteristics in Pakistan}
\label{table:struc_corr_pak}
\scalebox{0.8}{
\begin{tabular}{lcccc}
\toprule									
Variable	&	Coef.	&	Std. Error	&	Lower 95\%	&	Upper 95\%	\\
\cmidrule(lr){1-1}		\cmidrule(lr){2-2}		\cmidrule(lr){3-3}		\cmidrule(lr){4-4}		\cmidrule(lr){5-5}	
Intercept	&	0.9447	&	0.0022	&	0.9404	&	0.9490	\\
Distance	&	0.0052	&	0.0022	&	0.0009	&	0.0095	\\
Home income	&	0.0046	&	0.0028	&	-0.0008	&	0.0100	\\
Migrant Wage	&	-0.0758	&	0.0028	&	-0.0813	&	-0.0703	\\
Home HH size	&	0.0142	&	0.0033	&	0.0077	&	0.0208	\\
Operator Density	&	-0.0047	&	0.0022	&	-0.0090	&	-0.0004	\\
Home income std. dev.	&	-0.0038	&	0.0028	&	-0.0092	&	0.0016	\\
Migrant wage std. dev.	&	0.0480	&	0.0028	&	0.0425	&	0.0536	\\
Home HH size std. dev.	&	-0.0012	&	0.0036	&	-0.0077	&	0.0054	\\
\cmidrule(lr){1-5}									
Number of Corrs.	&	8256	&		&	F 	&	98.58	\\
SSR	&	326.3269	&		&	(Prob. >F) 	&	0.0000	\\
R$^2$	&	0.0873	&		&		&		\\
\bottomrule									
\end{tabular}
}
\begin{minipage}{0.57\textwidth}
\vspace{1pt} {\footnotesize Notes: results from regressing correlations in welfare ATEs according to the structural prior across all site pairs on the vector of differences in site characteristics for the corresponding pair. Numbers are rounded to the fourth decimal place.
\par}
\end{minipage}
\end{table}

\begin{table}[ht!]
\centering
\caption{Regressing Structural Correlations on Differences in Site Characteristics in India}
\label{table:struc_corr_ind}
\scalebox{0.8}{
\begin{tabular}{lcccc}
\toprule									
Variable	&	Coef.	&	Std. Error	&	Lower 95\%	&	Upper 95\%	\\
\cmidrule(lr){1-1}		\cmidrule(lr){2-2}		\cmidrule(lr){3-3}		\cmidrule(lr){4-4}		\cmidrule(lr){5-5}	
Intercept	&	0.9225	&	0.0003	&	0.9220	&	0.9231	\\
Distance	&	0.0017	&	0.0003	&	0.0011	&	0.0023	\\
Home income	&	0.0016	&	0.0004	&	0.0008	&	0.0023	\\
Migrant Wage	&	0.0004	&	0.0003	&	-0.0003	&	0.0010	\\
Home HH size	&	0.0012	&	0.0003	&	0.0005	&	0.0018	\\
Operator Density	&	-0.0802	&	0.0003	&	-0.0808	&	-0.0796	\\
Home income std. dev.	&	0.0060	&	0.0004	&	0.0052	&	0.0067	\\
Migrant wage std. dev.	&	-0.0024	&	0.0003	&	-0.0031	&	-0.0018	\\
Home HH size std. dev.	&	0.0011	&	0.0003	&	0.0004	&	0.0017	\\
\cmidrule(lr){1-5}									
Number of Corrs.	&	273,430	&		&	F 	&	9525.6627	\\
SSR	&	6402.9309	&		&	(Prob. >F) 	&	0.0000	\\
R$^2$	&	0.2180	&		&		&		\\
\bottomrule									
\end{tabular}
}
\begin{minipage}{0.6\textwidth}
\vspace{1pt} {\footnotesize Notes: results from regressing correlations in welfare ATEs according to the structural prior across all site pairs on the vector of differences in site characteristics for the corresponding pair. Numbers are rounded to the fourth decimal place.
\par}
\end{minipage}
\end{table}

\section{Robustness to Alternate Treatment Effects}
\label{sec:alternate_tes}

As a robustness check, we would like to see how the ranking of optimal corridor combinations changes in response to a more or less effective intervention.
Since the environment in Bangladesh is different now from how it was in 2015 when the \cite{Lee2018} experiment was carried out (for example, mobile money usage has increased nationally), we would like to see that our site selections are not too sensitive to the specific value of $\psi$ we estimated from the experimental microdata.
To do this, we suppose the point estimate for the parameter $\psi$ is larger or smaller than is estimated from the experimental data from \cite{Lee2018} while its estimated variance remains the same. This roughly assumes that the effectiveness of the experimental intervention was greater or lesser while sampling variability remained the same. 
Changing the structural parameter estimates in this way will lead to different specifications of the mean vector and variance matrix of the structural prior used for the site-selection algorithm, so we can view these results as a form of prior sensitivity or prior robustness exercise. 

Tables \ref{table:bangladesh_best_worst_psimag01}--\ref{table:bangladesh_best_worst_psimag3} display the best and worst combinations of two migration corridors in Bangladesh under the mixed prior with 0.5 weight on the model as the estimated treatment effect is magnified or diminished. The top site combination with our estimated value of $\psi$ always appears among the top five highest-welfare site combinations when $\psi$ is altered, along with between one and three other top sites from Table \ref{table:two_site_best_worst}. 

\begin{table}[ht!]
    \centering
    \caption{Welfare-Maximizing and Minimizing Experimental Site Combinations in Bangladesh; $\psi$ Scaled by 0.1}
    \label{table:bangladesh_best_worst_psimag01}
    \makebox[\textwidth][c]{
    \scalebox{0.8}{
    \begin{tabular}{llcllc}
    \toprule
    \multicolumn{3}{c}{Best Combinations}			&	\multicolumn{3}{c}{Worst Combinations}							\\
    \cmidrule(lr){1-3}	\cmidrule(lr){4-6}										
    Corridor 1	&	Corridor 2	&	Avg. Welfare	&	Corridor 1	&	Corridor 2	&	Avg. Welfare	\\
    \cmidrule(lr){1-1}	\cmidrule(lr){2-2}	\cmidrule(lr){3-3}				\cmidrule(lr){4-4}	\cmidrule(lr){5-5}	\cmidrule(lr){6-6}			
Dhaka-Kishoregonj	&	Dhaka-Noakhali	&	0.007	&	Dhaka-Bhola	&	Dhaka-Panchagarh	&	0.001	\\
Dhaka-Feni	&	Dhaka-Shariatpur	&	0.007	&	Dhaka-Gopalganj	&	Dhaka-Thakurgaon	&	0.001	\\
Dhaka-Bagerhat	&	Dhaka-Kishoregonj	&	0.007	&	Dhaka-Bhola	&	Dhaka-Thakurgaon	&	0.001	\\
Dhaka-Magura	&	Dhaka-Noakhali	&	0.007	&	Dhaka-Bhola	&	Dhaka-Madaripur	&	0.001	\\
Dhaka-Noakhali	&	Dhaka-Shariatpur	&	0.007	&	Dhaka-Bhola	&	Dhaka-Gopalganj	&	0.001	\\

        \bottomrule														
    \end{tabular}
    }
    }
\begin{minipage}{0.9\textwidth}
\vspace{1pt} {\footnotesize Notes: Avg. Welfare refers to the average across all candidate sites of the per-migrant welfare increase from site-level implementation recommendations, predicted under our preferred mixed prior with 0.5 weight on the structural model.
\par}
\end{minipage}
\end{table}

\begin{table}[ht!]
    \centering
    \caption{Welfare-Maximizing and Minimizing Experimental Site Combinations in Bangladesh; $\psi$ Scaled by 0.5}
    \label{table:bangladesh_best_worst_psimag05}
    \makebox[\textwidth][c]{
    \scalebox{0.8}{
    \begin{tabular}{llcllc}
    \toprule
    \multicolumn{3}{c}{Best Combinations}			&	\multicolumn{3}{c}{Worst Combinations}							\\
    \cmidrule(lr){1-3}	\cmidrule(lr){4-6}										
    Corridor 1	&	Corridor 2	&	Avg. Welfare	&	Corridor 1	&	Corridor 2	&	Avg. Welfare	\\
    \cmidrule(lr){1-1}	\cmidrule(lr){2-2}	\cmidrule(lr){3-3}				\cmidrule(lr){4-4}	\cmidrule(lr){5-5}	\cmidrule(lr){6-6}			
Dhaka-Kishoregonj	&	Dhaka-Noakhali	&	 0.014 	&	Dhaka-Gopalganj	&	Dhaka-Thakurgaon	&	 0.002 	\\
Dhaka-Bagerhat	&	Dhaka-Kishoregonj	&	 0.014 	&	Dhaka-Bhola	&	Dhaka-Thakurgaon	&	 0.002 	\\
Dhaka-Magura	&	Dhaka-Noakhali	&	 0.013 	&	Dhaka-Bhola	&	Dhaka-Panchagarh	&	 0.002 	\\
Dhaka-Narsingdi	&	Dhaka-Noakhali	&	 0.013 	&	Dhaka-Bhola	&	Dhaka-Madaripur	&	 0.001 	\\
Dhaka-Barguna	&	Dhaka-Kishoregonj	&	 0.013 	&	Dhaka-Bhola	&	Dhaka-Gopalganj	&	 0.001 	\\

    \bottomrule														
    \end{tabular}
    }
    }
    \begin{minipage}{0.9\textwidth}
\vspace{1pt} {\footnotesize Notes: Avg. Welfare refers to the average across all candidate sites of the per-migrant welfare increase from site-level implementation recommendations, predicted under our preferred mixed prior with 0.5 weight on the structural model.
\par}
\end{minipage}
\end{table}

\begin{table}[ht!]
    \centering
    \caption{Welfare-Maximizing and Minimizing Experimental Site Combinations in Bangladesh; $\psi$ Scaled by 1.5}
    \label{table:bangladesh_best_worst_psimag1}
    \makebox[\textwidth][c]{
    \scalebox{0.8}{
    \begin{tabular}{llcllc}
    \toprule
    \multicolumn{3}{c}{Best Combinations}			&	\multicolumn{3}{c}{Worst Combinations}							\\
    \cmidrule(lr){1-3}	\cmidrule(lr){4-6}										
    Corridor 1	&	Corridor 2	&	Avg. Welfare	&	Corridor 1	&	Corridor 2	&	Avg. Welfare	\\
    \cmidrule(lr){1-1}	\cmidrule(lr){2-2}	\cmidrule(lr){3-3}				\cmidrule(lr){4-4}	\cmidrule(lr){5-5}	\cmidrule(lr){6-6}			
Dhaka-Magura	&	Dhaka-Noakhali	&	 0.053 	&	Dhaka-Gopalganj	&	Gazipur-Panchagarh	&	 0.018 	\\
Dhaka-Kishoregonj	&	Dhaka-Noakhali	&	 0.052 	&	Dhaka-Madaripur	&	Gazipur-Panchagarh	&	 0.018 	\\
Dhaka-Faridpur	&	Dhaka-Noakhali	&	 0.052 	&	Dhaka-Bhola	&	Gazipur-Panchagarh	&	 0.018 	\\
Dhaka-Barguna	&	Dhaka-Kishoregonj	&	 0.052 	&	Dhaka-Bhola	&	Dhaka-Gopalganj	&	 0.016 	\\
Dhaka-Feni	&	Dhaka-Narsingdi	&	 0.051 	&	Dhaka-Bhola	&	Dhaka-Madaripur	&	 0.015 	\\

    \bottomrule														
    \end{tabular}
    }
    }
    \begin{minipage}{0.9\textwidth}
\vspace{1pt} {\footnotesize Notes: Avg. Welfare refers to the average across all candidate sites of the per-migrant welfare increase from site-level implementation recommendations, predicted under our preferred mixed prior with 0.5 weight on the structural model.
\par}
\end{minipage}
\end{table}

\begin{table}[ht!]
    \centering
    \caption{Welfare-Maximizing and Minimizing Experimental Site Combinations in Bangladesh; $\psi$ Scaled by 2.5}
    \label{table:bangladesh_best_worst_psimag3}
    \makebox[\textwidth][c]{
    \scalebox{0.8}{
    \begin{tabular}{llcllc}
    \toprule
    \multicolumn{3}{c}{Best Combinations}			&	\multicolumn{3}{c}{Worst Combinations}							\\
    \cmidrule(lr){1-3}	\cmidrule(lr){4-6}										
    Corridor 1	&	Corridor 2	&	Avg. Welfare	&	Corridor 1	&	Corridor 2	&	Avg. Welfare	\\
    \cmidrule(lr){1-1}	\cmidrule(lr){2-2}	\cmidrule(lr){3-3}				\cmidrule(lr){4-4}	\cmidrule(lr){5-5}	\cmidrule(lr){6-6}			
Dhaka-Magura	&	Dhaka-Noakhali	&	 0.110 	&	Dhaka-Gopalganj	&	Gazipur-Panchagarh	&	 0.079 	\\
Dhaka-Faridpur	&	Dhaka-Noakhali	&	 0.110 	&	Dhaka-Bhola	&	Gazipur-Panchagarh	&	 0.079 	\\
Dhaka-Kishoregonj	&	Dhaka-Noakhali	&	 0.110 	&	Dhaka-Madaripur	&	Gazipur-Panchagarh	&	 0.078 	\\
Dhaka-Narsingdi	&	Dhaka-Noakhali	&	 0.109 	&	Dhaka-Bhola	&	Dhaka-Madaripur	&	 0.077 	\\
Dhaka-Noakhali	&	Dhaka-Shariatpur	&	 0.109 	&	Dhaka-Bhola	&	Dhaka-Gopalganj	&	 0.077 	\\

    \bottomrule														
    \end{tabular}
    }
    }
    \begin{minipage}{0.9\textwidth}
\vspace{1pt} {\footnotesize Notes: Avg. Welfare refers to the average across all candidate sites of the per-migrant welfare increase from site-level implementation recommendations, predicted under our preferred mixed prior with 0.5 weight on the structural model.
\par}
\end{minipage}
\end{table}

\section{Specifying the Cost of Remitting Through Mobile Money}
\label{sec:gamma-def}

\begin{figure}[h]
    \centering
    \caption{Cost Schedules for Local Digital Money Transfer Services}
    \label{fig:gamma_calibrate}
    \begin{minipage}{0.5\textwidth}
    \includegraphics[width = \textwidth]{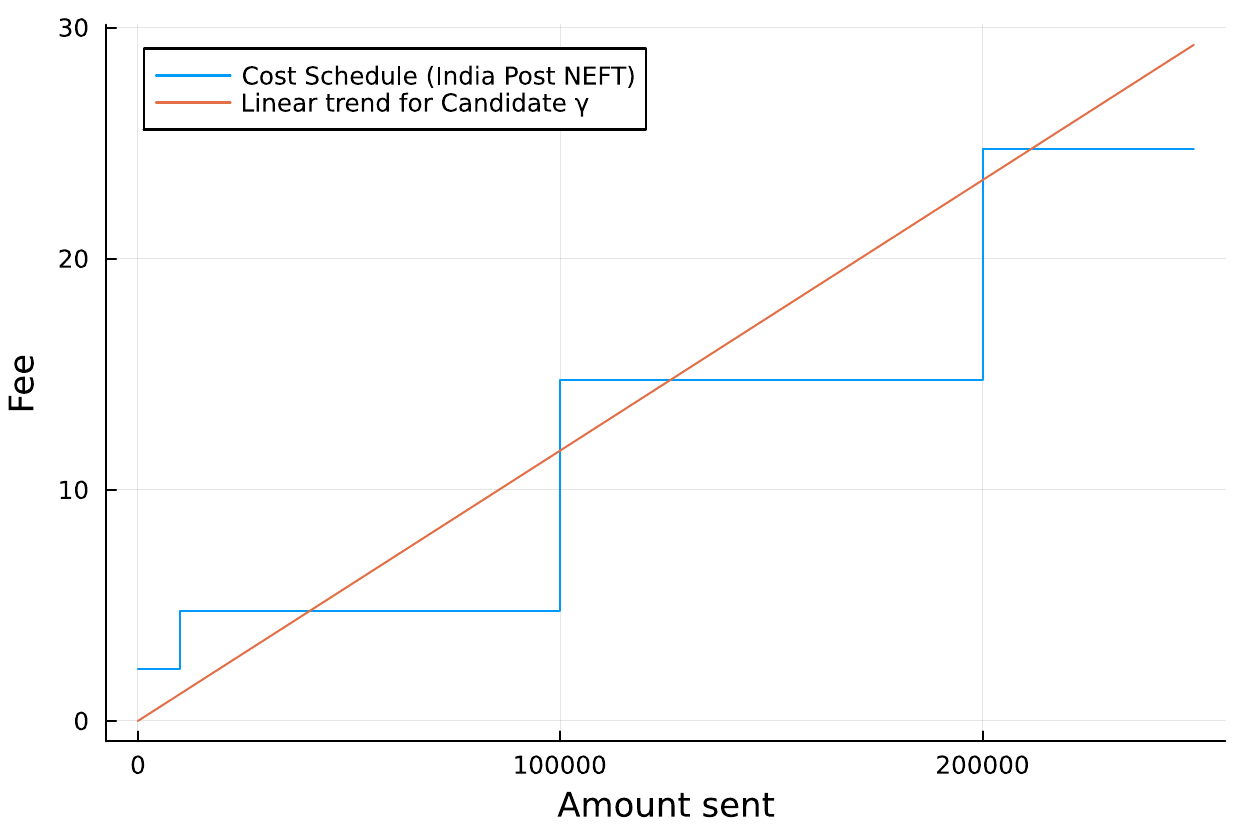}
    \end{minipage}\hfill
    \begin{minipage}{0.5\textwidth}
    \includegraphics[width = \textwidth]{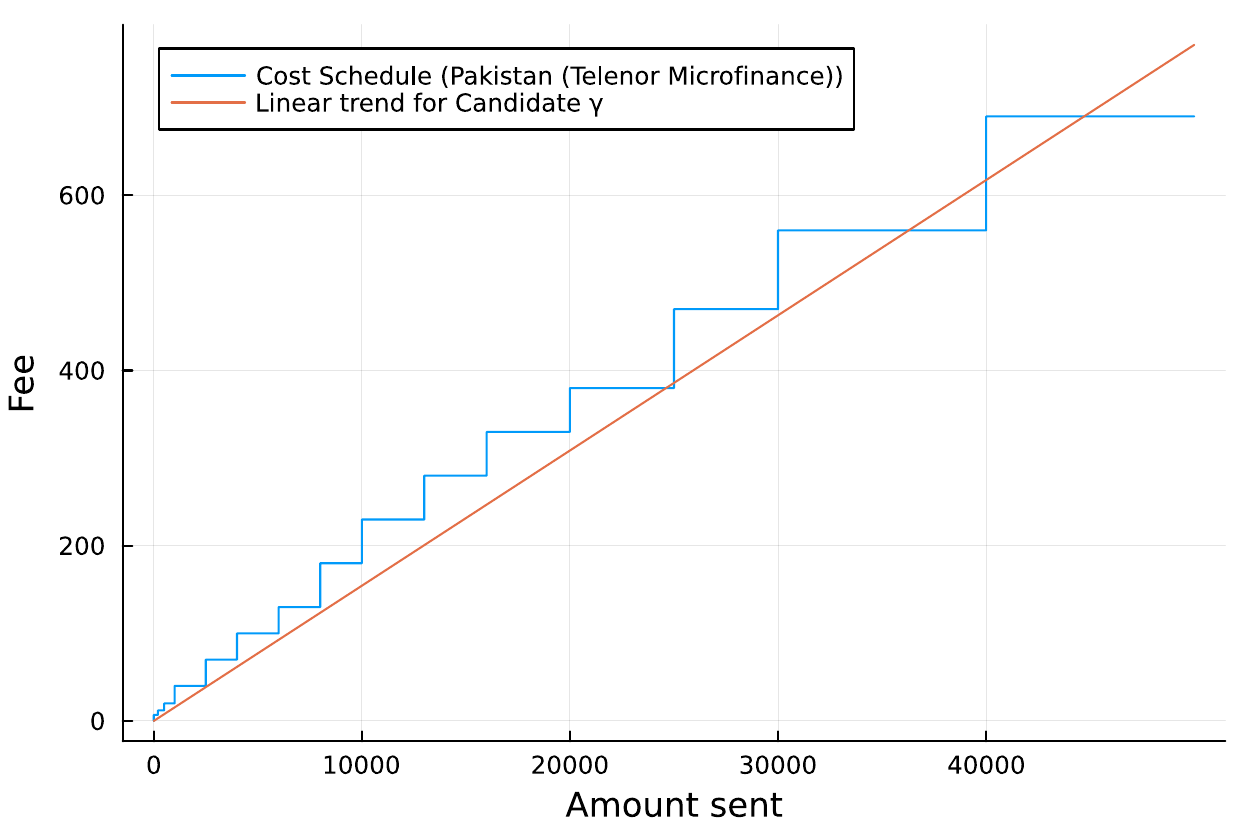}
    \end{minipage}
\end{figure}

The structural model laid out in Section \ref{sec:model} requires a cost of remitting through mobile money $\gamma$. For Bangladesh we use the figure from \cite{Lee2018}. For India and Pakistan we use cost schedules for digital money transfers using local services to calibrate a value for $\gamma$. For India we use the fee schedule for electronic fund transfers between savings accounts at India Post Offices and for Pakistan we use the cost of transferring funds digitally using Telenor Microfinance Bank services. The variable $\gamma$ is a scalar but the cost schedules above do not charge a flat fee for all transfers. We calibrate $\gamma$ as an approximate linear trend which represents the respective cost schedules. Figure \ref{fig:gamma_calibrate} shows the cost schedules for local digital money transfer options and superimposes a linear trend corresponding to the country specific $\gamma$ we use for the structural model. The value of $\gamma$ for Bangladesh, India and Pakistan are 0.02, 0.001 and 0.015, respectively.  

\section{Price Index Computation}

For each locality $l$ and a food product $j \in \{1, \ldots, J\}$, each country's expenditure survey dataset provides two key pieces of information for each household $i$ surveyed in this $(l,j)$ pair: quantity $q_{lji}$ and expenditure $\exp_{lji}$. Let $N_l$ be the total number of households surveyed in locality $l$. Then, for each product in each locality, we can obtain the average price $avg\_p_{lj} = \frac{\sum_{i \in l} exp_{lji}}{q_{lj}}$ and average consumption $avg\_q_{lj} = \frac{\sum_{i \in l} q_{lji.}}{N_l}$. 
 
For any migration corridor defined by an origin locality $l = o$ and destination locality $l = d$, define the food consumption basket at origin $o$ as the vector $q_o \equiv \left(avg\_q_{o1}, \dots, avg\_q_{oJ}\right)$. The corresponding price vector at origin is $p_o \equiv \left(avg\_p_{o1}, \dots, avg\_p_{oJ}\right)$ and the price vector at the destination $d$ is $p_d \equiv \left(avg\_p_{d1}, \dots, avg\_p_{dJ}\right)$. 
The Laspeyres price index we use is simply the the expenditure on the origin basket at origin prices (given by $p_o'q_o$) divided by expenditure on the origin basket at destination prices (given by $p_d'q_o$).

\end{document}